\documentclass{article}
\usepackage{graphicx}
\usepackage{amsmath}
\usepackage{amsfonts}
\usepackage{amssymb}
\usepackage{caption}
\usepackage{subcaption}
\usepackage{siunitx}
\usepackage[T1]{fontenc}
\usepackage{cite}
\usepackage{booktabs}
\usepackage[normalem]{ulem}
\usepackage[dvipsnames]{xcolor}

\usepackage{algorithm}
\usepackage{algpseudocode}

\usepackage[
left=1in,
right=1in,
bottom=0.75in,
top=1in,
footskip=0.5in,
]{geometry}

\newcommand{\CTCsub}{\textsc{ctc}}
\newcommand{\NNsub}{\textsc{nn}}

\usepackage{algorithm}
\usepackage{algpseudocode}
\title{Online Optimisation of Machine Learning Collision Models to Accelerate Direct Molecular Simulation of Rarefied Gas Flows}
\author{Nicholas Daultry Ball\thanks{Mathematical Institute, University of Oxford, Oxford OX2 6GG, UK} $^{,}$\thanks{Corresponding author, nicholas.daultryball@maths.ox.ac.uk}
\and
Jonathan F. MacArt\thanks{Department of Aerospace and Mechanical Engineering, University of Notre Dame, Notre Dame, IN 46556, USA}
\and 
Justin Sirignano\footnotemark[1]
}

\begin{document}
\maketitle
\begin{abstract}

We develop an online optimisation algorithm for in situ calibration of collision models in simulations of rarefied gas flows. The online optimised collision models are able to achieve similar accuracy to Direct Molecular Simulation (DMS) at significantly reduced computational cost for 1D normal shocks in argon across a wide range of temperatures and Mach numbers. DMS is a method of simulating rarefied gases which numerically integrates the trajectories of colliding molecules. It often achieves similar fidelity to Molecular Dynamics (MD) simulations, which fully resolve the trajectories of all particles at all times. However, DMS is substantially more computationally expensive than the popular Direct Simulation Monte Carlo (DSMC) method, which uses simple phenomenological models of the collisions. We aim to accelerate DMS by replacing the computationally costly Classical Trajectory Calculations (CTC) with a neural network collision model. A key feature of our approach is that the machine learning (ML) collision model is optimised online during the simulation on a small dataset of CTC trajectories generated in situ during simulations. The online Machine Learning DMS (ML-DMS) is able to reproduce the accuracy of MD and CTC-DMS for 1D normal shocks in argon at a significantly lower computational cost (by a factor of $\sim5$--$15$), at a wide range of physical conditions (Mach numbers $1.55\leq \textup{Ma}\leq 50 $, densities $\sim\qty{1e-4}{\kilogram\per\cubic\metre}$ to $\qty{1}{\kilogram\per\cubic\metre}$, and temperatures \qty{16}{\kelvin} to \qty{300}{\kelvin}). We also derive an online optimisation method for calibration of DSMC collision models given a model of the interatomic forces. In our numerical evaluations for 1D normal shocks, the online optimisation method matches or significantly improves the accuracy of VHS (Variable Hard Sphere) DSMC with respect to CTC-DMS (with a $\sim20 \times$ lower computational cost). 
\end{abstract}

\section{Introduction}
Many engineering applications, such as high altitude aerodynamics and micro-channel flows, involve non-continuum conditions where the Navier--Stokes equations of fluid dynamics are not valid, and a kinetic description is required. The Direct Simulation Monte Carlo (DSMC) method is a popular and accurate method for the numerical simulation of gas flows in these rarefied conditions \cite{bird_book,boyd_schwartzentruber_2017}. However, simulations usually rely on phenomenological models for intermolecular collisions, as well as for modelling processes such as the transfer of rotational and vibrational energy. By contrast, Molecular Dynamics (MD) simulations attempt to fully resolve the positions and velocities of all the atoms in the system, requiring only a model for the interatomic forces or potential energy surfaces (PES), in order to integrate the atomic trajectories. MD simulations of rarefied flows are now possible for many low-dimensional configurations such as the 1D shock using parallel computing resources and algorithms such as the Event-Driven/Time-Driven MD approach \cite{ED_TD_MD}. However, these approaches remain infeasible for complex, three-dimensional flows. DSMC models such as the widely used Variable Hard Sphere (VHS) model can be used to successfully reproduce the results of MD simulations, but in order to do so the model parameters must be manually adjusted with direct reference to the resulting shock profiles \cite{MDDilute}. 

Direct Molecular Simulation (DMS) has been proposed in order to incorporate fully-resolved integration of molecular collisions into the DSMC framework \cite{Koura_1997,Koura_N2_eq,Koura_N2_rel}, and a parallelised, GPU-accelerated implementation is presented in \cite{NORMAN2013153}. DMS relies only on a PES or interatomic potential, and in particular the method is much less sensitive than DSMC to the physical conditions of the simulation. In DMS, pairs of simulation particles, each representing a large number of physical molecules, are chosen to undergo collisions according to an acceptance-rejection algorithm, as is usual in DSMC, but the outcomes of collisions are resolved by direct integration of molecular trajectories, referred to as Classical Trajectory Calculation (CTC). This method saves a large part of the computational cost associated with MD, by requiring that the molecular trajectories be integrated only for those particles selected to collide, and is able to reproduce MD results accurately for 1D normal shocks \cite{MDDilute,NORMAN2013153}. DMS has also been used to simulate hypersonic flows around a cylinder \cite{DMS_cylinder} and double cone \cite{DMS_double_cone}. However, it is still significantly more computationally expensive than standard DSMC, where the outcome of a given collision is typically determined by a simple phenomenological model such as VHS. The objective of this paper is to reduce the computational cost of DMS by replacing the trajectory calculations with a collision model which is less costly to evaluate, while retaining the accuracy of the method. As an initial step towards the goal of efficient data driven DMS for rarefied flows in complex real world applications, this work introduces an online approach to training of collision models for DMS and DSMC, applied to the 1D normal shock problem in argon. We use an artificial Neural Network to model the collisions, but emphasise that the methodology does not depend on the particular model parametrisation used. (For example, our method is also applicable to widely-used DSMC collision models such as VHS.) The model parameters are obtained by training against CTC data, and the results benchmarked against a range of standard approaches. 

Machine learning is now widespread in the modelling of fluid flows \cite{turbulence_data}, with applications including methods for the solution of the governing PDEs with neural networks \cite{DGM, PINN}, data driven discovery of underlying physics \cite{learning_PDE}, and optimisation over closure models \cite{LES_closure_opt,sanderse_turbulence}.
Applications of machine learning to rarefied gas flows are not well developed as compared to continuum flows. Neural networks have been applied to reconstruct ab initio potential energy surfaces for $\textup{N}_2$--$\textup{N}_2$ collisions for use in DMS \cite{NNFit}, with improved computational efficiency compared to standard polynomial models. Machine learning has also been used to learn forces for use in MD simulations of $\textup{H}_2\textup{O}$ molecules \cite{MCDiffusion}. In both of these previous works machine learning is used to model the interactions between particles, but collisions must still be resolved with trajectory calculations. By contrast, we propose a method to entirely replace the trajectory calculations with a model trained to reproduce the dynamics for a given interatomic potential.

In \cite{Purdue_DSMC}, the authors used a polynomial approximation to the scattering angle of the Lennard--Jones (LJ) potential and applied the resulting model in DMS calculations. While in principle simpler than using a neural network model (with around 150 parameters), the construction of the approximation involved manual choices of regions for a piecewise polynomial model. The polynomial approximations are also obtained only for relative velocities less than a fixed maximum value, although this quantity is unbounded during the DMS procedure. The polynomial approximation approach is therefore not easily generalisable to other potentials. While the polynomial approximation was able to reproduce the value of the viscosity for a Lennard--Jones potential in subsonic Couette flow, the restrictions on the input values mean it may break down entirely in high temperature or high Mach number flows where large values of the collision energy are more likely, such as the shock. Parametrising our model with a neural network, trained online to ensure that training data is relevant to the given physical conditions, means the present approach is agnostic to the particular potential model used in DMS and has the potential for generalisation to collisions with rotational and vibrational degrees of freedom. The online training method also ensures the model encounters the correct range and distribution of the inputs for the physical conditions, with no manual parameter tuning required.

We first train neural networks to replace the trajectory integration procedure in DMS. Models receive as inputs the kinetic energy and impact parameter of a collision, and produce the resulting scattering angle. The use of such a collision model to replace CTC in the DMS procedure is referred to as ML-DMS. While neural networks trained offline on a given data set are shown to match very well to CTC-DMS and MD data for ``in-sample'' physical conditions, we also observe that they can fail to generalise well out-of-sample.

The standard approach in scientific machine learning is to train models based on experimental/numerical data for a a certain range of physical parameters, and hope for the resulting model to accurately generalize to out-of-sample physical regimes (on which the model was not trained). The size of the training dataset is often limited as compared to traditional uses of machine learning such as image recognition, due to the computational cost of generating high fidelity data and the cost of running experiments. Such ``offline'' training can suffer from overfitting to the training data and may fail to accurately generalize to out-of-sample physical regimes (or geometries) not included in the training dataset. Indeed, in this paper, we observe such overfitting for ML collision models trained offline for a fixed dataset (e.g. Mach 5) and then simulated out-of-sample on a new physical condition (e.g. Mach 10). To address this challenge for scientific ML models, we develop an online optimisation method which calibrates collision models to data generated in situ during the actual predictive simulation, thereby providing data relevant to the exact physical conditions at which the prediction is made. Training data is generated periodically during the simulation by evaluating a small fraction of collisions with CTC. The model parameters are continually updated using a stochastic gradient descent algorithm and are used to evaluate the remainder of the collisions. Our results, given in Section~\ref{sec:profiles}, demonstrate that online-optimised ML collision models outperform offline-trained ML collision models and are able to closely reproduce results obtained with full CTC-DMS.

\begin{figure}
    \centering
    \begin{subfigure}[t]{0.45\textwidth}
        \includegraphics[width=\textwidth]{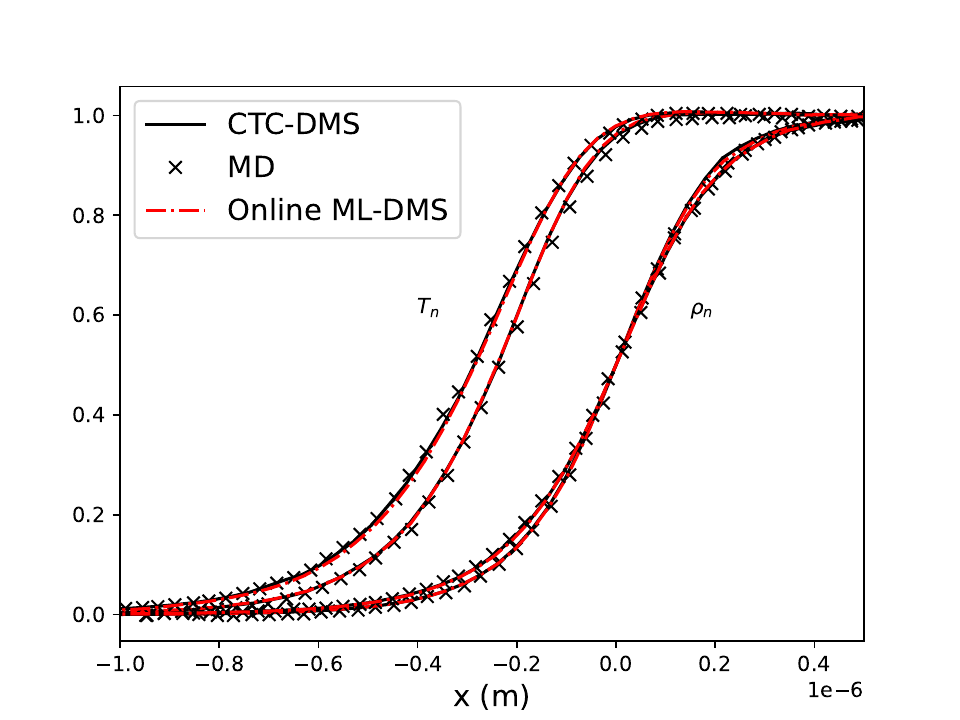}
        \caption{Mach numbers 5, 9 at $\rho_L=\qty{1}{\kilogram\per\cubic\metre}$, ${T_L=\qty{300}{\kelvin}}$.}
    \end{subfigure}
    \centering
    \begin{subfigure}[t]{0.45\textwidth}
        \includegraphics[width=\textwidth]{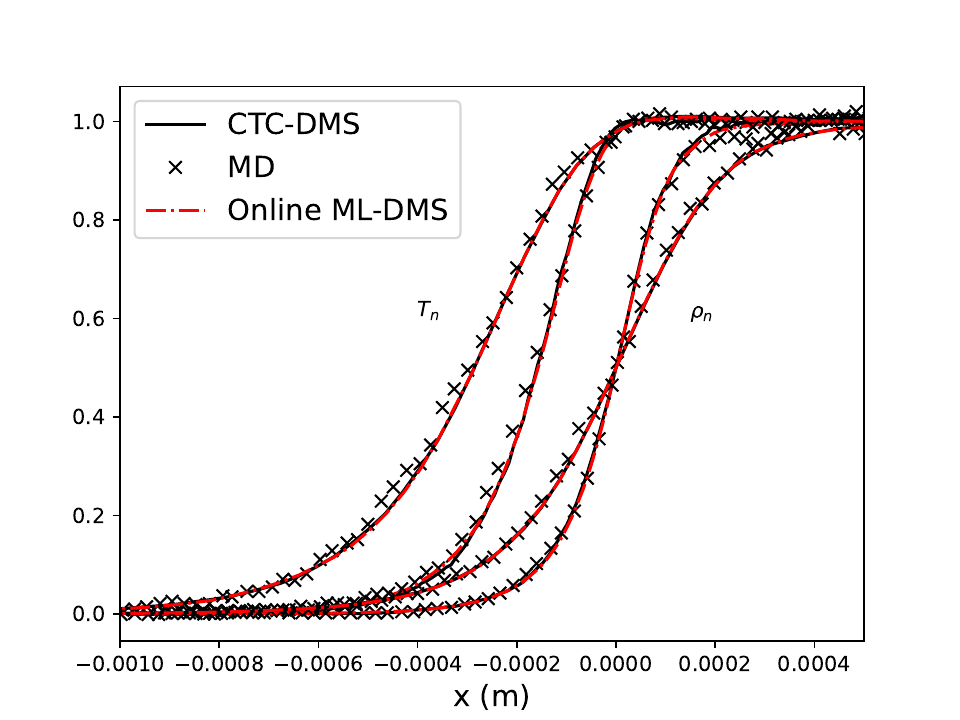}
        \caption{Mach numbers 7.183, 9 at $\rho_L=\qty{1e-3}{\kilogram\per\cubic\metre}$, $T_L=\qty{16}{\kelvin}$ and $T_L=\qty{300}{\kelvin}$ respectively.}
    \end{subfigure}
    \caption{Online ML-DMS simulations reproduce MD \cite{MDDilute} and CTC-DMS results.}
    \label{fig:intro}
\end{figure}

We develop and compare two approaches to the training of collision models during DMS. Firstly, models are trained on a subset of collisions during a relatively short initial period. The resulting model is used to evaluate the bulk of collisions during training and for the remainder of the simulation. This method is suitable if the inputs seen during training are representative of those during the rest of the simulation. In the case of the 1D shock, this is the case since the range of temperatures and densities seen in the shock is fixed by the Rankine--Hugoniot conditions. However, in more complicated scenarios this assumption may be unjustified if details of the steady state are not known ahead of time. In the second approach therefore, training steps are carried out periodically throughout the transient phase of the simulation, before averaging of the macroscopic variables begins. The model is thus trained on inputs which are guaranteed to be more representative of the final state than those encountered at initialisation. As the collision model becomes more accurate during simulation, the distribution of collisions used as training data will better reflect the true distribution, leading to further improvements in the training. Online ML-DMS reproduces CTC-DMS and MD results very closely with a significantly reduced computational cost, across a very wide range of Mach numbers ($1.55\leq \textup{Ma}\leq 50 $), densities ($\sim\qty{1e-4}{\kilogram\per\cubic\metre}$ to $\qty{1}{\kilogram\per\cubic\metre})$, and upstream temperatures (\qty{16}{\kelvin} to \qty{300}{\kelvin}). Figure~\ref{fig:intro} shows a subset of the full results set out in Section~\ref{sec:profiles}. The online training produces a very small overhead ($\sim 5\%$) compared to use of a pre-trained neural network model, and is around an order of magnitude faster than full CTC-DMS.

\begin{figure}
    \centering
    \begin{subfigure}[t]{0.45\textwidth}
        \includegraphics[width=\textwidth]{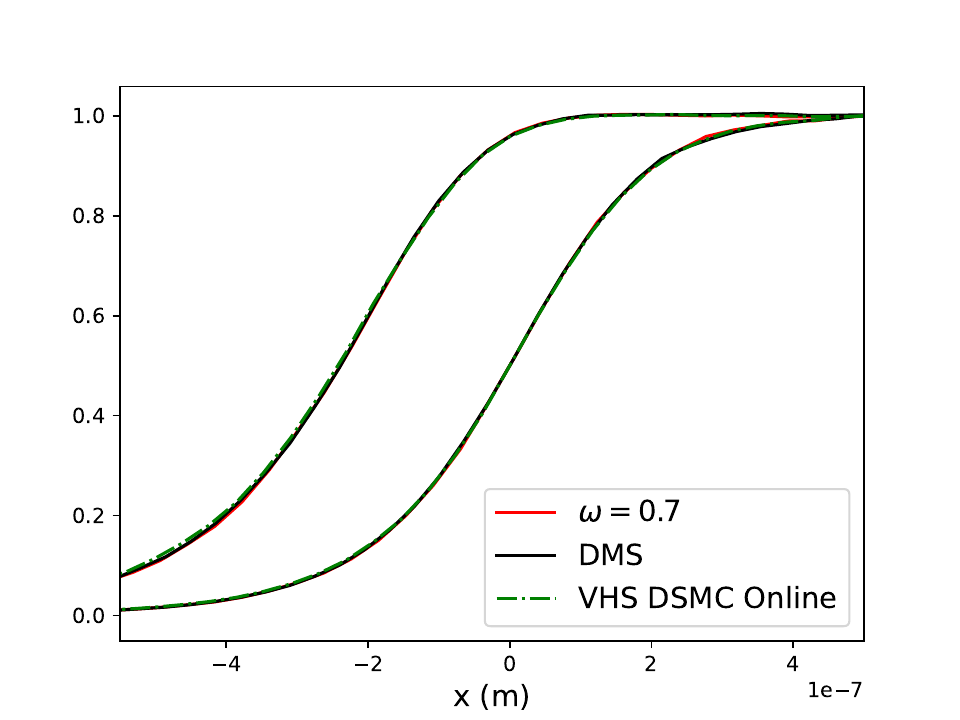}
        \caption{Mach 5 at $\rho_L=\qty{1}{\kilogram\per\cubic\metre}$, $T_L=\qty{300}{\kelvin}$.}
    \end{subfigure}
    \centering
    \begin{subfigure}[t]{0.45\textwidth}
        \includegraphics[width=\textwidth]{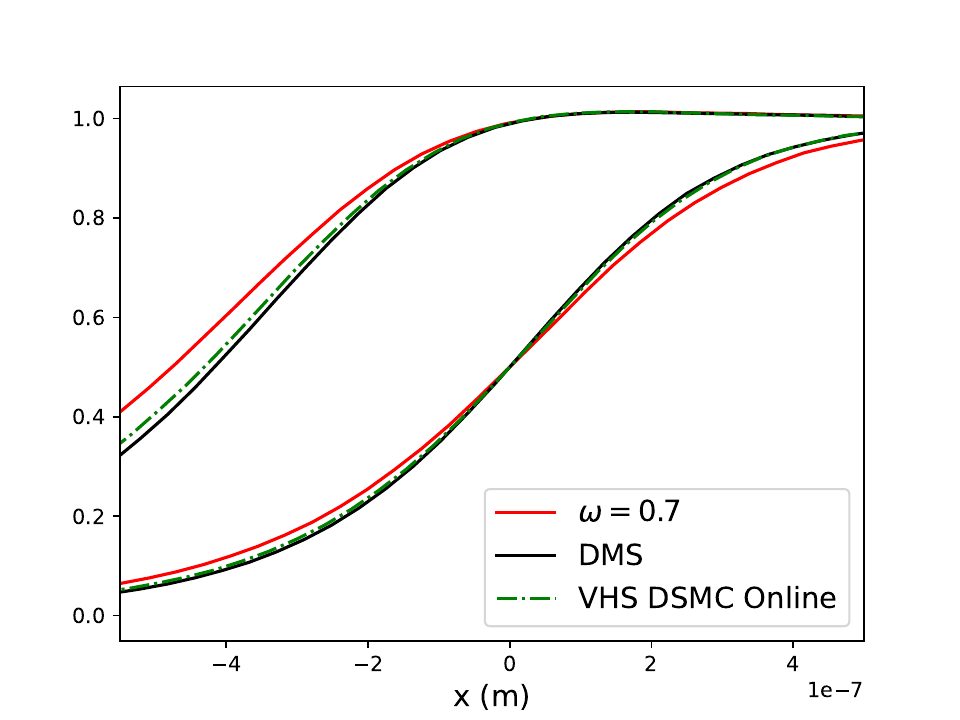}
        \caption{Mach 30 at $\rho_L=\qty{1}{\kilogram\per\cubic\metre}$, $T_L=\qty{300}{\kelvin}$.}
    \end{subfigure}
    \caption{Online optimisation of VHS DSMC reproduces or improves on DSMC with reference values. DSMC with both online training and fixed $\omega=0.7$ were carried out with reference parameters from \cite{boyd_schwartzentruber_2017} ($T_\mathrm{ref}=\qty{273}{\kelvin}$ and $d_\mathrm{ref}=\qty{3.974e-10}{\metre}$).}
    \label{fig:intro_VHS}
\end{figure}

Finally, we apply the online training procedure to DMSC for the calibration of parameters in the widely used VHS collision model, using the outcome of collisions obtained from CTC with the LJ potential. We derive a method for optimising the expected collision angle for a given DSMC step against trajectory data. The optimisation procedure is able to reproduce the VHS parameters found by manual comparison to CTC-DMS shock profiles, while requiring only $\sim 3$--$4$ times more computational time than a single DSMC simulation, and $\sim 20$ times less time than full CTC-DMS. Figure~\ref{fig:intro_VHS} shows example shock profiles obtained from CTC-DMS, online VHS and offline VHS DSMC using reference values from the literature, with full results for a wide range of Mach numbers in Section~\ref{sec:VHS_profiles}. We note that the online model training reproduces the reference values where they are accurate at low Mach numbers, and gives significantly improved accuracy relative to DMS at high Mach numbers.

A method of optimisation over DSMC using an adjoint gradient estimator has recently been developed in \cite{CAFLISCH2021110404,YANG2023112247}. This method could be used for offline optimisation of the VHS parameter through direct comparison with the final shock profile. However, in this case each gradient descent step would require one forward and one adjoint DSMC simulation to be carried out. Our approach, which directly calibrates the VHS parameter to CTC data, is significantly less computationally expensive.

We note that for a given PES the VHS model parameters can be fit to reproduce exactly the temperature exponent of the viscosity cross section at a given relative velocity, or Chapman--Enskog viscosity at a given temperature \cite{boyd_schwartzentruber_2017}. However, the reference temperature and viscosity data used for calibration must be chosen carefully, especially if a wide range of temperatures will be encountered \cite{boyd_schwartzentruber_2017}. Our approach therefore provides a principled and computationally efficient method of calibrating DSMC parameters based on microscopic dynamics, complementing the existing methods of calibration to viscosity data or visual comparison to experimental results. The method is generalisable to the case of simultaneous calibration of multiple parameters. It is also applicable to models for inelastic collisions, such as the rate of elastic relaxation in the widely used Larsen--Borgnakke model and the collision kernel proposed in \cite{djordic_CMT,Djordic_PRE}.

In Section \ref{sec:dms} we review the DSMC and DMS methods. The numerical scheme used for CTC is also described. In Section~\ref{sec:setup} we describe the 1D shock simulations which will be used to numerically evaluate
our methods and models. Our CTC-DMS implementation is also verified against MD results in the literature in Section~\ref{sec:validation}. We propose to replace the CTC trajectory simulation with a neural network collision model. Section \ref{sec:NN_sec} develops online optimisation methods for training ML collision models using CTC data generated in situ during simulations. Section~\ref{sec:results} presents a detailed comparison of numerical results for our method against the literature as well as benchmark DSMC and CTC-DMS models at a wide range of physical conditions. Finally, in Section~\ref{sec:VHS}, we apply a similar online training methodology to optimise over the parameter of the VHS model. This method produces shock profiles which match more closely to those obtained from CTC-DMS than DSMC with standard parameter values, with a computational cost much lower than DMS.

\section{DSMC and Direct Molecular Simulation}
\label{sec:dms}
The DSMC and DMS methods are stochastic particle methods, which approximate the distribution of real molecules with simulation particles, each representing a large number of physical molecules $W_p$. In the case of a monatomic gas such as argon with no internal degrees of freedom, the state of each simulation particle is specified by its velocity $v$ and position $x$. The spatial domain is divided into cells, within which simulation particles are chosen to collide randomly in a manner which represents the interactions of physical particles. For details on the justification of the stochastic approach, see \cite{boyd_schwartzentruber_2017,bird_book}. In both DSMC and DMS, the evolution of the simulation particles occurs in discrete timesteps of length $\Delta t$. Each timestep consists of $C$ collision steps and a convection step. DMS and DSMC differ in the cross section used to select which particles collide and how these collisions are evaluated (i.e., the post-collision velocities).

For both methods, in a collision step, pairs of particles in the same cell are selected to collide. This portion of the algorithm follows the standard null-collision method as described in \cite{boyd_schwartzentruber_2017}. In each spatial cell, with volume $V_\mathrm{cell}$, $N_c$ pairs of computational particles are selected uniformly at random to be considered for a collision. The rate of collisions for particles with relative velocity $g$ depends on the cross section $\sigma(g)$, which is chosen differently between DSMC and DMS. Let the velocities of the $N$ particles in a cell be $\{v_i\}_{i=1}^N$, and define $\Delta v_\mathrm{max}=2\max_i \lvert v_i -\frac{1}{N}\sum_{j=1}^N v_j\rvert$ as an upper bound on the relative velocity over all pairs of particles, and $\Sigma=\sigma(\Delta v_\mathrm{max})\Delta v_\mathrm{max}$. Then $N_c$ is given by \begin{equation}
    N_c = \frac{N(N-1)W_p\Sigma\Delta t}{2 C V_\mathrm{cell}}.
\end{equation}  
The $N_c$ pairs are referred to as being chosen for a virtual collision. Each virtual collision pair $(i,i')$, is then selected to undergo a real collision with probability
\begin{equation}
    \frac{\sigma(\lvert v_i-v_{i'}\rvert)\lvert v_i-v_{i'}\rvert}{\Sigma}.\label{eqn:null-coll}
\end{equation}
This is carried out via acceptance-rejection. If a pair is accepted to undergo a real collision, its post-collisional relative velocity must be calculated. Given a new value of the relative velocity $g'$ between the two particles, the conservation of momentum through the collision gives the post-collision velocities of the particles as
\begin{align}
    v'_i&=\frac{v_i + v_{i'}}{2} + \frac{g'}{2},\\
    v'_{i'}&=\frac{v_i + v_{i'}}{2} - \frac{g'}{2}.
\end{align} 
After the collision steps are carried out and the particle velocities updated, the particle positions are updated according to 
\begin{equation}
    x_i(t+\Delta t) = x_i(t) + v_i \Delta t.
\end{equation}
The steps described above are common between DSMC and DMS. The methods differ in the choice of cross section $\sigma$ and the method of calculating the post-collision relative velocity.

\subsection{VHS DSMC}
The VHS model is commonly used for collisions in DSMC simulations \cite{bird_book}. It is a phenomenological model which is not intended to accurately describe the microscopic collision dynamics. The model is instead designed to be simple, while incorporating the experimentally observed fact that the cross section for collisions decreases with the relative velocity.

The VHS cross section is given by the hard sphere expression $\pi d^2$, with a diameter $d$ varying as a function of the relative velocity: \begin{equation}
    \sigma(g)=\pi d_\mathrm{ref}^2\left(\frac{g_\mathrm{ref}}{g}\right)^{2\omega-1}.
\end{equation}
The cross section therefore decreases with the relative velocity, with the dependence given by the parameter $\omega$. Taking $\omega=\frac{1}{2}$ corresponds to a hard sphere gas with a fixed diameter, and $\omega=1$ gives the Maxwell molecule model. Defining $\nu=\omega-\frac{1}{2}$, the reference velocity is given in terms of a reference temperature $T_\mathrm{ref}$ by 
\begin{equation} 
    g_\mathrm{ref}^{2\nu}=\frac{1}{\Gamma(2-\nu)}\sqrt{\frac{2 k_B T_\mathrm{ref}}{m}}^{2\nu},
\end{equation} 
with $m$ being the reduced mass of a colliding pair.
If a pair of particles is accepted for a real collision, the scattering is then assumed to be isotropic, with the direction of the post collision relative velocity chosen uniformly on the unit sphere. Explicitly, this is
\begin{align}
    g'_x&=g\cos\phi\sin\theta,\\
    g'_y&=g\sin\theta\sin\phi,\\
    g'_z&=g\cos\theta,
\end{align}
with $\theta=\arccos\left(2R_1-1\right)$ and $\phi=2\pi R_2$, for $R_1$, $R_2$ drawn independently, uniformly from $[0,1]$.

\subsection{DMS}
DMS differs from DSMC in that the dynamics of each collision are intended to represent the microscopic dynamics with a given interaction potential. For CTC-DMS this is carried out by integrating the classical equations of motion. 

A collision between two argon atoms is uniquely specified by the collision energy $e$, which is a function of the relative velocity, and the impact factor $b$, which can be understood as the perpendicular distance between the initial particle trajectories in the centre of mass frame. Since in a stochastic method such as DMS the positions assigned to simulation particles do not correspond to those of underlying physical particles, the impact parameter must be chosen at random.

For argon, without any internal degrees of freedom, the collision amounts to obtaining a collision angle $\chi\in[0,\pi]$ through which the relative velocity is rotated. Details of the CTC procedure for doing so are provided in Section~\ref{sec:CTC}.

In argon the collision dynamics are completely specified by a single function of two variables. Therefore, if this function could be exhaustively tabulated, the DMS procedure could be implemented with a lookup table similar to those used to model chemical processes in reacting or combusting flows \cite{PIERCE_MOIN_2004}. We do not explore this approach here, since it would not be feasible to generalise to molecules with internal degrees of freedom. For example, in diatomic molecules, even neglecting vibrational modes, there are seven inputs for each trajectory calculation, making the resulting table impractically large. For example to use 50 values of each input would require $50^7$ table entries, and several terabytes of memory. Therefore, it is necessary in practice to calculate the trajectory outcomes during the simulation, or to use a coarse-grained ``state-to-state'' method which involves additional modelling assumptions and may introduce inaccuracies \cite{DMS_cylinder,Panesi_coarse_grain_1,Panesi_coarse_grain_2}.

Once $\chi$ has been calculated, the post collision relative velocity $g'$ is obtained from the pre-collision value $g$ as follows \cite[Equation 2.22]{bird_book}:\begin{align}
\label{eqn:collision}
    g'_x &= \cos(\chi)g_x + \sin(\chi)\sin(\epsilon)\left(g_y^2+g_z^2\right)^{\frac{1}{2}},\\
    g'_y &= \cos(\chi)g_y + \sin(\chi)\frac{\lvert g\rvert g_z\cos(\epsilon)-g_xg_y\sin(\epsilon)}{(g_y^2+g_z^2)^{\frac{1}{2}}},\\
    g_z'&= \cos(\chi)g_z - \sin(\chi)\frac{\lvert g\rvert g_y\cos(\epsilon)+g_xg_z\sin(\epsilon)}{(g_y^2+g_z^2)^{\frac{1}{2}}},
    \label{eqn:post_collision}
\end{align}
where $\epsilon$ is an angle which may be chosen uniformly at random due to the rotational symmetry of the collision around the axis of the relative velocity.

Therefore, for complete specification of the algorithm it is necessary to choose a model for the cross section appearing in Equation~\ref{eqn:null-coll}, a method of selecting the impact parameter $b$ for each collision, and finally a function $(e,b)\mapsto\chi$.

The collision cross section governs the rate at which pairs of molecules at a given relative velocity are selected to undergo a collision. Since in CTC-DMS the collision outcomes are determined by the details of the trajectory calculation, the cross section is chosen to be conservatively large. The cross section used in this paper is determined by the maximum impact parameter resulting in an appreciable collision, 
\begin{align}
    \sigma(g)&=\pi b_\mathrm{max}^2(g),\\
    b_\mathrm{max}(g) &= Ag^{B},
\end{align}
where the power law parametrisation of the maximum impact parameter, with $A=69$, $B=-\frac{1}{3}$ as used in \cite{NORMAN2013153}, was found to minimise the number of collisions resulting in a very small change in relative velocity. 
As in \cite{NORMAN2013153} the impact parameter for each collision is then chosen to be consistent with the cross section, as $b=R^{\frac{1}{2}}b_\mathrm{max}$, where $R$ is a standard uniform random variable. 

The CTC-DMS procedure used as a benchmark in this paper is completely specified for reference in Algorithm~\ref{alg:DMS}. Details of the CTC calculations to specify the function $\chi_\CTCsub(e,b)$ are given in Section~\ref{sec:CTC}. In general, we refer to DMS using an ML trajectory model $\chi_\NNsub(e,b;\theta)$ as ML-DMS (see Section~\ref{sec:ML_DMS}), where $\theta$ are parameters which must be trained/calibrated. We emphasise that the choice of a neural network model for the collisions is only made as a flexible nonlinear parametric model which is easy to train. A large class of alternative collision models -- including widely-used DSMC collision models such as VHS -- could be used within the framework of this paper and their parameters can be trained/calibrated with our online optimisation method.

\begin{algorithm}
\caption{Direct Molecular Simulation}
\label{alg:DMS}
    \begin{algorithmic}
    \Require $\Delta t$, $T$, $n_\mathrm{cells}$, collision steps per DSMC step $C$, physical particles per simulation particle $W_p$
    \Require cross section $\sigma(v)$, max impact parameter $b_\mathrm{max}(v)$
        \State $\{v_i\}\gets \text{Initial velocities}$
        \State $\{x_i\}\gets \text{Initial positions}$
        \While{$t<T$}
            \State{$c\gets 0$}
            \While{$c<C$}
                \For{\text{each cell}}
                    \State $N\gets$ number of particles in cell
                    \State $V_c \gets$ volume of cell
                    \State {$\Delta v_\mathrm{max}\gets 2\max_i \lvert v_i -\frac{1}{N}\sum_j v_j\rvert$ within the cell } 
                    \State $\Sigma\gets\Delta v_\mathrm{max}\sigma(\Delta v_\mathrm{max})$
                    \State $N_c\gets \frac{N(N-1)W_p \Delta t}{2V_{c}C}$
                    \State Uniformly select $N_c$ virtual collision pairs
                \EndFor
                \For{each virtual pair $(i,i')$}
                        \State{Draw $R_1,R_2\sim \mathcal{U}[0,1]$}
                        \State $\sigma_{i}=\lvert v_i-v_{i'}\rvert\sigma(\lvert v_i-v_{i'}\rvert)$
                        \If{$\sigma_{i}>R_1\Sigma$}
                        \State Collision $i$ accepted
                        \State $e_i \gets \frac{\lvert v_i-v_{i'}\rvert^2\mu}{2}$
                        \State $b_i\gets \sqrt{R_2}b_\mathrm{max}(\lvert v_i-v_{i'}\rvert)$
                        \EndIf
                    \EndFor
                \For{each accepted collision pair $(j,j')$}
                    
                    \State $\chi_j\gets \chi_\CTCsub(e_j,b_j)$ or $\chi_\NNsub(e_j,b_j;\theta)$

                    \State Use $\chi_j$ to update $v_{j,}, v_{j'}$ according to Equation~\ref{eqn:post_collision}
                \EndFor
                \State{$c\gets c+1$}
            \EndWhile
            \State{$x_i \gets x_i + v_i \Delta t$} 
            \State $t\gets t+\Delta t$
        \EndWhile
    \end{algorithmic}
\end{algorithm}

\subsubsection{Trajectory Calculations} \label{sec:CTC}
The CTC-DMS method calculates $\chi_\CTCsub(e,b)$ through numerical integration of a trajectory with a given potential. Trajectory calculations are carried out to solve Newton's equations of motion,\begin{align}
    \dot{x_i}&=v_i,\\
    m_i\dot{v_i}&=\frac{\partial\phi(\lvert x_1-x_2\rvert)}{\partial r},
\end{align} for $i=1,2$ and a given interatomic potential $\phi(r)$. We use the velocity Verlet integrator \cite{verlet}, with update at each time step given by
\begin{align}
    x_i(t+\Delta t)&=x_i(t)+v_i(t)\Delta t \pm \frac{f_1}{2m_i}\Delta t ^2,\\
    v_i(t+\Delta t) &=v_i(t) \pm\frac{f_1+f_2}{2m_i}\Delta t.
\end{align} The opposite sign in $\pm$ is chosen for each particle to conserve momentum. The forces are those estimated at the start and end of the timestep \begin{align}
    f_1 &= \frac{\partial\phi(\lvert x_1(t)-x_2(t)\rvert)}{\partial r},\\
    f_2 &= \frac{\partial\phi(\lvert x_1(t+\Delta t)-x_2(t+\Delta t)\rvert)}{\partial r}.
\end{align} The timestep was taken as  $\Delta t=\qty{1}{\femto\second}$, which was observed to be sufficient to conserve energy during the collision. Each collision often involves several thousand Verlet timesteps, whereas for DSMC collisions require only a small number of operations. For this reason, CTC-DMS simulations require several orders of magnitude more computational time than DSMC. Simulations were initialised and impact parameters chosen according to the procedure below, following \cite{NORMAN2013153}. 
\begin{itemize}
    \item One particle was initially placed at the origin, with the other particle at $x$-position $D_\mathrm{cutoff}=4\sigma_\mathrm{LJ}$, displaced by a distance $b$ from the $x$-axis at an angle $\epsilon$ chosen uniformly at random.
    \item The particles were given a relative velocity aligned along the $x$-axis of $g = \sqrt{{2e}/{\mu}}$, with $\mu={m_\mathrm{Ar}}/{2}$ the reduced mass of the collision pair.
    \item The simulation was terminated when the particles again reached a distance of $D_\mathrm{cutoff}$ apart, or when the trajectory time reached the DSMC timestep size. The collision angle was then calculated as the angle between the $x$-axis and the final velocity of the particle initially placed at the origin.
\end{itemize}

In our DMS simulations, trajectory calculations use the Lennard--Jones potential\begin{equation}
    \phi(r) = 4\epsilon_\mathrm{LJ}\left(\left(\frac{\sigma_\mathrm{LJ}}{r}\right)^{12}-\left(\frac{\sigma_\mathrm{LJ}}{r}\right)^{6}\right),
\end{equation} with the same parameter values as were used in \cite{NORMAN2013153,MDNobleMixtures} (see Table~\ref{table:LJParams}).

\begin{table}
\centering
\begin{tabular}{ cccc } 
 \toprule
 &&&\\[-1em] & $\epsilon_\mathrm{LJ}$ (K) & $\sigma_\mathrm{LJ}$ (\AA) & Molecular Mass (Au) \\ 
 \midrule
 &&&\\[-1em] Ar & 119.18 & 3.42 & 39.9\\ 
 \bottomrule
\end{tabular}
\caption{Parameters for Lennard--Jones potential used in trajectory calculations} \label{table:LJParams}
\end{table}
\subsection{Experimental Setup: Physical Conditions and Numerical Simulation Details}
\label{sec:setup}
The DSMC and DMS methods were implemented for the 1D normal shock in argon. The physical conditions are specified by the upstream density $\rho_L$, upstream temperature $T_L$ and Mach number, with downstream boundary conditions given by the Rankine--Hugoniot conditions. The number of simulation particles is kept constant in time. In order to enforce the boundary conditions, any particles leaving the domain are regenerated from a one-sided Maxwellian with moments corresponding to the shock conditions.

Unless otherwise stated, for all simulation results presented in this paper, $n_\mathrm{cells}=100$ spatial cells are used over a physical domain of length $40\lambda_L$, where $\lambda_L$ is the upstream mean free path, with $10^{6}$ simulation particles. The time step size is ${\Delta t_0=\qty{5e-12}{\second}}$ for simulations at $\rho_L=\qty{1}{\kilogram\per\cubic\meter}$, and otherwise scaled in inverse proportion to the density as $\Delta t=\frac{\qty{1}{\kilogram\per\cubic\meter}}{\rho}\Delta t_0$. Simulations are initialised with a jump between upstream and downstream equilibrium distributions determined by the Rankine--Hugoniot conditions, and run for 1,000 timesteps before averaging of the macroscopic variables is carried out for a further 500 timesteps to produce the final results. In order to reduce statistical scatter in the profiles, simulations at the lowest Mach number of 1.55, and those at cold temperatures ($T_L=\qty{16}{\kelvin}$) are instead carried out with $2\times10^{6}$ particles, and averaging of the macroscopic variables begins after 2,000 timesteps, proceeding for a further 1,000 steps. 
This simulation length is chosen conservatively in order to produce smooth and stable shock profiles, and to allow consistent comparison between simulations with different methods. It may be possible to obtain smooth shock profiles with a shorter simulation time, however we note that for a Mach 5 shock at $T_L=\qty{300}{\kelvin}$, $\rho_L=\qty{1}{\kilogram\per\cubic\metre}$, a transient stage of 500 timesteps followed by 250 averaging steps is found to be insufficient for a smooth profile in the region just downstream of the shock.

\subsection{DMS Verification}
\label{sec:validation}
To verify the implementation, CTC-DMS results are compared to MD from \cite{MDDilute}, at free stream densities of $\rho_L=\qty{1}{\kilogram\per\cubic\metre}$ and $\rho_L= \qty{1e-3}{\kilogram\per\cubic\metre}$. Temperature and density profiles for these conditions are provided in Section~\ref{sec:profiles} (Figures~\ref{fig:moderate} and \ref{fig:rarefied}) in the context of the performance evaluation of the proposed ML-DMS models. The density and temperature profiles of the present CTC-DMS calculations match very closely with MD results. 

\begin{figure}
    \centering
    \includegraphics[width=0.6\linewidth]{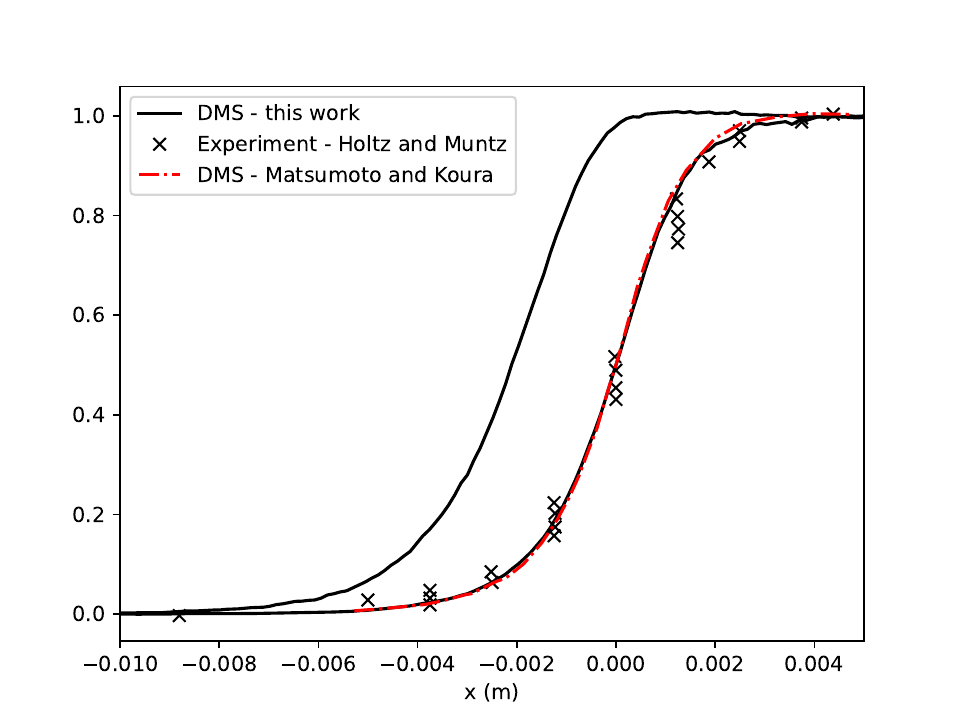}
    \caption{Comparison of density and temperature profiles obtained from DMS against the experiments of Holtz and Muntz \cite{Holtz_Muntz} and the DMS of Koura and Matsumoto \cite{Koura_argon}, carried out at Mach 7.183 with upstream number density $n_L=\qty{1.144e-21}{\per\cubic\metre}$, $T_L=\qty{16}{\kelvin}$. The upstream mean free path used to compare against the non-dimensionalised results from \cite{Koura_argon} was estimated as $\lambda_L=\qty{4.35e-4}{\metre}$. }
    \label{fig:holtz}
\end{figure}

CTC-DMS results are compared directly with experimental data in Figure~\ref{fig:alsmeyer}. For higher Mach numbers, most obviously Mach 9, the DMS does not match exactly to the available experimental data. We instead observe slight discrepancies versus experimental density profiles, namely a thinner shock at high Mach numbers, however this is likely to be a result of the particular intermolecular potential used. Similar discrepancies between CTC-DMS with the Lennard--Jones potential and experimental profiles were reported in \cite{Koura_argon} when comparing to the experimental measurements of Holtz and Muntz \cite{Holtz_Muntz}. For additional validation of the present implementation, Figure~\ref{fig:holtz} shows a comparison at these experimental conditions, corresponding closely to the profiles reported in \cite{Koura_argon}. In summary, our CTC-DMS simulations have excellent agreement with MD simulations from the literature (for the same Lennard--Jones potential) and good (although not exact) agreement with experimental data (likely due to the Lennard--Jones potential being an approximation and also, perhaps, noisy data from experiments).

Simulations were also carried out with more particles and timesteps, in order to confirm the convergence of the method, with no change in the results.

\section{Neural Network Collision Models}
\label{sec:NN_sec}

\subsection{DMS with Neural Network} \label{sec:ML_DMS}
Before presenting the online optimisation method for collision models -- which is one of the main contributions of this paper -- we first describe the architecture of the neural network collision model $\chi_\NNsub$ and the method of offline training which serves as a benchmark. The inputs for all neural network models in this paper are taken to be the non-dimensional collision energy $e^*$ and impact parameter $b^*$. We generated a dataset of Ar--Ar collisions with $e^*\in[0,100]$, $b^*\in[0,5]$, to be used in offline training. These ranges of the input values were chosen to include the vast majority of collisions observed in a Mach 5 shock at $T_L=\qty{300}{\kelvin}$.

\begin{figure}
    \centering
    \begin{subfigure}[t]{0.45\textwidth}
    \centering
        \includegraphics[width=\textwidth]{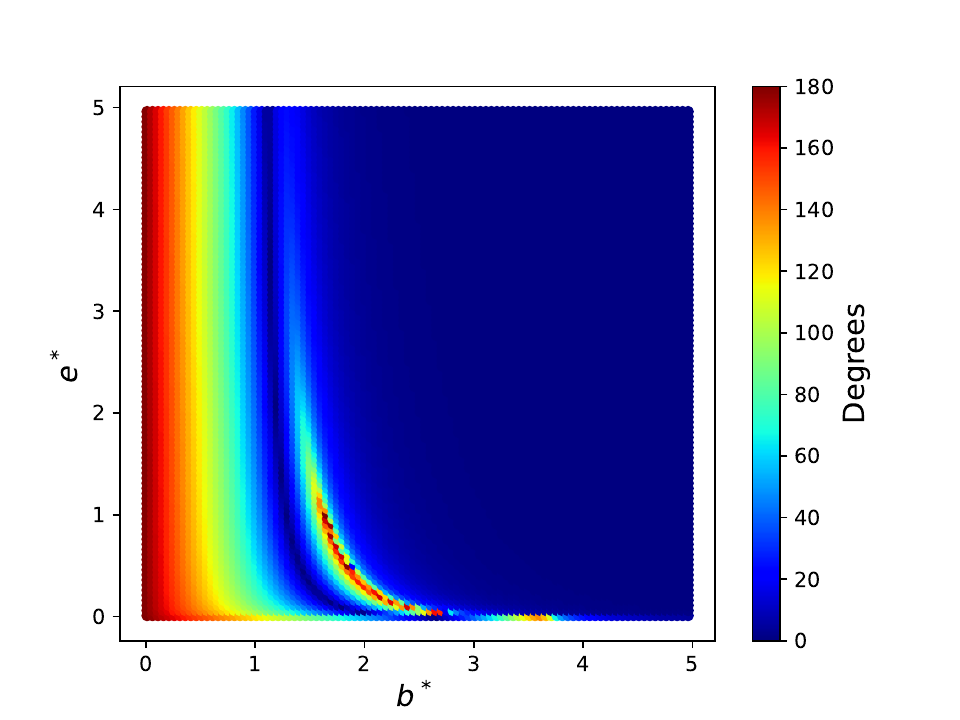}
    \caption{``Low energy'' collision dataset}
    \label{fig:enter-label}
    
    \end{subfigure}
     \begin{subfigure}[t]{0.45\textwidth}
    \centering
        \includegraphics[width=\textwidth]{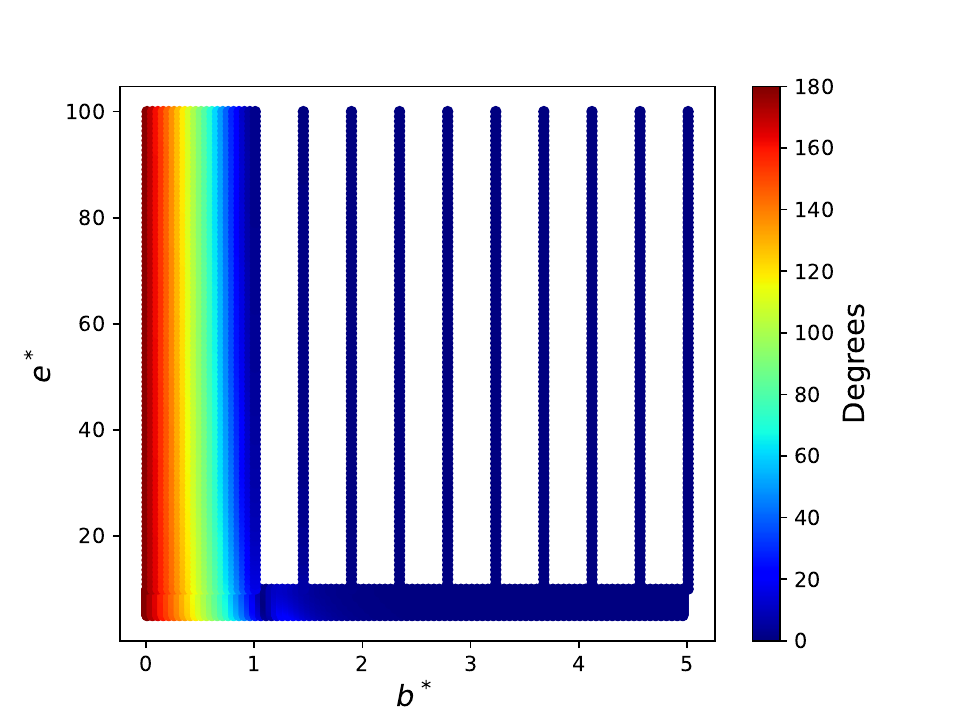}
    \caption{``High energy'' collision dataset. Blank portions of the figure were not used in training.}
    \label{fig:enter-label}
    \end{subfigure}
    
    \begin{subfigure}[t]{0.45\textwidth}
    \centering
        \includegraphics[width=\textwidth]{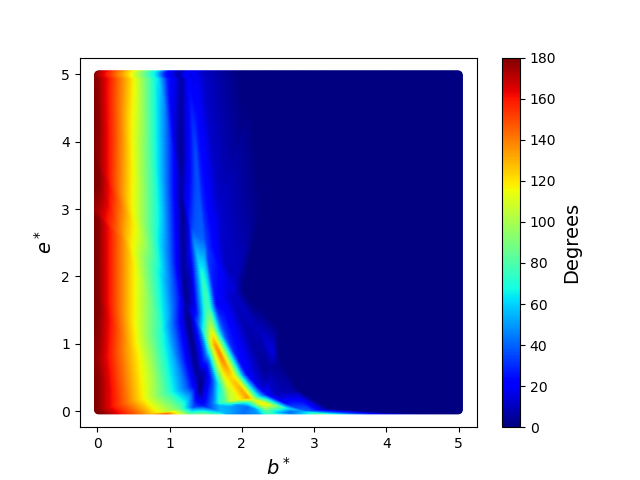}
        \caption{``Low energy'' model output}
    \end{subfigure}
\centering
    \begin{subfigure}[t]{0.45\textwidth}
    \centering
        \includegraphics[width=\textwidth]{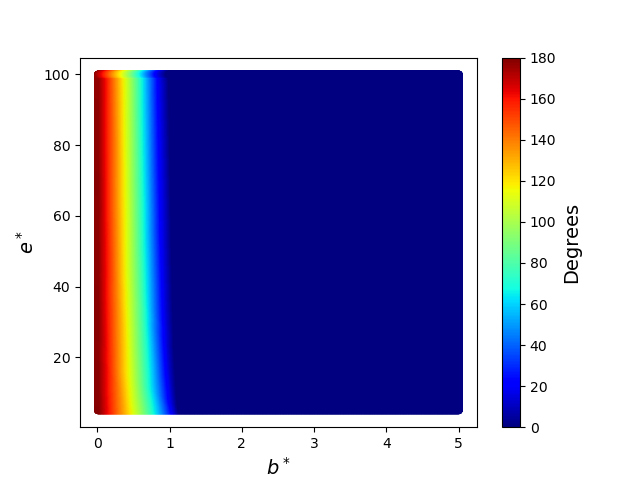}
        \caption{``High energy'' model output}
    \end{subfigure}
    \centering

    \caption{Reference data and model outputs for both low and high energy neural network models. The isolated band of high scattering angles in (a) corresponds to long lived collisions. It is around this region that the dependence on the initial conditions and collision energy is most important. Note that the trained low energy model captures the behaviour in this region quite well, despite the relative lack of training data in this region.}
    \label{fig:argon_NN_out}
\end{figure}

The offline benchmark model consists of a neural network with three fully connected layers, each with 50 units and ReLU activation functions. The neural network architecture used for offline-trained models is a function of $e^*$, $b^*$, and learned parameters $\theta$. $e^*$ can be very large, commonly up to several hundred, and is unbounded due to the stochastic nature of the simulation. The low energy regime, $0\leq e^*\leq 5$, contains a region of high sensitivity to the impact parameter $b$, shown in Figure~\ref{fig:argon_NN_out}(a), while the dependence in the high energy regime (Figure~\ref{fig:argon_NN_out}(b)) is much simpler. Therefore, separate sets of neural network parameters are trained for each regime. The collision angle $\chi$ used in offline ML-DMS was taken to be a piecewise function of $e^*$, with distinct sets of learned parameters $\theta=\{\theta_1,\theta_2\}$ in the two regimes: \begin{equation}
    \chi_\NNsub^\mathrm{offline}(e^*,b^*;\theta)=\begin{cases}
        \chi_{\NNsub}(e^*,b^*;\theta_1) & e^*\leq 5, \\
        \chi_{\NNsub}(e^*,b^*;\theta_2) & e^* > 5. \\
    \end{cases}
\end{equation}

The inputs and outputs of the network were normalised to lie within the interval $[0,1]$ to facilitate training. The training was carried out with the RMSProp algorithm, taking 4,500 and 150 epochs for the low and high energy datasets respectively, with a mini-batch size of 5 and a learning rate of $10^{-4}$. The neural network outputs are compared with the underlying data in Figures~\ref{fig:argon_NN_out}(c--d), showing that the model output closely agrees with the output of trajectory calculations. 

Shock profiles obtained from ML-DMS with the offline neural network collision model are shown in Figures~\ref{fig:moderate} ($\rho_L=\qty{1}{\kilogram\per\cubic\metre}$) and \ref{fig:rarefied} ($\rho_L=\qty{1e-3}{\kilogram\per\cubic\metre}$). Since the training data were selected for a Mach 5 shock at $\qty{300}{\kelvin}$, as expected the profiles match closely to CTC-DMS results in this case and for the cold shock at Mach 7.183, $T_L=\qty{16}{\kelvin}$. However, the temperature profile deviates significantly from the CTC-DMS at Mach 9, where the temperature of the gas and energy of the collisions are higher. The inability of the model to generalise well outside of the restricted training data set motivates the development of the online training method in the next section.

\subsection{Online Training}
\label{sec:NN}
While the neural network model trained offline shows good accuracy on shocks up to Mach 7, at a range of densities and temperatures, the results deviate significantly from the MD results in the temperature profile at Mach 9 (see Figure~\ref{fig:moderate}). Since at higher Mach numbers, collisions at high energy become more frequent, this suggests that the neural network is encountering data outside of the training range and failing to extrapolate correctly (i.e. generalisation error). It would be possible to solve this problem by training a new model for higher energy collisions, but each new model would also encounter problems at ever higher Mach numbers. Another drawback of the offline model used in the previous section is the use of a separately trained network parameters for inputs in two distinct regions.

Therefore we instead develop a new online algorithm to train a single neural network model $\chi_\NNsub(e^*,b^*;\theta)$ \emph{online} for each simulation, on CTC training data generated in situ during the simulation. In contrast to the offline model above, the online trained network uses a single set of parameters for all inputs. The online training ensures that the training data is representative of the physical conditions encountered during the current predictive simulation, in contrast to the standard approach to scientific machine learning which relies on a dataset which must be curated in advance. Trained models may also be re-used in situations where the collision parameters are expected to fall inside the learned region, such as shocks at similar Mach numbers and temperatures, even when the geometry or other aspects of the flow are different. 

Since the amount of training data required is very small compared to the total number of simulated collisions during DMS, with the model being trained on less than $0.1\%$ of the number of collisions evaluated during the simulation, this method still results in significant computational speed up versus a full CTC-DMS simulation. In particular, our results demonstrate that, given a small set of CTC trajectories generated during the simulation, the trained model is able to successfully generalise to the remaining collisions, leading to accurate overall simulation predictions. 

Our online algorithm is also able to better replicate the flexibility of DMS with respect to changes in the physical parameters. One of the advantages of DMS is that the effects of changes in the PES or interatomic potential may be investigated easily by adjusting the potential model's parameters, without changes in the simulation code. Although in the offline ML-DMS method, this would require re-training a new model on a new dataset, the online method handles the required retraining automatically, while still requiring a fraction of the computational resources of DMS. 

We present and compare two approaches for online model training. In the first method, the model is trained on a subset of collisions during a relatively short initial period, taken to be the first $n_\mathrm{steps}=20$ timesteps. The training period was chosen to allow some time for the distributions upstream and downstream of the shock to mix, and yields good results in practice. This method is expected to be suitable in the case that the distribution of inputs observed during this period is representative of those during the rest of the simulation. This is the case for the 1D shock since the initial conditions reflect the range of temperatures and velocities expected in the final steady state. However, in more complicated scenarios this assumption may be unjustified if details of the statistical steady state are not known ahead of time. In this case, it is desirable to continue model training throughout the simulation to ensure inputs are sampled from the full range required. In the second approach therefore, training steps are carried out periodically throughout the transient phase of the simulation. The model then sees a range of inputs which are guaranteed to be more representative of the final statistical distribution than those encountered at initialisation.

In both online training algorithms, up to $N_\mathrm{train}=54,000$ of the collisions at each training step are simulated using CTC, and the neural network trained for 100 epochs on this data. An epoch consists of one pass over the training dataset, each with up to 216 minibatch updates. At each step those collisions used for training are updated using the CTC procedure. During steps where training does not take place, all collisions are evaluated using the neural network.
 
 The online optimisation samples a random subset of particles, generates the CTC trajectory data, and then (in parallel to the ML-DMS) updates the collision model parameters via stochastic gradient descent steps. During each training epoch the collisions in the dataset are shuffled randomly and partitioned into minibatches of size m. One gradient descent step is taken for each minibatch. Given the CTC procedure as $\chi_\CTCsub(e,b)$, and current neural network parameters $\theta_k$, the loss function is evaluated at the $k$-th gradient descent step as 
 \begin{equation}
     L(\theta_k)=\sum_{j=1}^m(\chi_\CTCsub(e_j,b_j)-\chi_\NNsub(e_j,b_j;\theta_k))^2.
 \end{equation} 
The neural network parameters are then updated according to\begin{equation}
    \theta_{k+1}=\theta_k-\alpha_k\eta_k,
\end{equation}
where $\eta_k = \nabla_{\theta} L(\theta_k)$ is the gradient of the loss function over the minibatch, evaluated using automatic differentiation. The RMSProp algorithm is used to select an adaptive learning rate $\alpha_k$ for each minibatch, with the base learning rate for each epoch additionally scheduled to decay according to\begin{equation}
    \alpha=\alpha_0\frac{A}{B+Cl},
\end{equation} where $l$ is the total number of training epochs so far. Values of $A=B=400$, $C=1$ were found to give good results. The initial learning rate was set to $\alpha_0=\qty{1e-3}{}$, and a minibatch size of $m=250$ was found to make the training process reliable (see Section~\ref{sec:training} for details of the hyperparameter selection). 

In summary, at a single step during the online training, the parameters $\theta$ and particle data $V$, $X$ are updated according to:
\begin{itemize}
    \item Collision pairs chosen according to DSMC algorithm.
    \item If training occurs, for each epoch, for each minibatch:\begin{itemize}
        \item[$\diamond$] Loss function $L$ evaluated using CTC procedure and NN model with parameters $\theta$.
        \item[$\diamond$] Gradient $\eta=\nabla_\theta L$ evaluated using automatic differentiation.
        \item[$\diamond$] $\theta \gets \theta-\alpha\eta.$
    \end{itemize}
    \item $V \gets V(\theta)$ using Neural Network model output and Equation~\ref{eqn:post_collision}.
    \item $X\gets X + \Delta t V.$
\end{itemize}
 
The general online training procedure for DMS is described in detail in Algorithm~\ref{alg:DMS_train}. Within this algorithm the two online methods outlined here vary as follows:
\begin{itemize}
    \item Method 1: Training is carried out for 100 epochs on each of the first 20 timesteps. In Algorithm~\ref{alg:DMS_train} this corresponds to $\verb|TRAIN IF |(t<20 \Delta t )\verb| AND | (c=0)$.
    \item Method 2: Training is carried out for 50 epochs every 20th DSMC timestep, before averaging begins. In Algorithm~\ref{alg:DMS_train} this corresponds to $\verb|TRAIN IF | (\left \lfloor \frac{t}{\Delta t}\right \rfloor \mod 20 = 0)\verb| AND |(c=0)$
\end{itemize}

\begin{algorithm}
\caption{Direct Molecular Simulation with Online Training}
\label{alg:DMS_train}
    \begin{algorithmic}
    \Require $\Delta t$, $T$, $n_\mathrm{cells}$, collision steps per DSMC step $C$, physical particles per simulation particle $W_p$
    \Require cross section $\sigma(v)$, max impact parameter $b_\mathrm{max}(v)$
        \State $\{v_i\}\gets \text{Initial velocities}$
        \State $\{x_i\}\gets \text{Initial positions}$
        \While{$t<T$}
            \State{$c\gets 0$}
            \While{$c<C$}
                \For{\text{each cell}}
                    \State $N\gets$ number of particles in cell
                    \State $V_c \gets$ volume of cell
                    \State {$\Delta v_\mathrm{max}\gets 2\max_i \lvert v_i -\frac{1}{N}\sum_j v_j\rvert$ within the cell } 
                    \State $\Sigma\gets\Delta v_\mathrm{max}\sigma(\Delta v_\mathrm{max})$
                    \State $N_c\gets \frac{N(N-1)W_p \Delta t}{2V_{c}C}$
                    \State Uniformly select $N_c$ virtual collision pairs
                \EndFor
                \For{each virtual pair $(i,i')$}
                        \State{Draw $R_1,R_2\sim \mathcal{U}[0,1]$}
                        \State $\sigma_{i}=\lvert v_i-v_{i'}\rvert\sigma(\lvert v_i-v_{i'}\rvert)$
                        \If{$\sigma_{i}>R_1\Sigma$}
                        \State Collision $i$ accepted
                        \State $e_i \gets \frac{\lvert v_i-v_{i'}\rvert^2\mu}{2}$
                        \State $b_i\gets \sqrt{R_2}b_\mathrm{max}(\lvert v_i-v_{i'}\rvert)$
                        \EndIf
                    \EndFor
                
                    \If{Train model this step}
                        \For{each training epoch}
                            \State{Partitition accepted pairs up to max number $N_\mathrm{train}$ into minibatches of size $m$}
                            \For{each minibatch}
                                \State{$L\gets \sum_j\left(\chi_\NNsub(e_j,b_j;\theta) - \chi_\CTCsub(e_j,b_j)\right)^2$}
                                \State Evaluate $\nabla_\theta L$ with automatic differentiation
                                \State Choose adaptive learning rate $\alpha$
                                \State {$\theta \gets \theta - \alpha \nabla_\theta L$}
                            \EndFor
                        \EndFor
                    \EndIf
                \For{each accepted collision pair $(j,j')$}
                    \If{Train model this step \textbf{and} $j<N_\mathrm{train}$}
                        \State $\chi_j\gets \chi_\CTCsub(e_j,b_j)$
                        \Else
                            \State $\chi_j\gets \chi_\NNsub(e_j,b_j;\theta)$ 
                         \EndIf
                    \State Use $\chi_j$ to update $v_{j,}, v_{j'}$ according to Equation~\ref{eqn:post_collision}
                \EndFor
                \State{$c\gets c+1$}
            \EndWhile
            \State{$x_i \gets x_i + v_i \Delta t$} 
            \State $t\gets t+\Delta t$
        \EndWhile
    \end{algorithmic}
\end{algorithm}

In a typical simulation the transient period lasts for 1,000 timesteps, so in this case both methods give the model roughly the same length of training in total. 
\section{Numerical Results}
\label{sec:results}
\subsection{Hyperparameter Selection}
\label{sec:training}
\begin{figure}
    \centering
    \begin{subfigure}[b]{0.45\textwidth}
        \includegraphics[width=\textwidth]{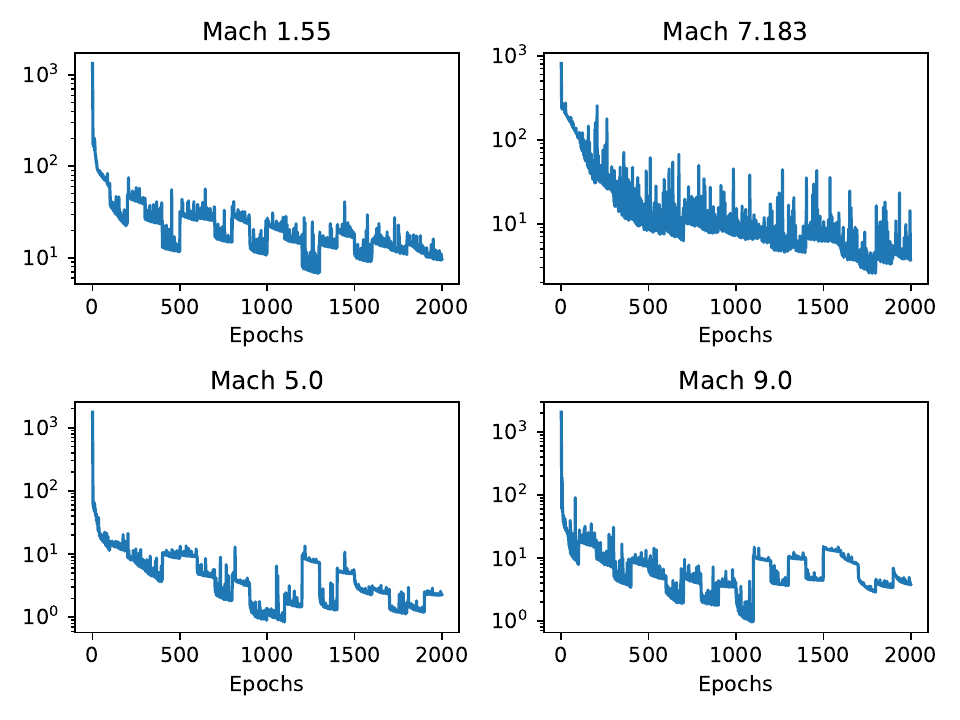}
    \caption{$\rho_L=\qty{1}{\kilogram\per\cubic\metre}$}
    \end{subfigure}
    \centering
    \begin{subfigure}[b]{0.45\textwidth}
        \includegraphics[width=\textwidth]{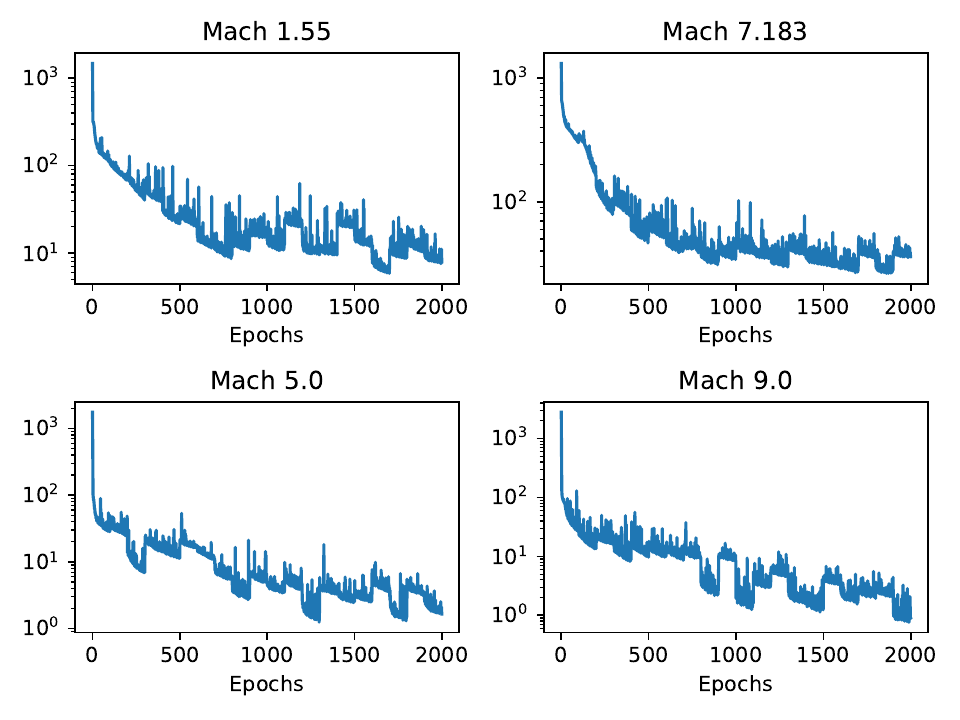}
    \caption{$\rho_L=\qty{1e-3}{\kilogram\per\cubic\metre}$}
    \end{subfigure}
    \centering
    \begin{subfigure}[b]{0.45\textwidth}
        \includegraphics[width=\textwidth]{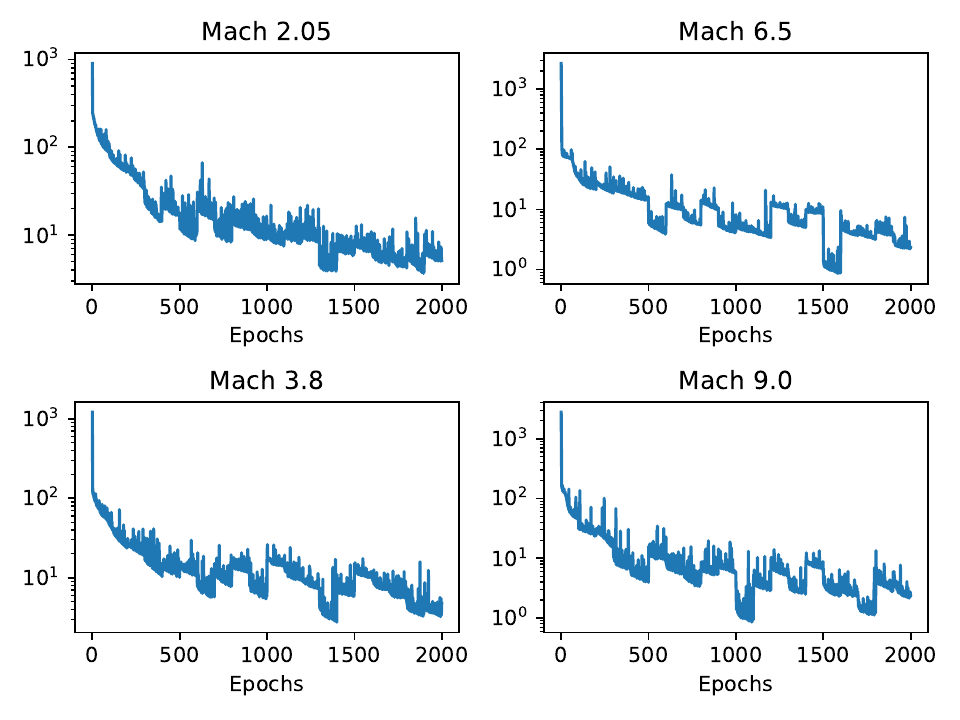}
    \caption{Alsmeyer conditions}
    \end{subfigure}
    \caption{Evolution of the loss function during online training, with a learning rate of $\qty{1e-3}{}$ and 100 epochs per timestep.  }
    \label{fig:online_training}
\end{figure}

Hyperparameters for the online training were selected in order to give robust and accurate results on a subset of the testing cases which appeared most sensitive to the details of the training. Figures~\ref{fig:online_training} shows the loss functions and learning rates during online training (using the first method of training during an initial period of 20 timesteps) for the cases displayed in the main text. For all cases, the loss declines sharply over the first hundred or so epochs of training, and for most, reaches a plateau well before the end of the training period. However, the full training time seems to be more necessary at cold temperatures and low Mach numbers. This behaviour is expected since these cases will have a higher proportion of collisions occurring at low energies, where the collision angle is more sensitive to the inputs. Training time could be reduced by adapting the length of the training period to the simulation parameters. However, in our experiments we have consistently chosen to train for 20 steps in order to give a single robust choice of hyperparameters. Hyperparameters were selected based on training performance at Mach numbers up to Mach 9. Simulations at Mach 10-50 can therefore be viewed as out-of-sample since the training procedure was not tuned for these cases.

\begin{figure}
    \centering
    \includegraphics[width=0.6\linewidth]{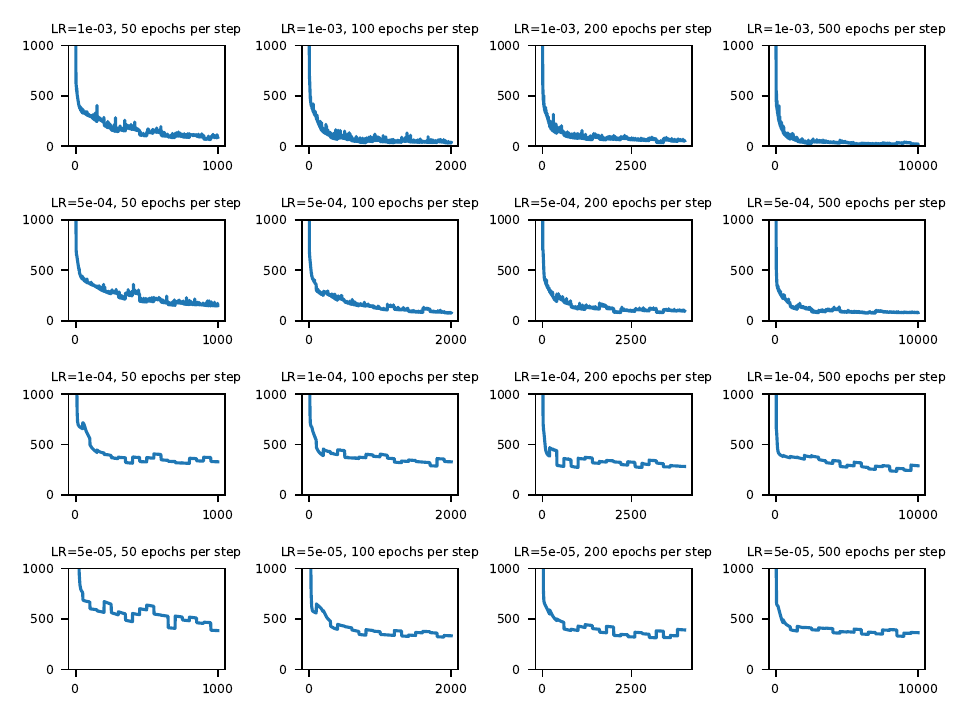}
    \caption{Effect of changes in learning rate and epochs per time step on the evolution of the loss function for a fixed mini-batch size of 250, during online training at Mach 7.183. A learning rate of $\qty{1e-3}{}$ with 100 epochs per DSMC timestep was chosen to give a good balance between robust training and computational efficiency. }
    \label{fig:parameter_sweep}
\end{figure}

Figure~\ref{fig:parameter_sweep} shows the effect of changes in the initial learning rate and the number of training epochs per step, motivating the choice of training hyperparameters used in the simulations. The plots were generated at a Mach number of 7.183 in the rarefied density case, since it was this simulation which seemed to show the most sensitivity to the training parameters. From this plot an initial learning rate of $\alpha_0=\qty{1e-3}{}$ and 100 epochs per time step were selected to give a good balance between reduction in the loss, and time spent during training. 

\subsection{Shock Profiles}
\label{sec:profiles}
\begin{figure}
    \centering
    \begin{subfigure}[b]{0.45\textwidth}
        \includegraphics[width=\textwidth]{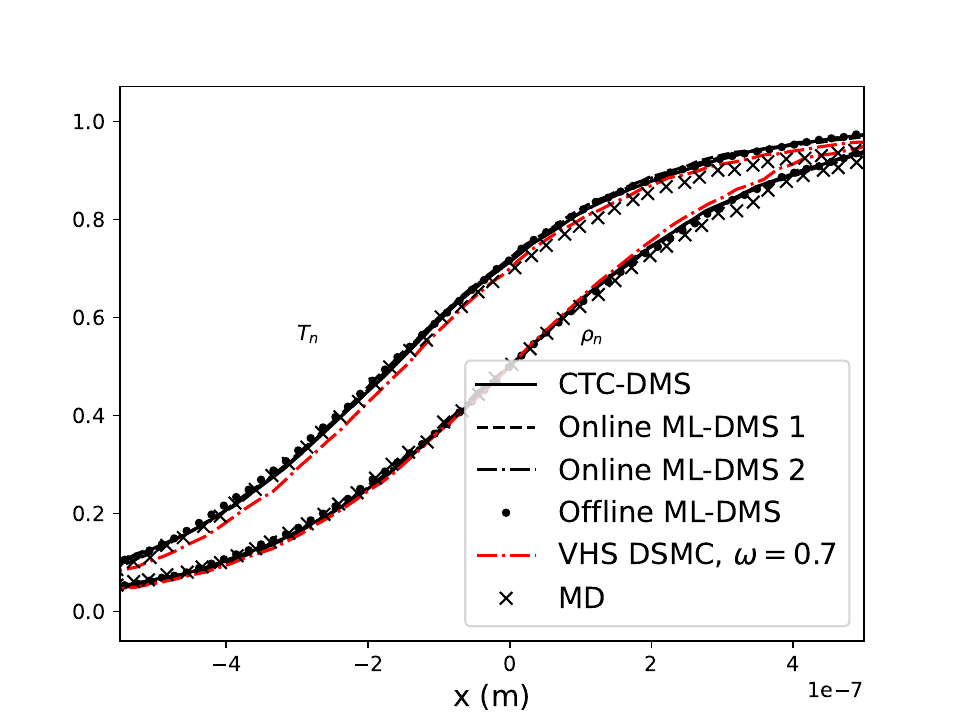}
    \caption{Mach 1.55, $T_L=\qty{300}{\kelvin}$}
    \end{subfigure}
    \centering
    \begin{subfigure}[b]{0.45\textwidth}
        \includegraphics[width=\textwidth]{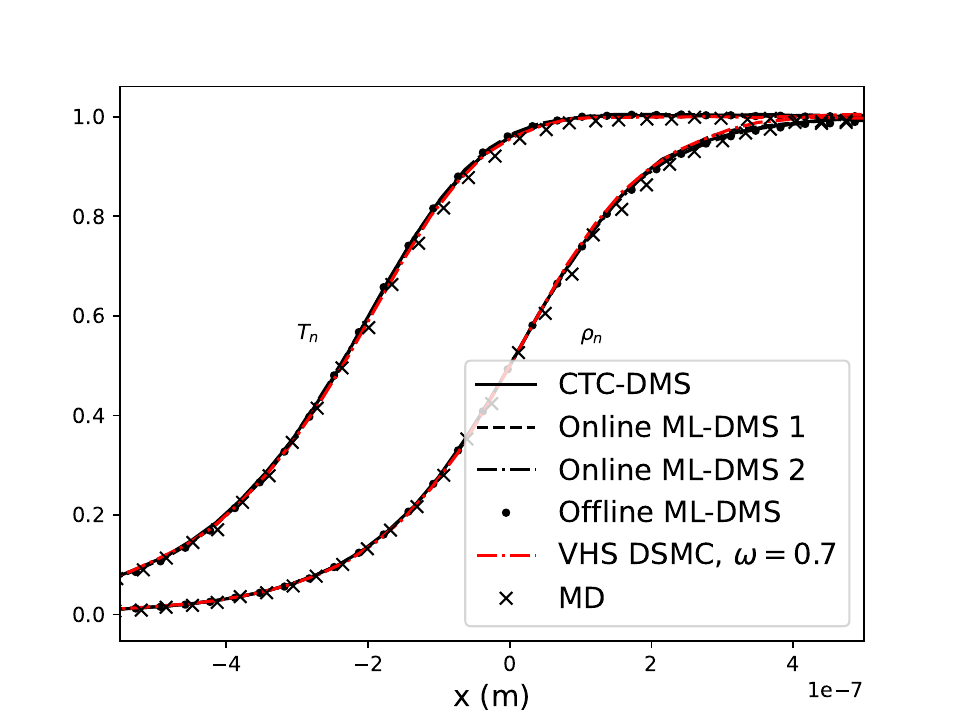}
    \caption{Mach 5, $T_L=\qty{300}{\kelvin}$}
    \end{subfigure}
    \centering
    \begin{subfigure}[b]{0.45\textwidth}
        \includegraphics[width=\textwidth]{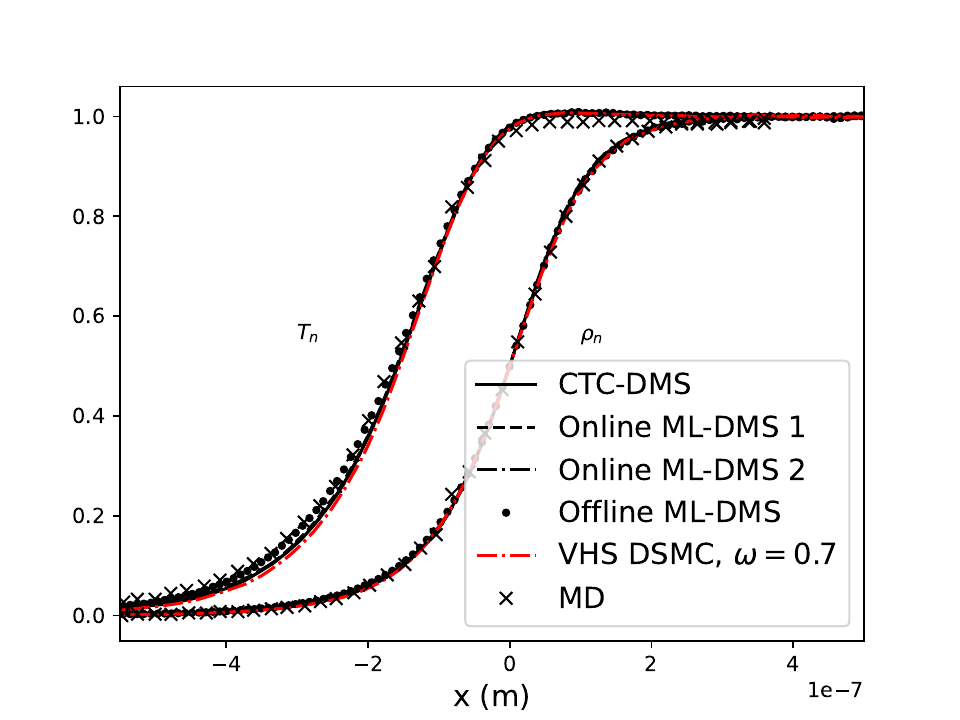}
    \caption{Mach 7.183, $T_L=\qty{16}{\kelvin}$}
    \end{subfigure}
    \centering
    \begin{subfigure}[b]{0.45\textwidth}
        \includegraphics[width=\textwidth]{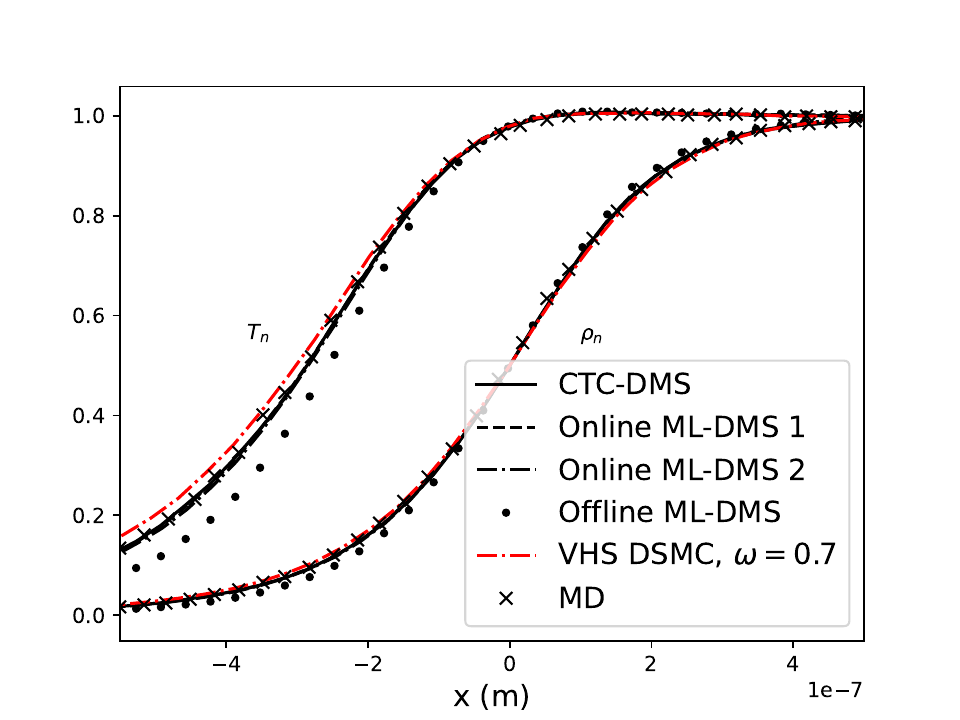}
    \caption{Mach 9, $T_L=\qty{300}{\kelvin}$}
    \end{subfigure}
    \caption{Shock profiles at $\rho_L=\qty{1}{\kilogram\per\cubic\metre}$. MD data points are from \cite{MDDilute}.}
    \label{fig:moderate}
\end{figure}

\begin{figure}
    \centering
    \begin{subfigure}[b]{0.45\textwidth}
        \includegraphics[width=\textwidth]{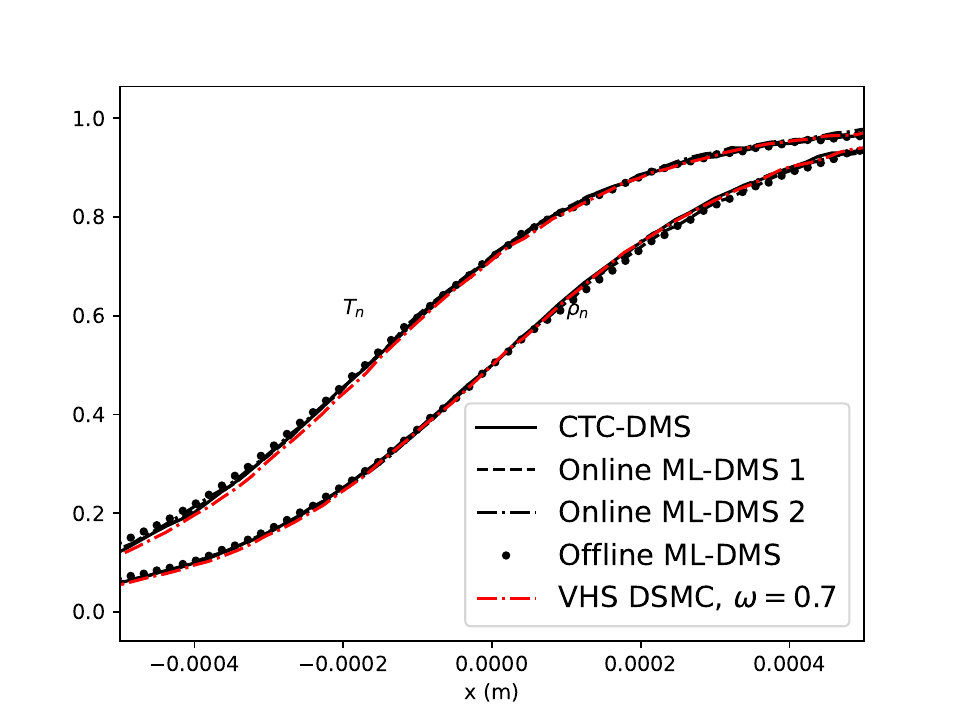}
    \caption{Mach 1.55, $T_L=\qty{300}{\kelvin}$}
    \end{subfigure}
    \centering
    \begin{subfigure}[b]{0.45\textwidth}
        \includegraphics[width=\textwidth]{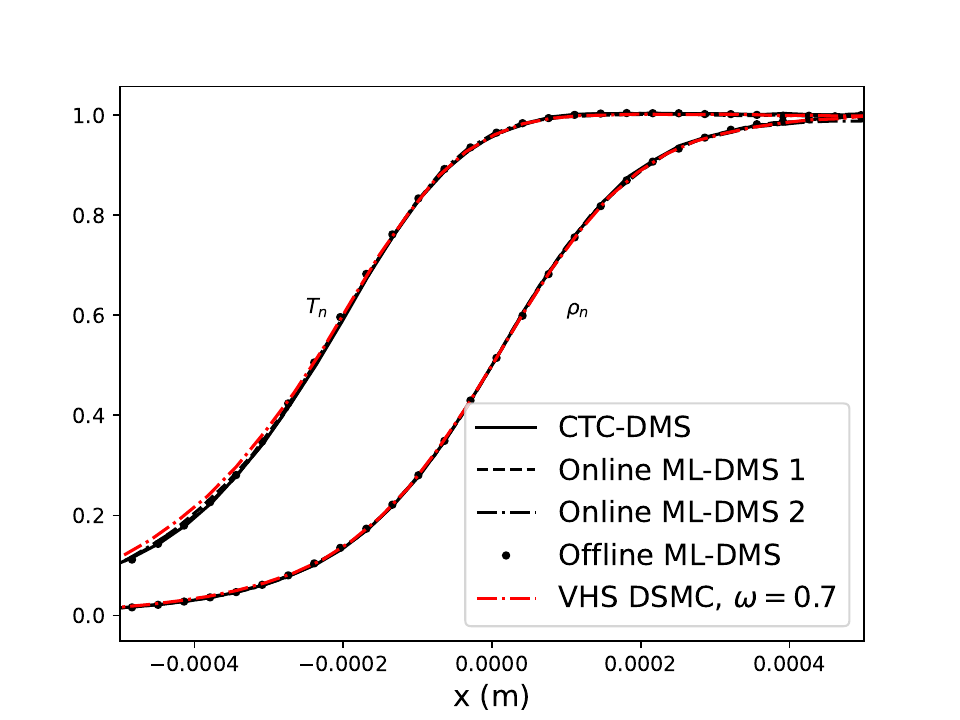}
    \caption{Mach 5, $T_L=\qty{300}{\kelvin}$}
    \end{subfigure}
    \centering
    \begin{subfigure}[b]{0.45\textwidth}
        \includegraphics[width=\textwidth]{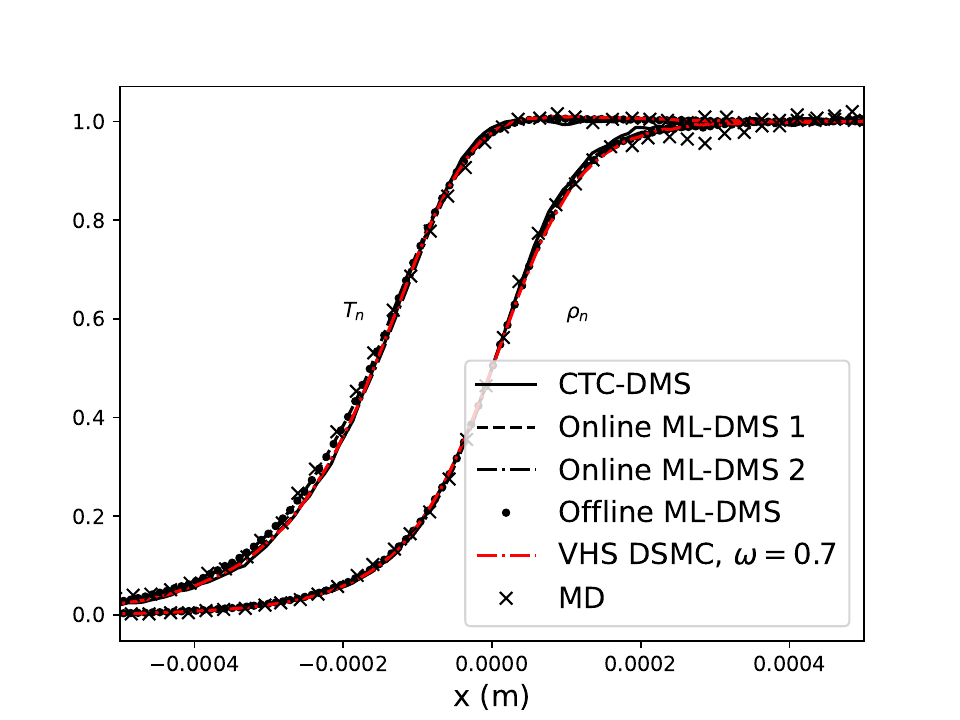}
    \caption{Mach 7.183, $T_L=\qty{16}{\kelvin}$}
    \end{subfigure}
    \centering
    \begin{subfigure}[b]{0.45\textwidth}
        \includegraphics[width=\textwidth]{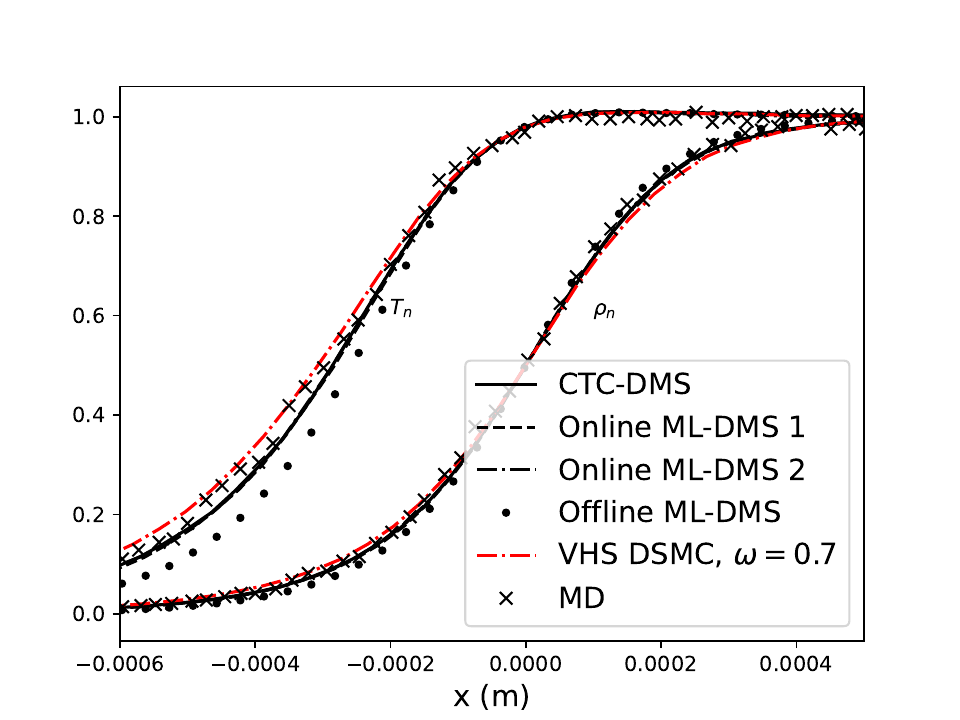}
    \caption{Mach 9, $T_L=\qty{300}{\kelvin}$}
    \end{subfigure}
    \caption{Shock profiles at $\rho_L=\qty{1e-3}{\kilogram\per\cubic\metre}$. MD data points are from \cite{MDDilute}.}
    \label{fig:rarefied} 
\end{figure}

The models proposed above have been evaluated at a wide range of physical conditions, including for a number of cases where MD and experimental reference data is available from the literature.
Figure~\ref{fig:moderate} shows density and temperature profiles for shocks at upstream Mach numbers $1.55,5,9$, with $\rho_L=\qty{1}{\kilogram\per\cubic\metre}$ and $T_L=\qty{300}{\kelvin}$, corresponding to a moderate density shock at room temperature, as well as a much colder example at Mach 7.183 and $T_L=\qty{16}{\kelvin}$. Figure~\ref{fig:rarefied} shows the same Mach numbers and temperatures at significantly more rarefied conditions ($\rho_L=\qty{1e-3}{\kilogram\per\cubic\metre}$). All three ML-DMS methods are compared against CTC-DMS obtained with our code, as well as MD results taken from \cite{MD_argon}, which gives details of MD simulations run at these conditions. 

All shock profiles are additionally compared to DSMC results obtained using the VHS model, with parameters $\omega=0.7$ and  $d_\mathrm{ref}=3.974\,\text{\r{A}}$, $T_\mathrm{ref}=\qty{273}{\kelvin}$. These values of these parameters were observed to give the best fit to Alsmeyer's experimental data in \cite{boyd_schwartzentruber_2017}. We observe that these VHS parameters fit the density profiles of MD and CTC-DMS, but are less accurate for the temperature profile at higher Mach numbers (see Figure~\ref{fig:profiles-out-of-sample_moderate_300} for upstream Mach numbers 15--50). Other choices of $\omega$ and the reference values could be chosen by hand for different physical conditions to give closer agreement with the DMS and MD profiles, as in \cite{MDDilute}. However, a priori this requires DMS/MD data, which can be computationally expensive to simulate and will likely be unavailable for complex geometries. Our aim is to give a comparison to a DSMC model which could be chosen a priori from the literature without any calibration to datasets which may not be available. 


 The profiles obtained from the offline neural network show close agreement between full trajectory calculation and the neural network models at Mach 5, although there are some discrepancies in the temperature profiles at Mach 9, motivating the introduction of the online training procedure. In contrast to the offline case, density and temperature profiles from both online models are in close agreement with both CTC-DMS and MD results at all Mach numbers tested. This shows the advantage of the online method in reducing overfitting and generalising to new prediction cases. 
 
\begin{figure}
    \centering
    \begin{subfigure}[b]{0.45\textwidth}
        \includegraphics[width=\textwidth]{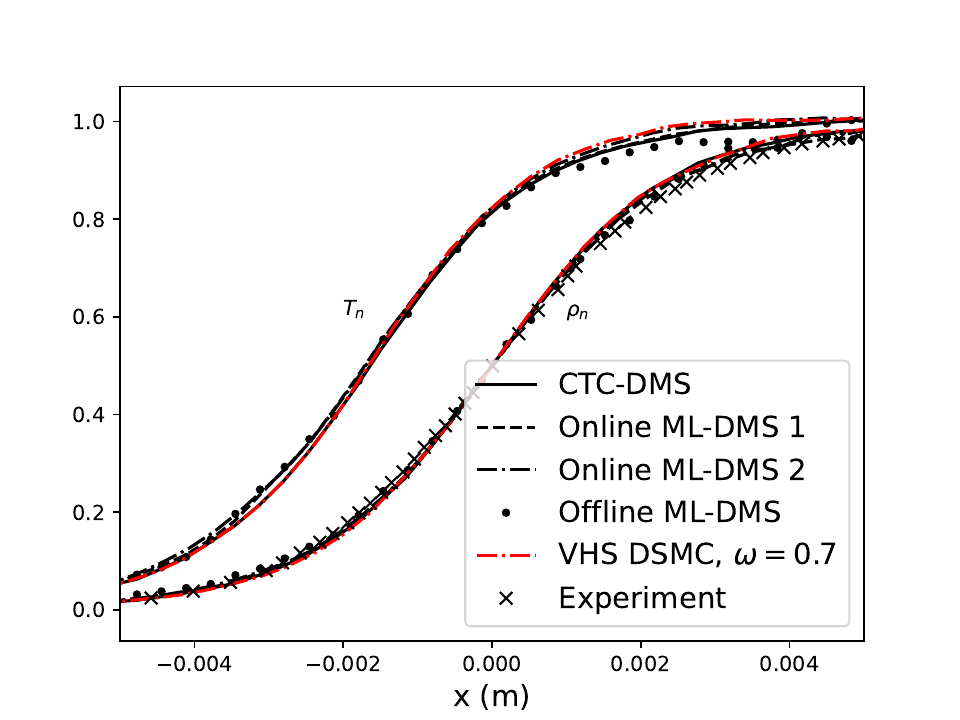}
    \caption{Mach 2.05}
    \end{subfigure}
    \centering
    \begin{subfigure}[b]{0.45\textwidth}
        \includegraphics[width=\textwidth]{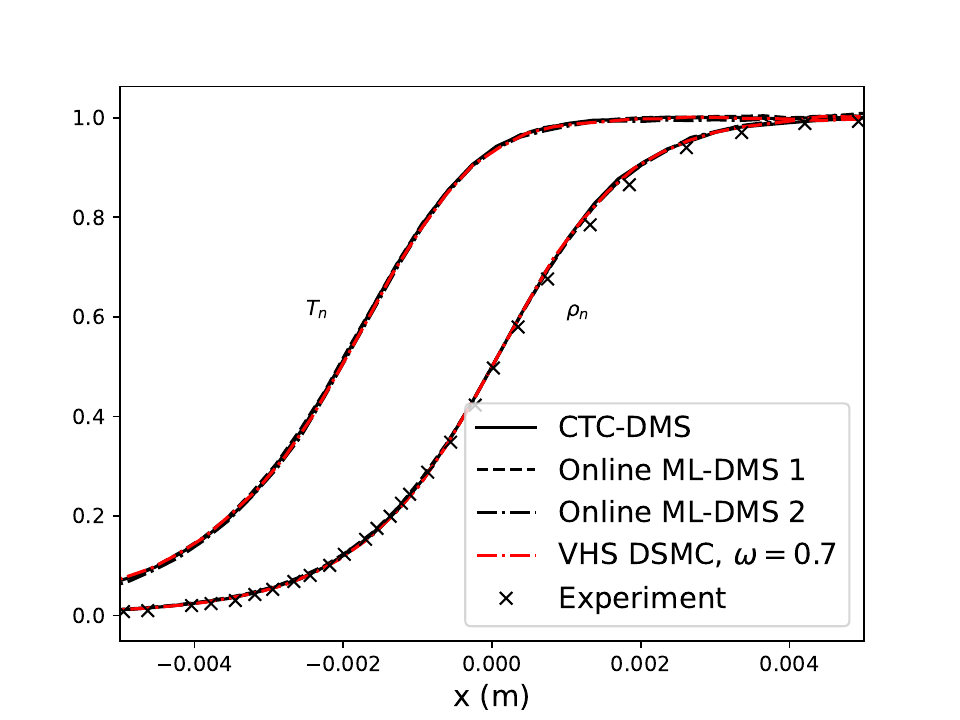}
    \caption{Mach 3.8}
    \end{subfigure}
    \centering
    \begin{subfigure}[b]{0.45\textwidth}
        \includegraphics[width=\textwidth]{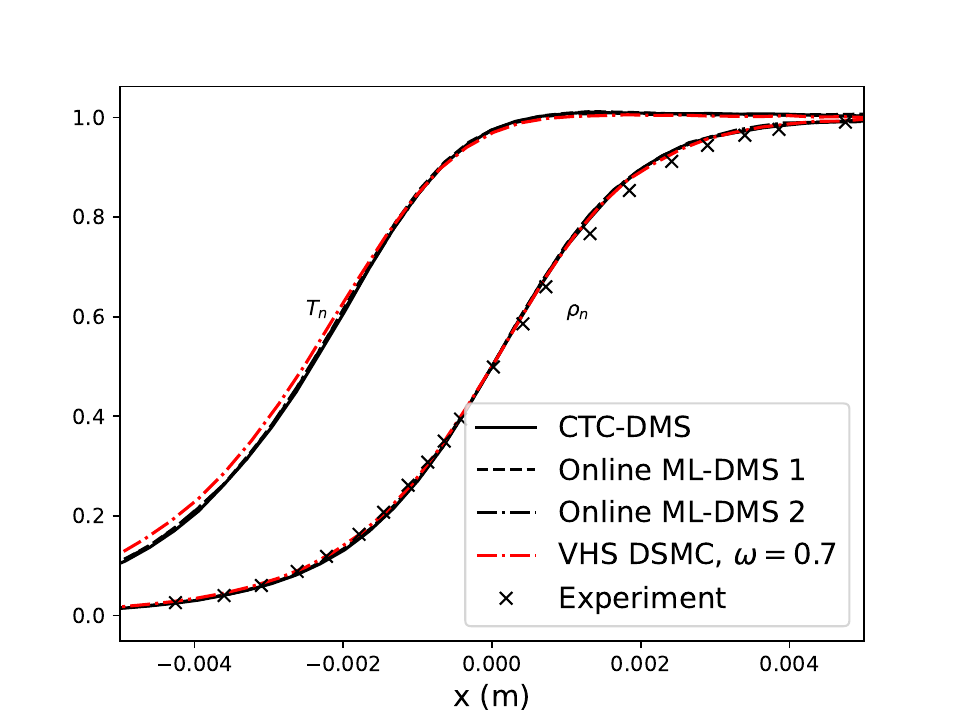}
    \caption{Mach 6.5}
    \end{subfigure}
    \centering
    \begin{subfigure}[b]{0.45\textwidth}
        \includegraphics[width=\textwidth]{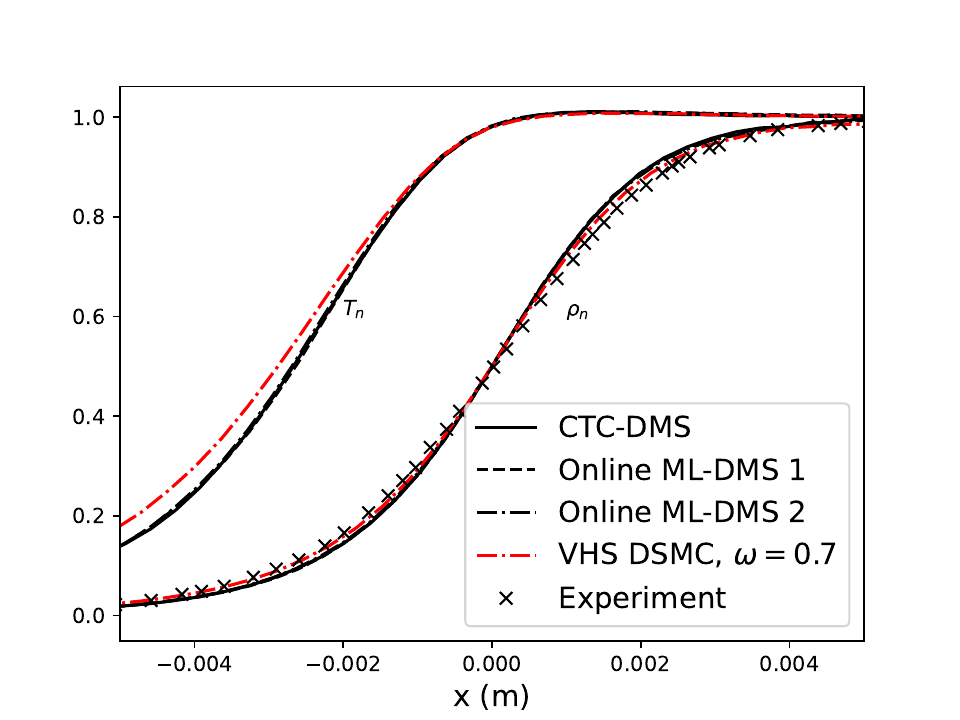}
    \caption{Mach 9}
    \end{subfigure}
    \caption{Shock profiles at $\rho_L=\sim\qty{1e-4}{\kilogram\per\cubic\metre}$, $T_L=\qty{300}{\kelvin}$. Experimental data points are taken from Alsmeyer \cite{Alsmeyer_1976}.}
    \label{fig:alsmeyer}
\end{figure}

The results of training using the online algorithm have been further compared to experimental data from Alsmeyer \cite{Alsmeyer_1976}. Figure \ref{fig:alsmeyer} shows comparisons between experimental shock profiles and those obtained from the online Neural Network approach, at freestream pressure of $p_L=\qty{6.667}{\pascal}$ and temperature $T_L=\qty{300}{\kelvin}$ and Mach numbers 2.05, 3.8, 6.5, 9. At Mach 6.5 and, especially, Mach 9, the DMS results begin to deviate from the experimental profiles. Since the DMS has been shown to be capable of reproducing MD results using the LJ potential where this is available, and a similar discrepancy was observed in \cite{Koura_argon} as discussed in Section~\ref{sec:validation}, we expect that this is due to the particular model used for interatomic interactions. In all cases the online trained models are in very close agreement with the results of CTC-DMS, showing that the trained ML-DMS model is capable of reproducing the results of CTC.

\subsubsection{Out-of-Sample Evaluation}
\begin{figure}
    \centering
    \begin{subfigure}[b]{0.45\textwidth}
        \includegraphics[width=\textwidth]{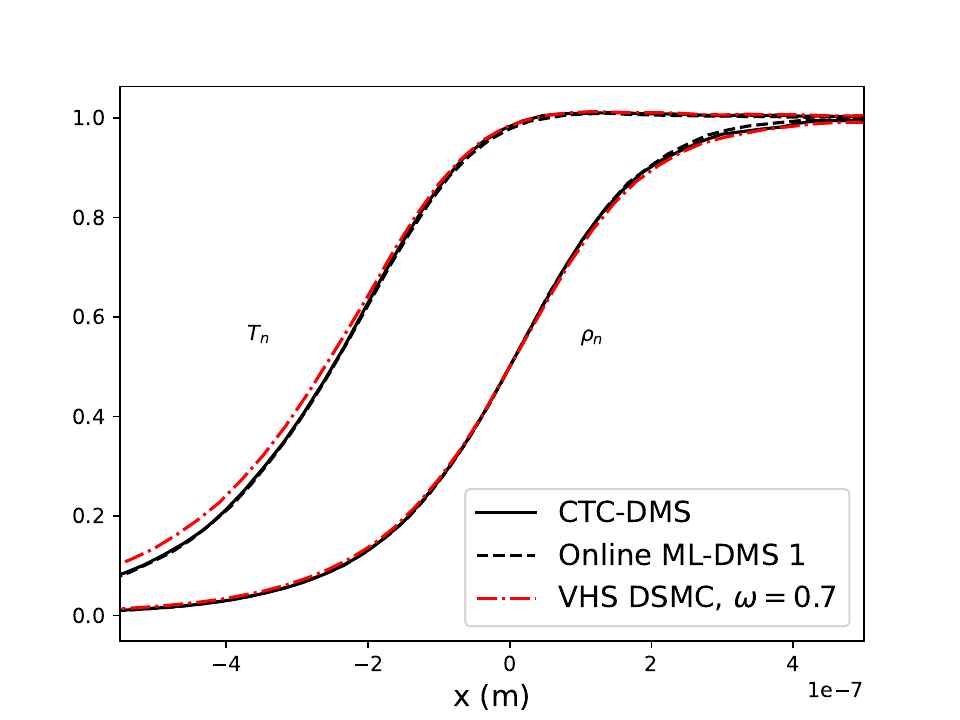}
    \caption{Mach 10}
    \end{subfigure}
    \centering
    \begin{subfigure}[b]{0.45\textwidth}
        \includegraphics[width=\textwidth]{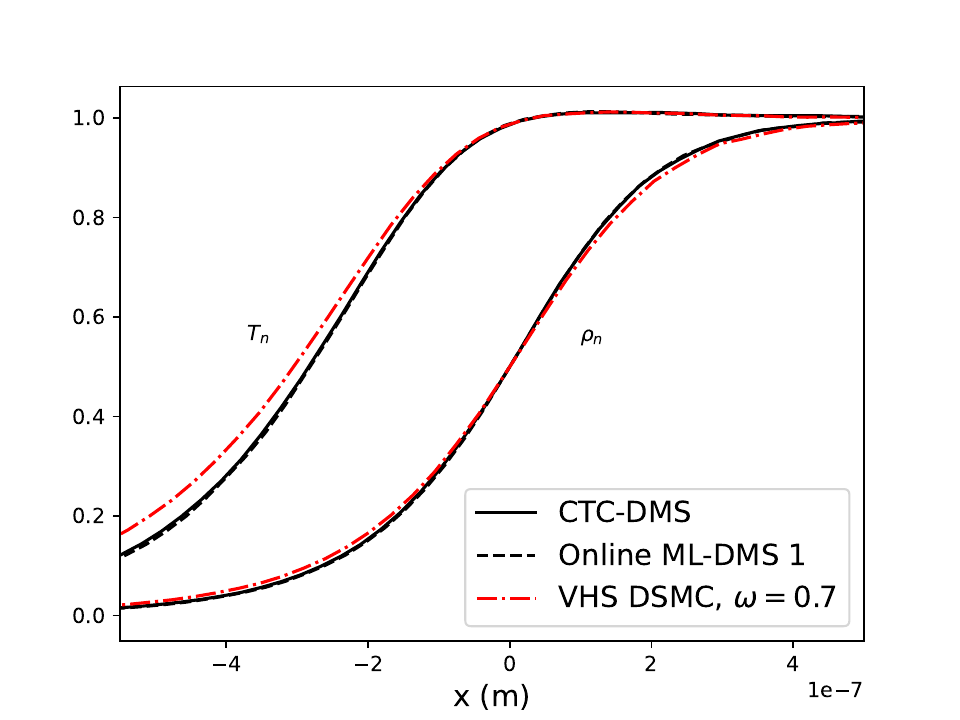}
    \caption{Mach 15}
    \end{subfigure}
    \centering
    \begin{subfigure}[b]{0.45\textwidth}
        \includegraphics[width=\textwidth]{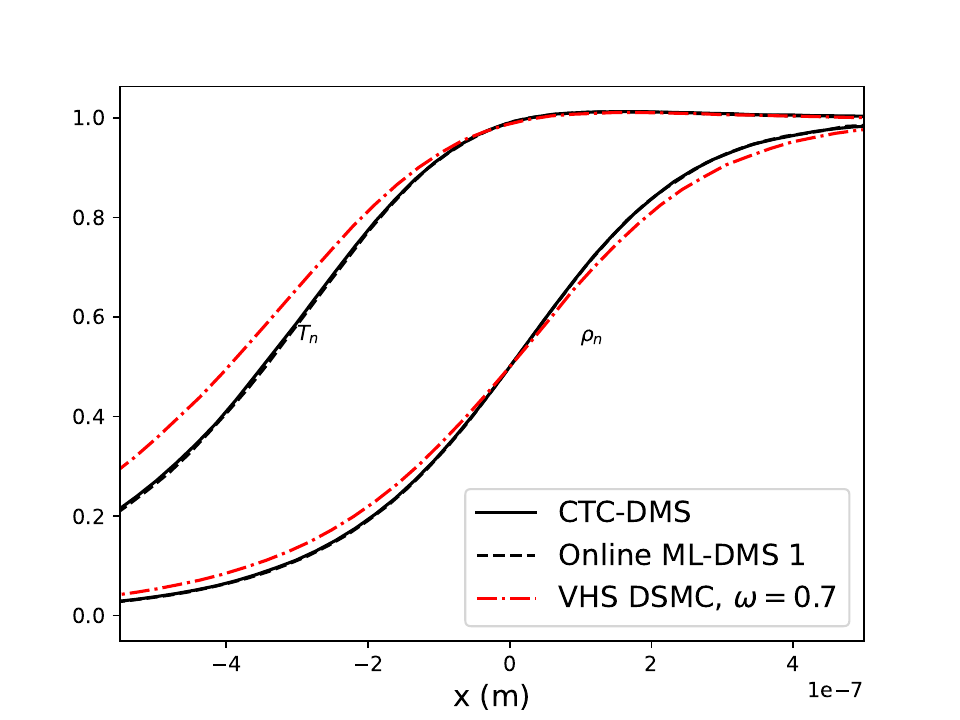}
    \caption{Mach 30}
    \end{subfigure}
    \caption{Shock profiles for ``out-of-sample'' testing at $\rho_L=\qty{1}{\kilogram\per\cubic\metre}$, and intermediate temperature $T_L=\qty{100}{\kelvin}$. }
    \label{fig:profiles-out-of-sample_moderate}
\end{figure}

\begin{figure}
    \centering
    \begin{subfigure}[b]{0.45\textwidth}
        \includegraphics[width=\textwidth]{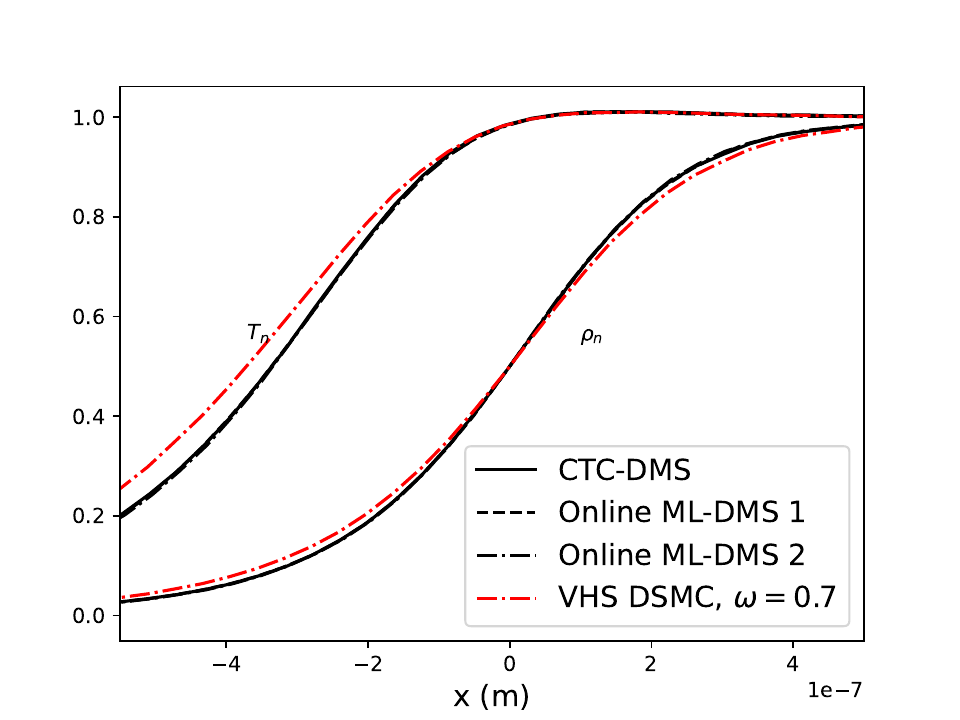}
    \caption{Mach 15}
    \end{subfigure}
    \centering
    \begin{subfigure}[b]{0.45\textwidth}
        \includegraphics[width=\textwidth]{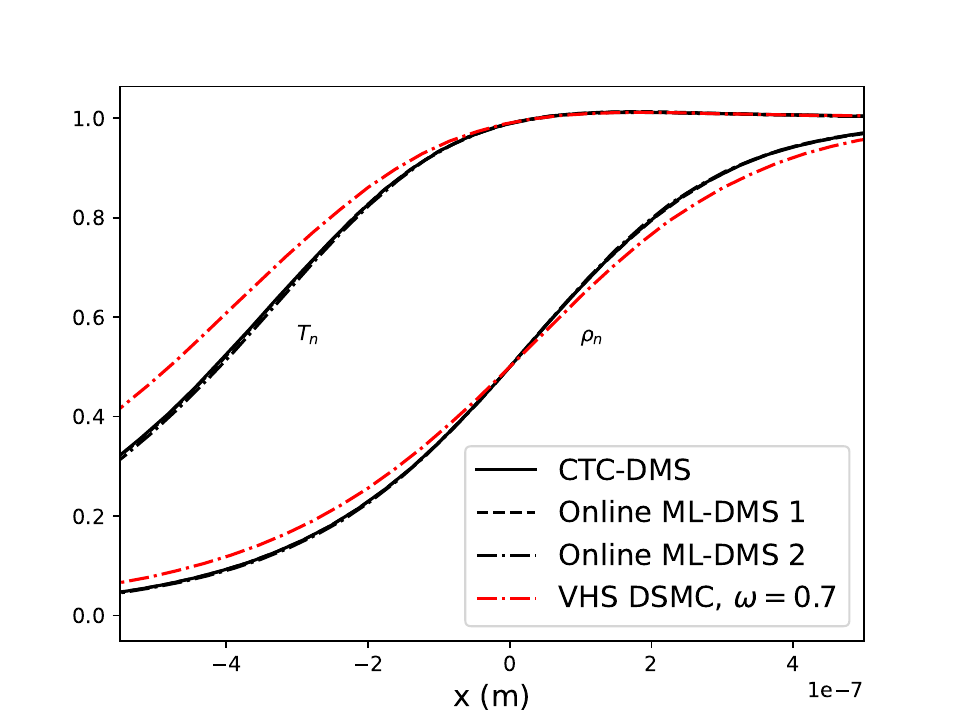}
    \caption{Mach 30}
    \end{subfigure}
    \centering
    \begin{subfigure}[b]{0.45\textwidth}
        \includegraphics[width=\textwidth]{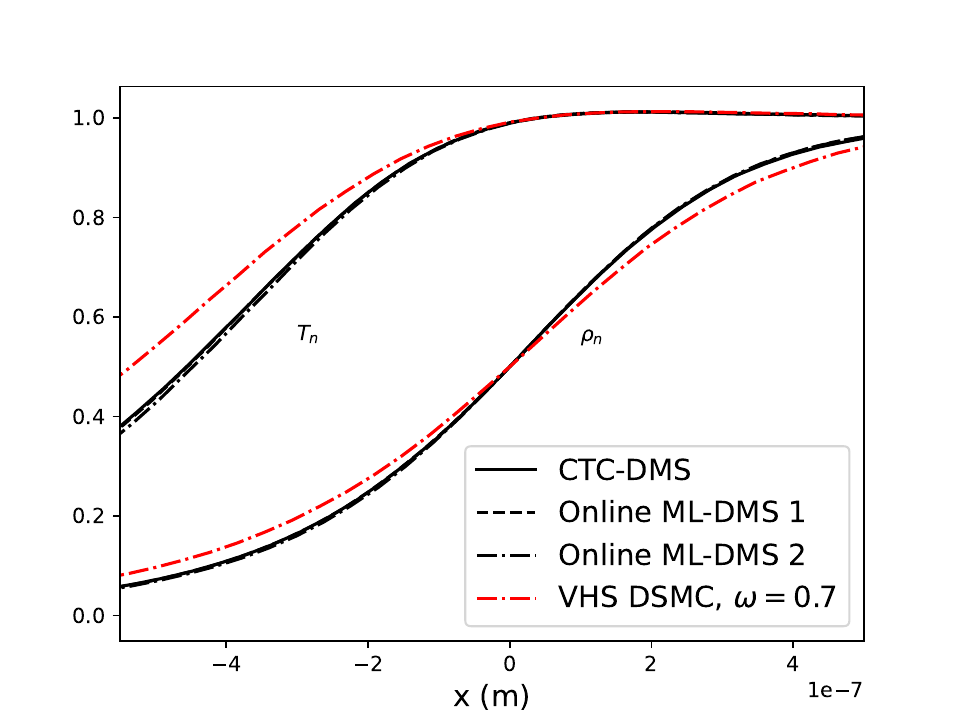}
    \caption{Mach 40}
    \end{subfigure}
    \centering
    \begin{subfigure}[b]{0.45\textwidth}
        \includegraphics[width=\textwidth]{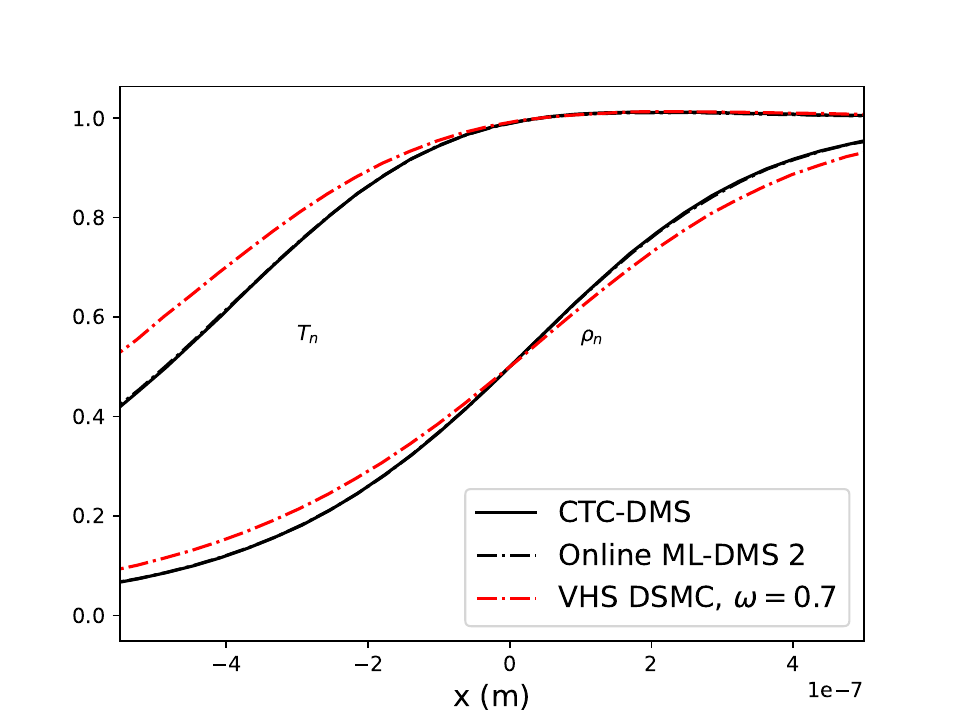}
    \caption{Mach 50}
    \end{subfigure}
    \caption{Shock profiles for ``out-of-sample'' testing at $\rho_L=\qty{1}{\kilogram\per\cubic\metre}$, $T_L=\qty{300}{\kelvin}$ }
    \label{fig:profiles-out-of-sample_moderate_300}
\end{figure}

While the online nature of the training means that application to any new situation can be seen as ``out-of-sample'', we wish to evaluate the performance of the online ML-DMS on physical conditions for which the hyperparameters were not tuned. The training hyperparameters and model architectures used in the online algorithms above were selected based on performance on the given physical conditions ($T_L=\qty{16}{\kelvin}$, $\qty{300}{\kelvin}$) and Mach numbers ($1.55\leq \textup{Ma}\leq 9$). To ensure that these hyperparameters are sufficiently general, and have not been overfit to the scenarios used for testing, the performance of both online ML-DMS algorithms has further been evaluated on test cases at higher Mach numbers $\mathrm{Ma}=10$ to $50$ and at an intermediate temperature $T_L=\qty{100}{\kelvin}$, at freestream density $\qty{1}{\kilogram\per\cubic\metre}$. The procedure is observed to perform well both during training and in the final shock profile. Shock profiles for this case are compared with CTC-DMS in Figure~\ref{fig:profiles-out-of-sample_moderate}, for high Mach numbers at $\qty{100}{\kelvin}$, and in Figure~\ref{fig:profiles-out-of-sample_moderate_300} at $\qty{300}{\kelvin}$. Excellent agreement is observed between CTC-DMS and both online training methods. Additionally, at these high Mach numbers, the discrepancy between VHS DSMC and DMS becomes more significant. The online ML-DMS simulation closely matches the CTC-DMS solutions while VHS DSMC is less accurate.  

\begin{figure}
    \centering
    \begin{subfigure}[b]{0.45\textwidth}
        \includegraphics[width=\textwidth]{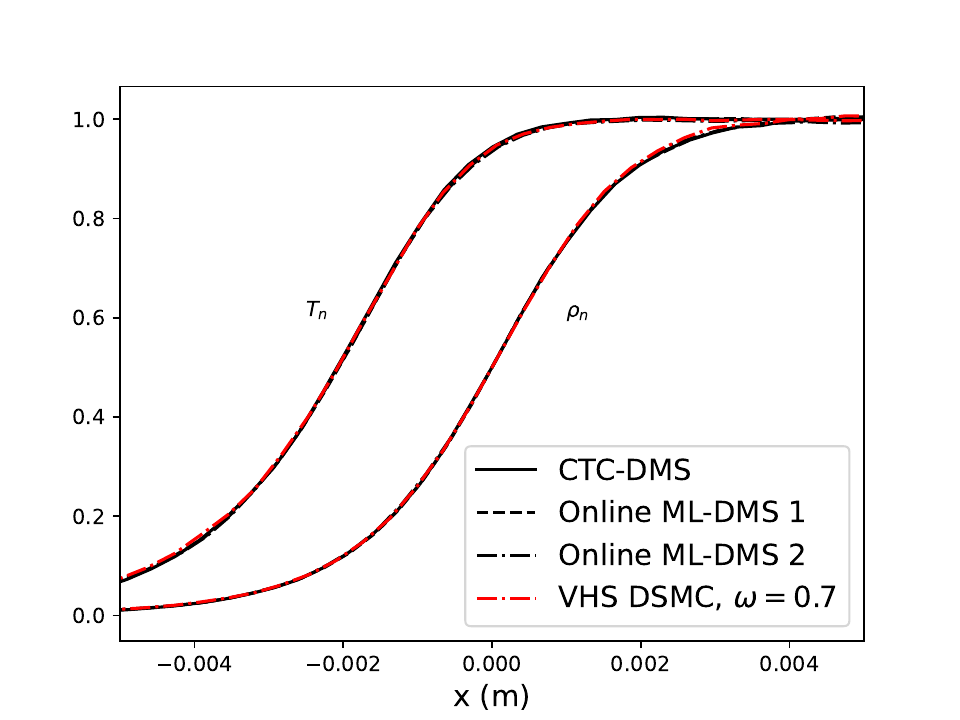}
    \caption{Mach 4}
    \end{subfigure}
    \centering
    \begin{subfigure}[b]{0.45\textwidth}
        \includegraphics[width=\textwidth]{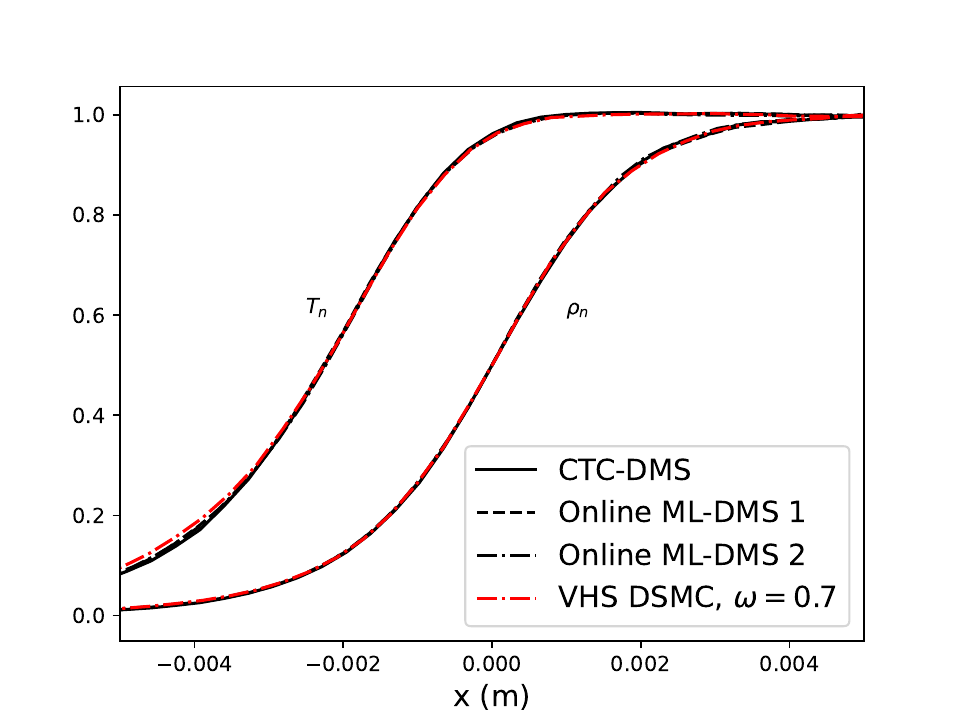}
    \caption{Mach 5}
    \end{subfigure}
    \centering
    \begin{subfigure}[b]{0.45\textwidth}
        \includegraphics[width=\textwidth]{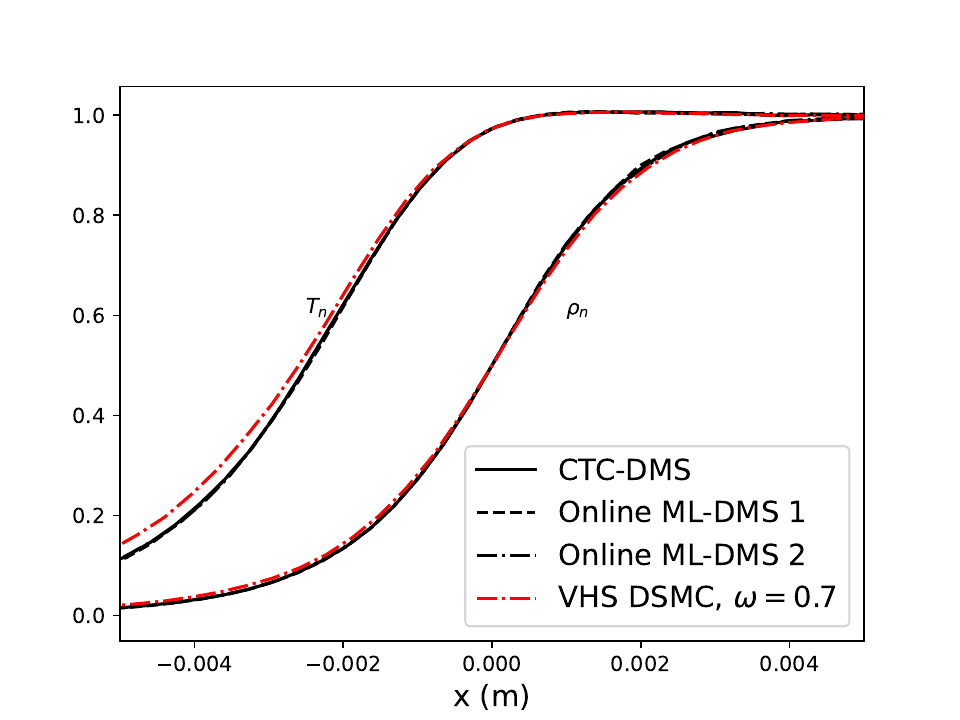}
    \caption{Mach 7}
    \end{subfigure}
    \centering
    \begin{subfigure}[b]{0.45\textwidth}
        \includegraphics[width=\textwidth]{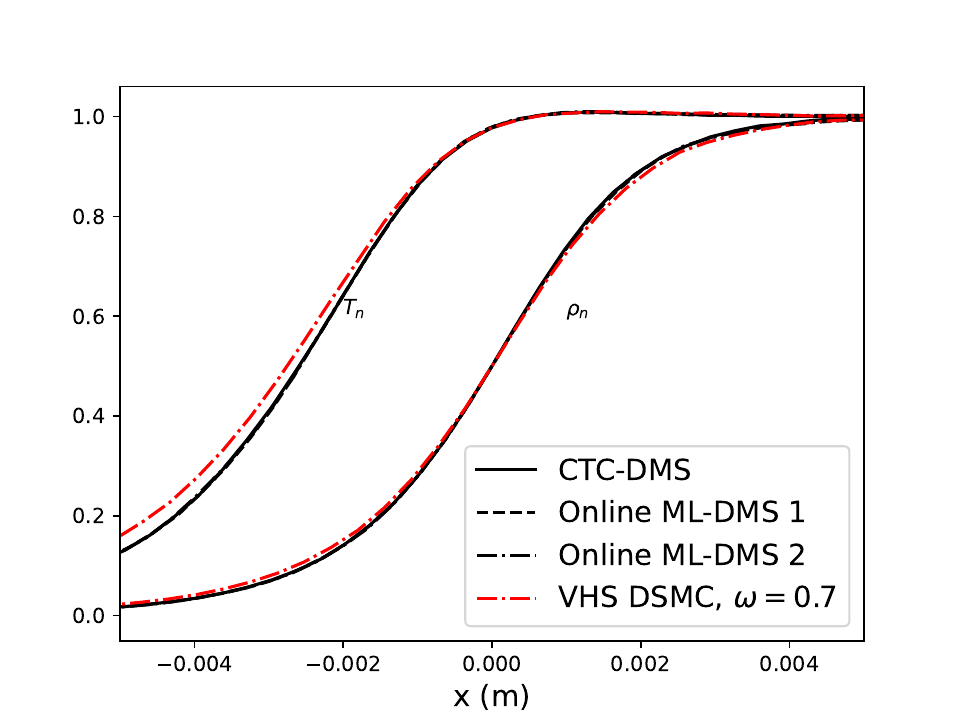}
    \caption{Mach 8}
    \end{subfigure}
    \caption{Shock profiles at Alsmeyer experimental conditions: $\rho_L=\sim\qty{1e-4}{\kilogram\per\cubic\metre}$, $T_L=\qty{300}{\kelvin}$.}
    \label{fig:alsmeyer_extra}
\end{figure}

\begin{figure}
    \centering
    \begin{subfigure}[b]{0.45\textwidth}
        \includegraphics[width=\textwidth]{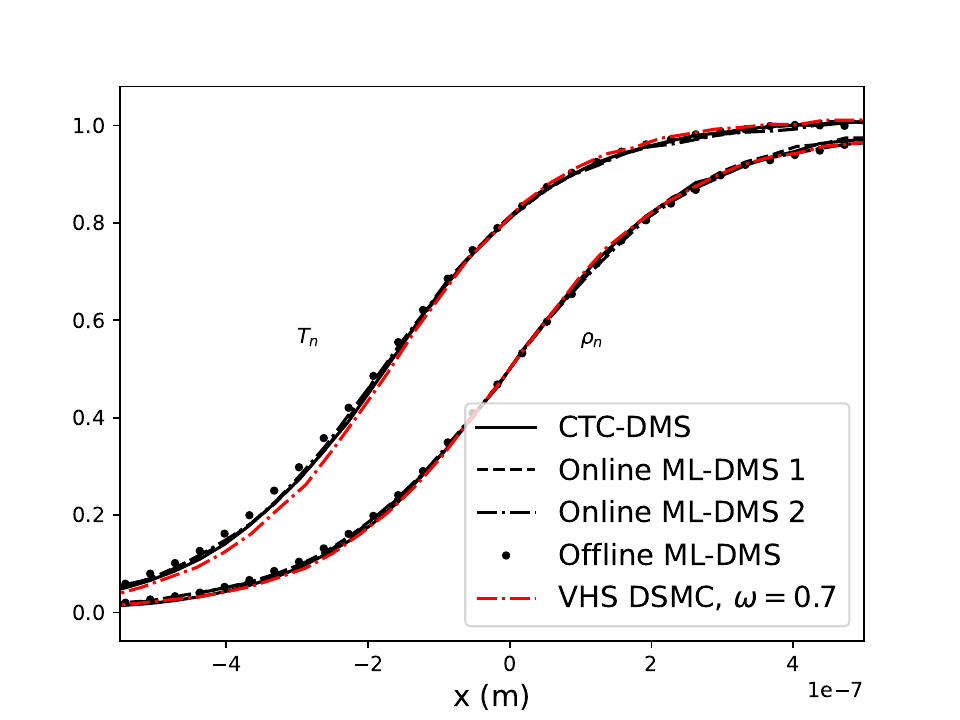}
    \caption{Mach 2}
    \end{subfigure}
    \centering
    \begin{subfigure}[b]{0.45\textwidth}
        \includegraphics[width=\textwidth]{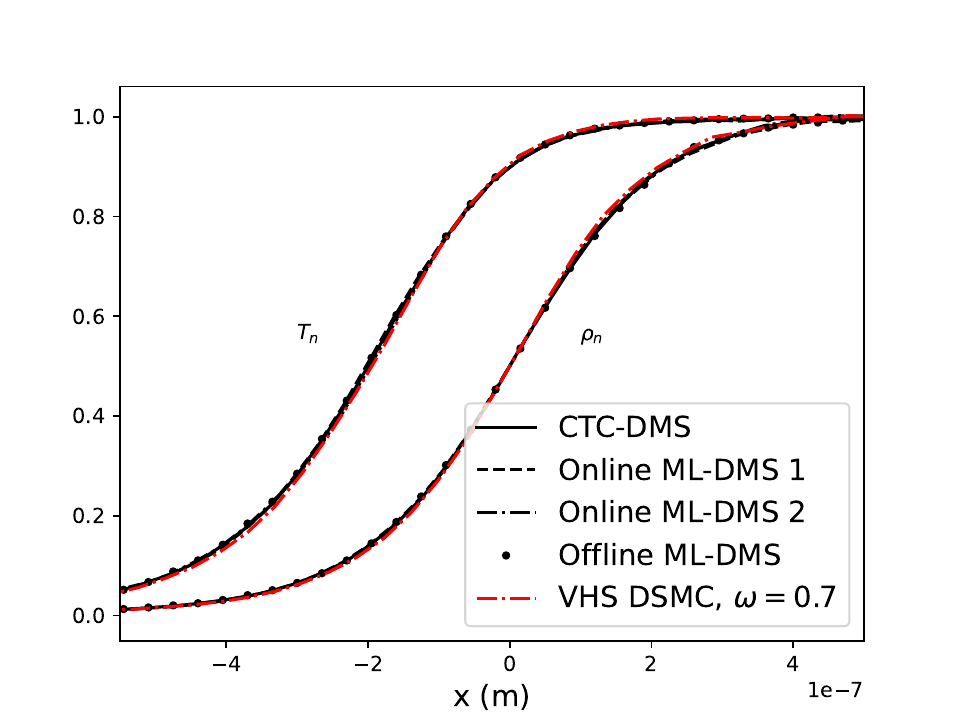}
    \caption{Mach 3}
    \end{subfigure}
    \centering
    
    \begin{subfigure}[b]{0.45\textwidth}
        \includegraphics[width=\textwidth]{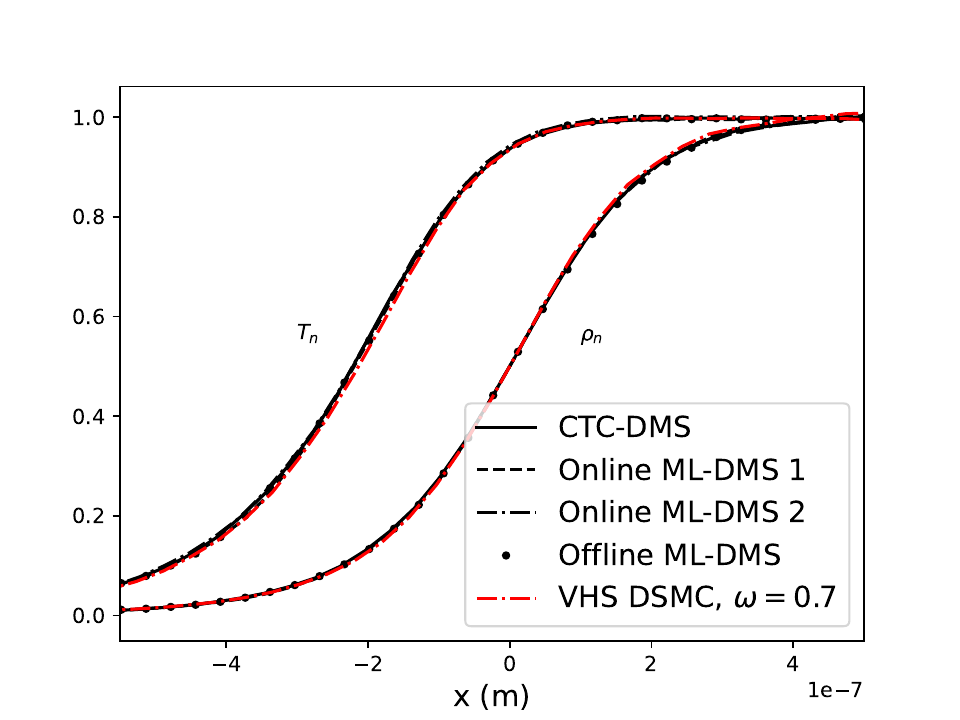}
    \caption{Mach 4}
    \end{subfigure}
    \centering
    \begin{subfigure}[b]{0.45\textwidth}
        \includegraphics[width=\textwidth]{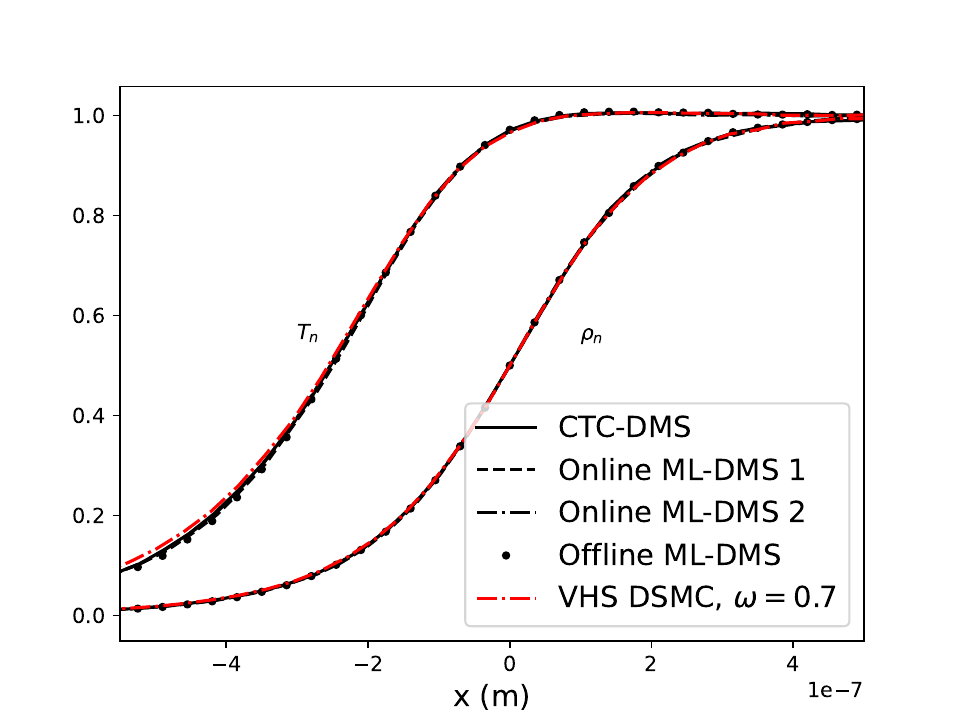}
    \caption{Mach 6}
    \end{subfigure}
    \centering
    \begin{subfigure}[b]{0.45\textwidth}
        \includegraphics[width=\textwidth]{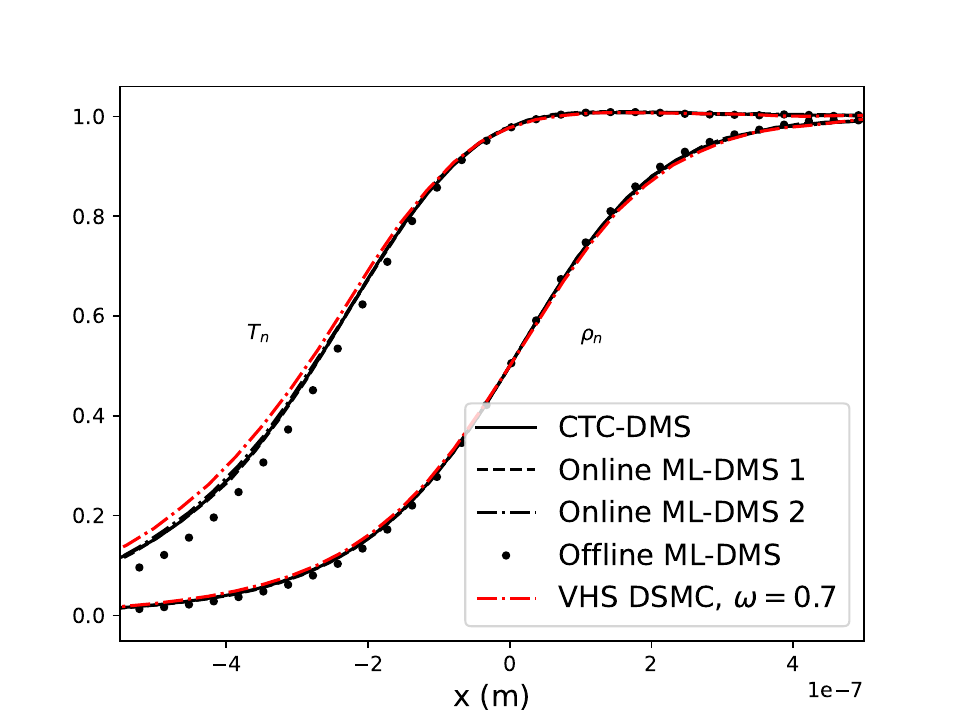}
    \caption{Mach 8}
    \end{subfigure}
    \caption{Shock profiles at moderate density conditions: $\rho_L=\qty{1}{\kilogram\per\cubic\metre}$, $T_L=\qty{300}{\kelvin}$.}
    \label{fig:moderate_extra}
\end{figure}

\begin{figure}
    \centering
    \begin{subfigure}[b]{0.45\textwidth}
        \includegraphics[width=\textwidth]{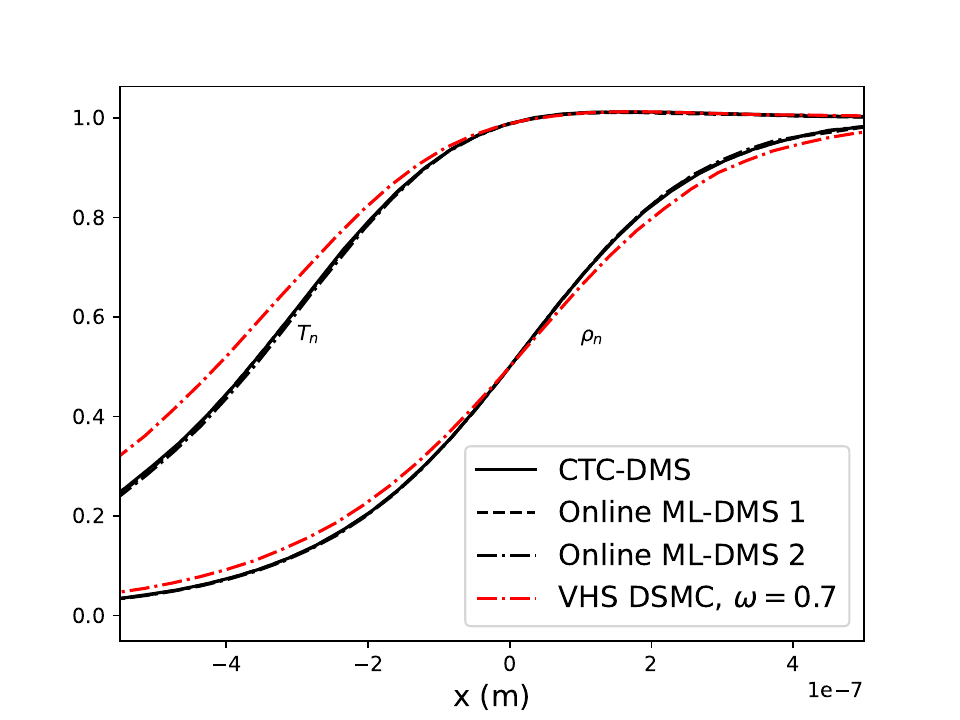}
    \caption{Mach 20}
    \end{subfigure}
    \centering
    \begin{subfigure}[b]{0.45\textwidth}
        \includegraphics[width=\textwidth]{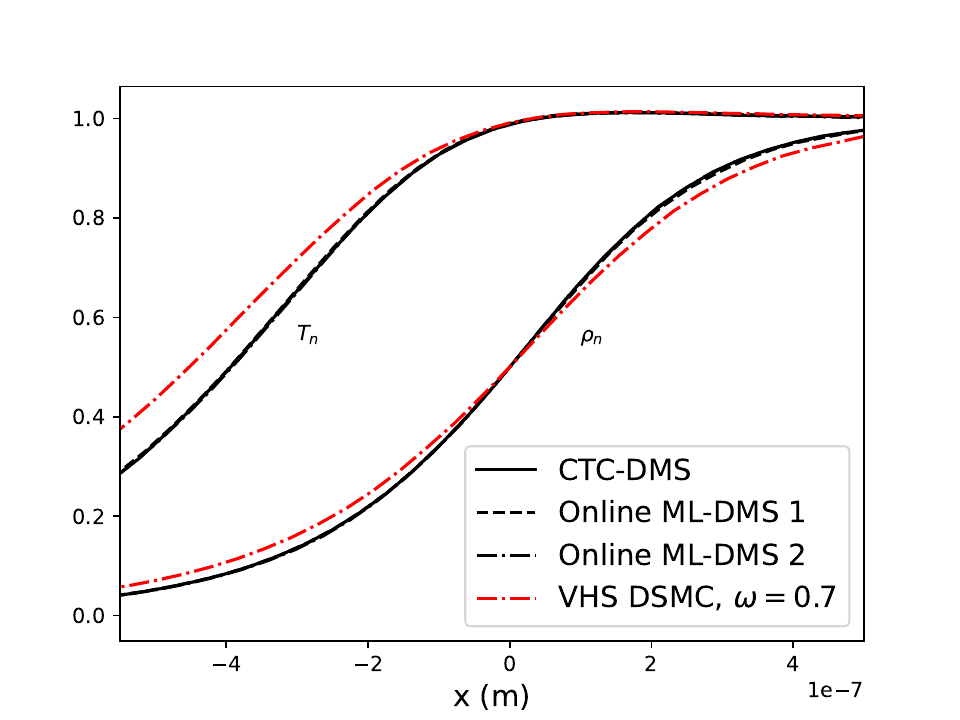}
    \caption{Mach 25}
    \end{subfigure}
    \centering
    
    \begin{subfigure}[b]{0.45\textwidth}
        \includegraphics[width=\textwidth]{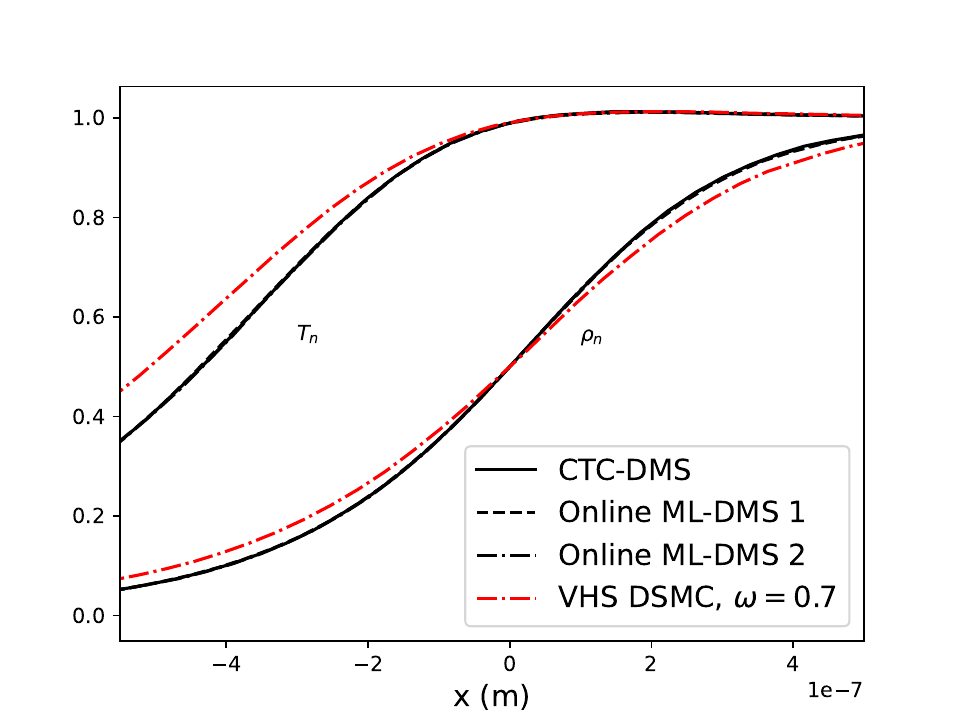}
    \caption{Mach 35}
    \end{subfigure}
    \centering
    \begin{subfigure}[b]{0.45\textwidth}
        \includegraphics[width=\textwidth]{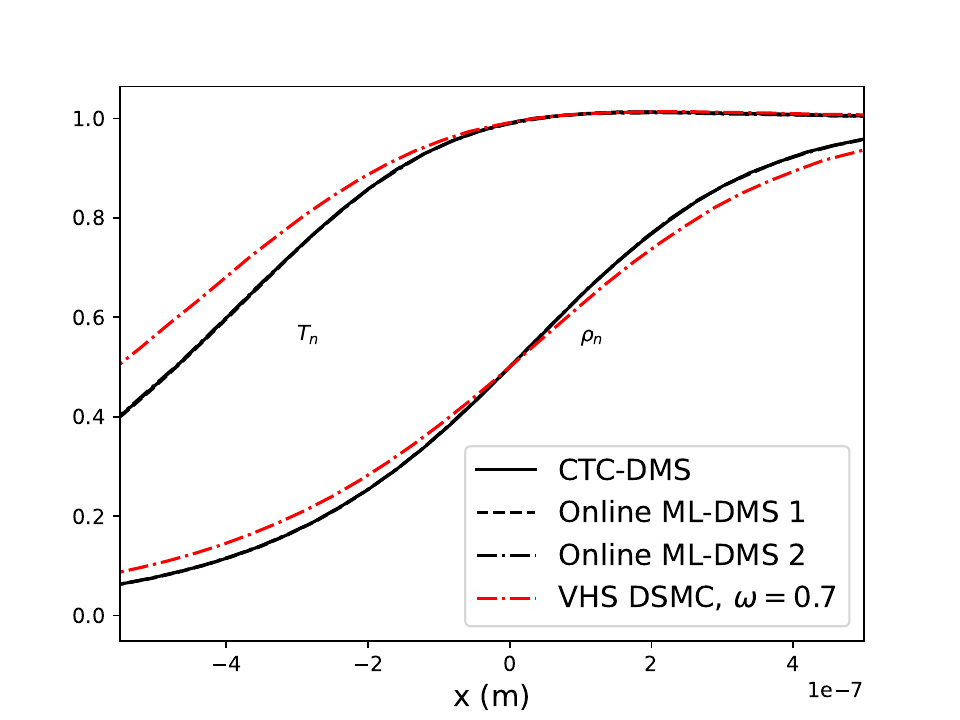}
    \caption{Mach 45}
    \end{subfigure}
    \caption{High Mach number shock profiles at moderate density conditions: $\rho_L=\qty{1}{\kilogram\per\cubic\metre}$, $T_L=\qty{300}{\kelvin}$.}
    \label{fig:moderate_extra_high}
\end{figure}
\begin{figure}
    \centering
    \begin{subfigure}[b]{0.45\textwidth}
        \includegraphics[width=\textwidth]{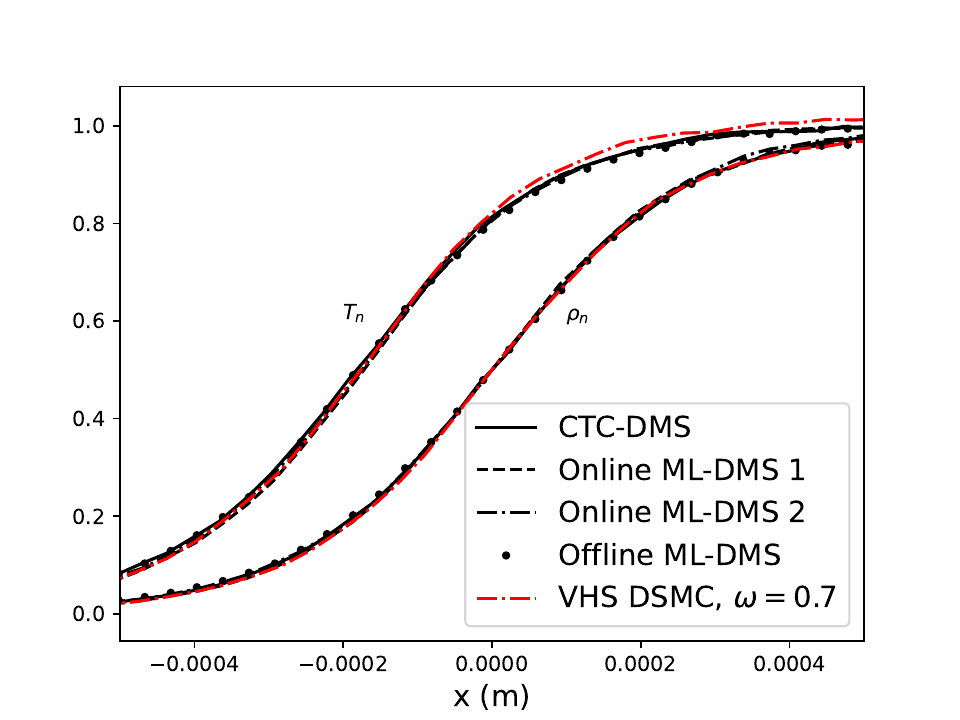}
    \caption{Mach 2}
    \end{subfigure}
    \centering
    \begin{subfigure}[b]{0.45\textwidth}
        \includegraphics[width=\textwidth]{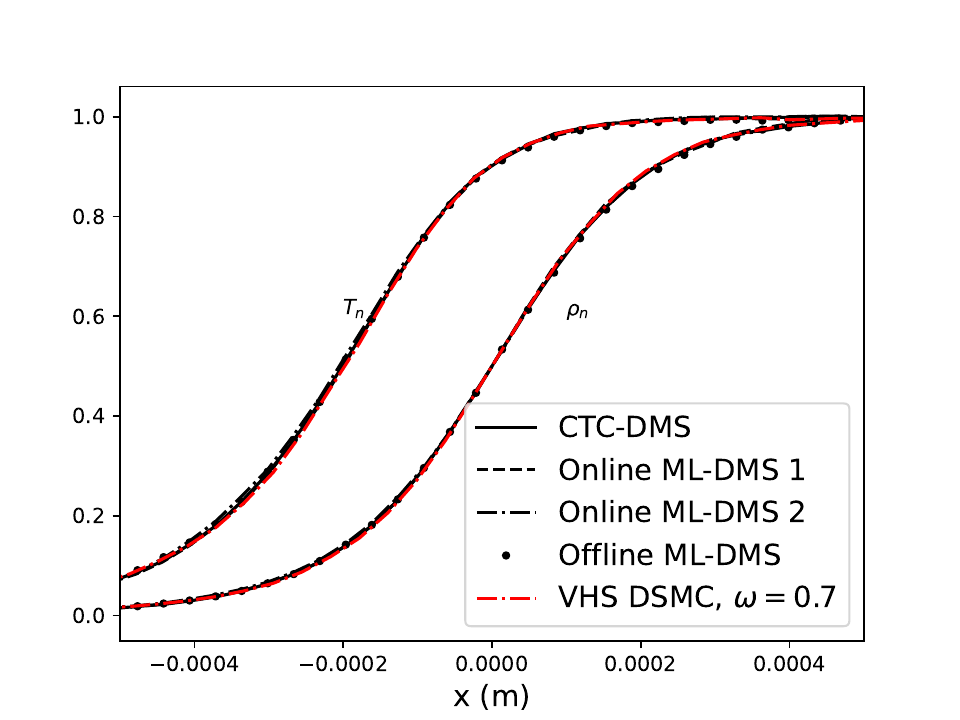}
    \caption{Mach 3}
    \end{subfigure}
    \centering
    
    \begin{subfigure}[b]{0.45\textwidth}
        \includegraphics[width=\textwidth]{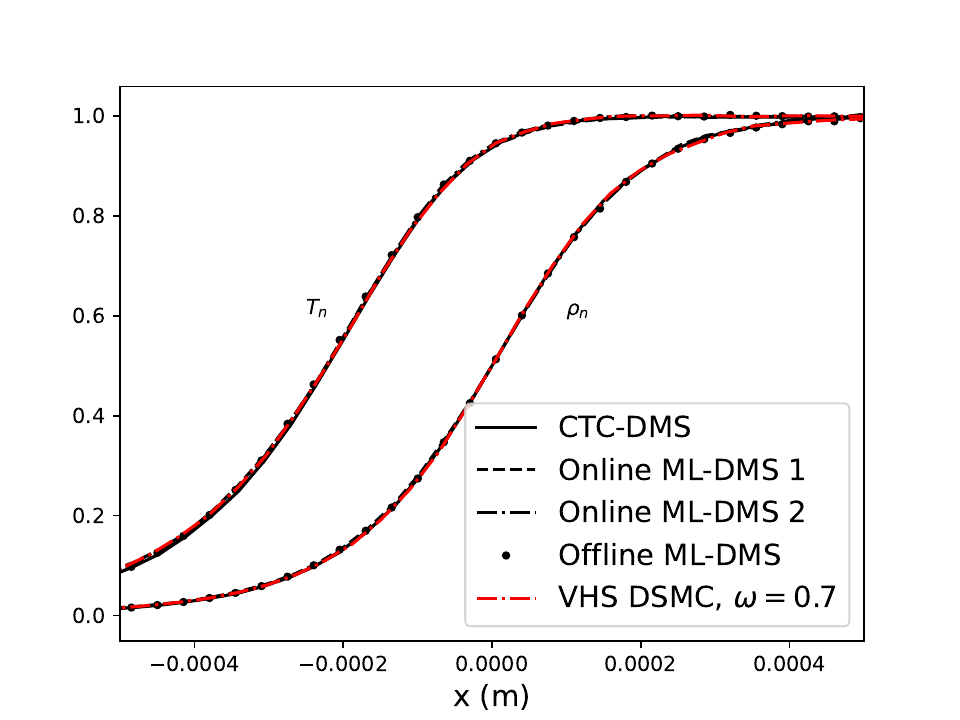}
    \caption{Mach 4}
    \end{subfigure}
    \centering
    \begin{subfigure}[b]{0.45\textwidth}
        \includegraphics[width=\textwidth]{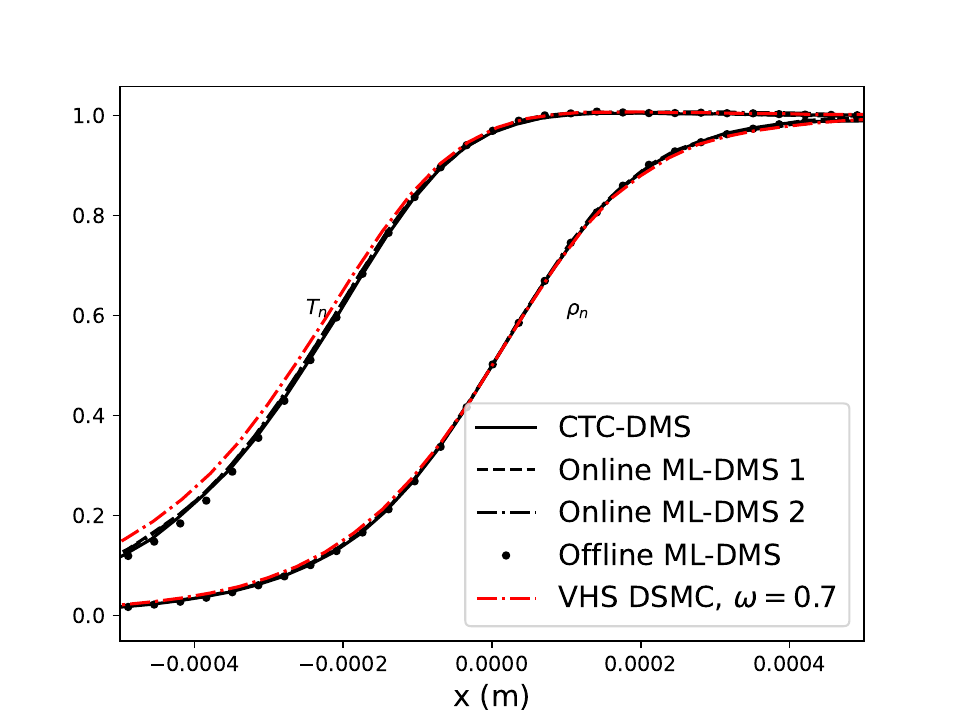}
    \caption{Mach 6}
    \end{subfigure}
    \centering
    \begin{subfigure}[b]{0.45\textwidth}
        \includegraphics[width=\textwidth]{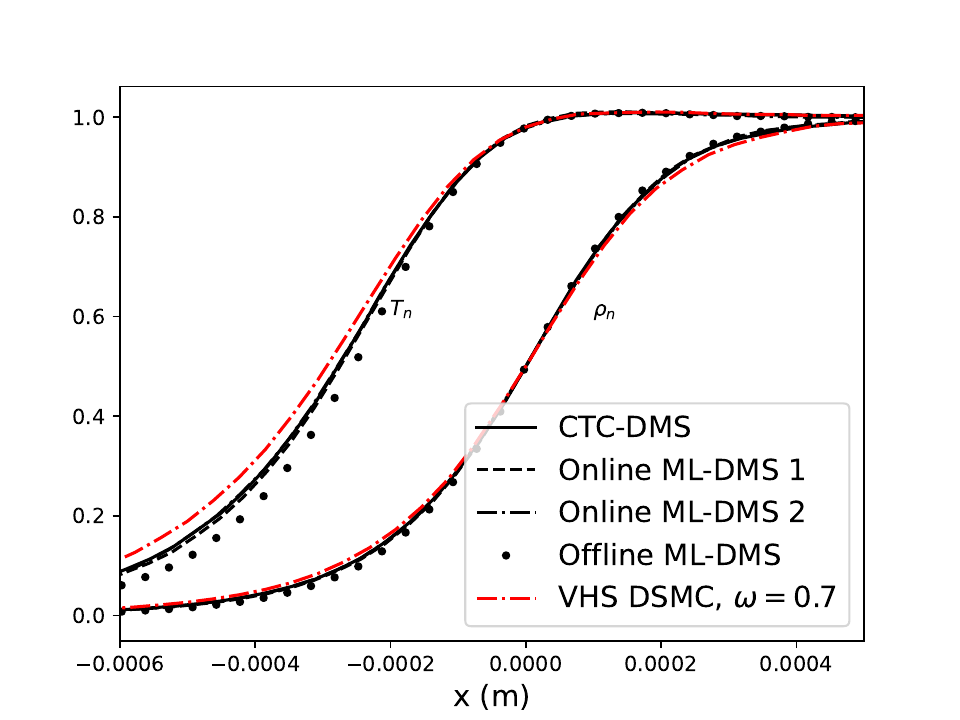}
    \caption{Mach 8}
    \end{subfigure}
    \caption{Shock profiles at rarefied conditions: $\rho_L=\qty{1e-3}{\kilogram\per\cubic\metre}$, $T_L=\qty{300}{\kelvin}$.}
    \label{fig:rarefied_extra}
\end{figure}

\subsection{Computational Efficiency}
A comparison of the computational cost of a single timestep across the methods considered in this work can be found in Table~\ref{table:comp_cost}, showing a reduction in computational time between CTC-DMS and ML-DMS of $\sim15 \times$ on a GPU. CPU simulations used an Intel 8360Y CPU. GPU simulations were carried out on a single Nvidia A100. Cross device portability of the code was achieved using the PyTorch library. The classical trajectory calculation subroutine was accelerated using PyTorch's Just-In-Time (JIT) compilation to TorchScript, but there was no additional code optimisation. Therefore, significant increases in computational speed are likely possible for all methods considered, on both CPU and GPU. 
\begin{table}
\centering
\begin{tabular}{ ccc } 
 \toprule
  Method & Walltime CPU (s) & Walltime GPU (s) \\ 
 \midrule
 DSMC & 1.45& 0.14\\ 
 DMS &  2848.2 & 32\\
 NN Offline & 17.0 &  1.8\\ 
 NN Online (training step)&  362.0 & 45\\
 NN Online (inference only)& 16.0 & 1.6 \\
 NN Online (average)& 18.5 & 2.2\\
 VHS Online (training) & 298.3 & 0.7\\
 VHS Online (average) & 199.2 & 0.53 \\
 \bottomrule
\end{tabular}
\caption{Comparison of average walltime for one DSMC timestep, for simulations at Mach 5, $\rho_L=\qty{1}{\kilogram\per\cubic\metre}$. Due to the larger cross section used, in DMS and Neural Network simulations each timestep includes 15 collision steps, whereas for DSMC each timestep includes only one collision step. The VHS online optimisation algorithm is described in Section~\ref{sec:VHS}. {For online simulations, average values are representative for the whole length of the simulation, where training steps are more computationally expensive.} } \label{table:comp_cost}
\end{table}

For example, the online training could be terminated once the loss function reaches a pre-determined level, saving computational time otherwise spent on unnecessary training. Training could alternatively be carried out in parallel on a second processor, with the most up-to-date model used at each timestep of the simulation. This parallel implementation is similar to the second online method proposed above, and should limit the time required for the simulation to that of evaluating the network. In order to simplify the implementation and facilitate comparison of the methods we do not investigate these potential optimisations here. It is also likely that a custom, optimised GPU implementation of both the trajectory calculations (such as that set out in \cite{NORMAN2013153}) and the underlying DSMC procedure would give significant improvements in computational efficiency.

With the present implementation, DMS simulations of 1500 timesteps, with $10^6$ particles, took around 14 hours to complete on the GPU. In contrast, the online neural network simulations took under an hour on the same machine. The online simulations required only around 5\% more total computational time than the offline method, for a significant improvement in accuracy.

\section{Online Optimisation of the VHS Model}
\label{sec:VHS}
Previous work on MD simulations of normal shock waves has found that VHS DSMC is able to reproduce experimental and MD results at a range of physical conditions, provided that the VHS parameters are properly chosen, with appropriate values of $\omega=0.7$ for a normal shock at $\qty{300}{\kelvin}$ and $\omega=0.81$ at $\qty{16}{\kelvin}$ reported in \cite{MDDilute}. Whereas the appropriate parameters must either be chosen by hand as in \cite{MDDilute}, or calibrated to viscosity data, we aim to develop an online optimisation method to calibrate DSMC parameters to DMS CTC data generated in situ during the DSMC prediction. The accuracy of our online DSMC optimisation method is evaluated for the VHS model in a 1D shock.

The goal of the online optimisation method is to select DSMC parameters from DMS CTC trajectory data such that the DSMC simulation will match the shock profiles obtained from DMS for a given interatomic potential. An objective function for the online optimisation must be constructed. Direct comparison of the DMS CTC collision model and DSMC collision model is typically not possible due to the different cross sections used in the acceptance-rejection steps of DMS and DSMC. We therefore minimise the conditional expectation of the discrepancy between collision outcomes for VHS and CTC given the velocities and positions of the particles.

\subsection{DSMC Online Optimisation}
\subsubsection{Uniformly Sampled Estimator}
We wish to find VHS parameters to minimise the magnitude of the expected difference in the outcome for a single particle between collision models, written as 
\begin{equation}
    L(\omega_t)=\frac{1}{2}\sum_{i=1}^N\left(\mathbb{E}_t[\chi_{\textsc{vhs}, \omega_t}^i(t) ]-\mathbb{E}_t[\chi^i_\CTCsub(t) ]\right)^2.
    \label{eqn:loss}
\end{equation}
Here $\chi_{X}^i$ represents the collision angle for particle $i$ by method $X$, which is a random variable depending on 
the choice of virtual collision partners, the acceptance-rejection step which determines whether a real or virtual collision takes place, and the choice of CTC impact parameter. Expectations $\mathbb{E}_t$ are taken conditional on the positions and velocities of all particles at the current time step.

In order to use stochastic gradient descent to find an optimal parameter $\omega^*$, we evaluate the gradient of this loss with respect to $\omega$ as 
\begin{equation}
    \nabla_\omega L = \sum_{i=1}^N\left(\mathbb{E}_t[\chi_{\textsc{vhs}, \omega_t}^i(t)]-\mathbb{E}_t[\chi^i_\CTCsub]\right)\nabla_\omega \mathbb{E}_t[\chi_{\textsc{vhs}, \omega_t}^i(t)].
\end{equation}
We have 

\begin{equation}
    \mathbb{E}_t[\chi_{{\textsc{vhs}, \omega_t}}^i]=\sum_{j\neq i} \frac{\pi}{2}P(\text{real collision} | \text{virtual collision between i, j}) \times P(\text{virtual collision between i, j}),\label{eqn:expec}
\end{equation}
since, by symmetry, the expectation of the collision angle given that a real collision takes place is $\frac{\pi}{2}$. Let $g_{ij}=\lvert v_i-v_j\rvert$. Taking $\Sigma$ as an upper bound for the cross section $\sigma(g)g$ over the cell, from the Nanbu--Babovsky DSMC algorithm \cite{bird_book}\begin{equation}
P(\text{real collision} | \text{virtual collision between i, j})=\frac{\sigma_{\omega_t}(g_{ij})g_{ij}}{\Sigma},
\end{equation}
and
\begin{eqnarray}
    P(\text{virtual collision between i, j}) &=& \frac{2N_c}{N} \times \frac{2N_c-1}{N-1} \times \frac{1}{2N_c-1} \notag \\
    &=&\frac{2N_c}{N(N-1)} \notag \\
    &=& \frac{W_p\Sigma\Delta t}{V_\mathrm{cell}}.
    \label{eqn:p_virtual}
\end{eqnarray}
In Equation~\ref{eqn:p_virtual}, the first term is the probability that particle $i$ is selected as one of the $2N_c$ particles for virtual collision. The second term is the probability for $j$ to be selected given that $i$ has been. The final term is the probability of $j$ to be selected as the collision partner for $i$. Therefore

\begin{equation}
    \mathbb{E}_t[\chi_{\textsc{vhs}, \omega_t}^i(t)]=\sum_{j\neq i}\frac{\pi}{2}\sigma_{\omega_t}(g_{ij})g_{ij}\frac{W_p\Delta t}{V_\mathrm{cell}}.
    \label{eqn:coll}
\end{equation}
As required, this quantity is dimensionless and physically represents the expected number of collisions for a single physical particle in one time step, multiplied by $\frac{\pi}{2}$. We do not wish to calculate this sum as it would be costly to compute $N(N-1)$ cross sections, one for each pair of particles in the simulation. We can instead view the sum as an expectation over uniformly sampled pairs. Therefore we have an un-biased, one-sample, Monte Carlo estimator for the expectation\begin{equation}
    \hat{E}^i_{\omega_t}=(N-1)\frac{\pi}{2}\sigma_{\omega_t}(g_{ik})g_{ik}\frac{W_p\Delta t}{V_\mathrm{cell}},
\end{equation}
where particle $k$ is sampled uniformly at random from the $N-1$ potential collision pairs. We now directly evaluate the gradient of the expectation\begin{equation}
    \nabla_\omega \mathbb{E}_t[\chi_{\textsc{vhs}, \omega_t}^i]= \sum_{j\neq i}\frac{\pi}{2}\nabla_\omega(\sigma_{\omega_t}(g_{ij}))g_{ij}\frac{W_p\Delta t}{V_\mathrm{cell}},\label{eqn:expec_grad}
\end{equation}
and similarly obtain an un-biased estimate for the gradient as\begin{equation}
    \hat{\eta}^i=(N-1)\frac{\pi}{2}\nabla_\omega(\sigma_{\omega_t}(g_{ik}))g_{ik}\frac{W_p\Delta t}{V_\mathrm{cell}},\label{eqn:uniform}
\end{equation}
where particle $k$ is chosen uniformly, and the derivative can be evaluated with automatic differentiation.

Using these expressions, we obtain an estimator for the term\begin{equation}
    \mathbb{E}_t[\chi_{\textsc{vhs}, \omega_t}^i]\nabla_\omega \mathbb{E}_t[\chi_{\textsc{vhs}, \omega_t}^i]=\frac{\pi^2}{4}\frac{W_p^2\Delta t^2}{V_c^2}\sum_{j\neq i}\sum_{k \neq i} \sigma_{\omega_t}(g_{ik})g_{ik}\nabla_\omega(\sigma_{\omega_t}(g_{ij}))g_{ij}.
\end{equation}
We again obtain an unbiased estimate by picking two collision partners $j,k$ uniformly, with the possibility that $k=j$
\begin{equation}
    \widehat{E \eta} = \frac{\pi^2}{4}\frac{W_p^2\Delta t^2}{V_c^2} (N-1)^2 \sigma_{\omega_t}(g_{ik})g_{ik}\nabla_\omega(\sigma_{\omega_t}(g_{ij}))g_{ij}=\hat{E}\hat{\eta}.
    \label{eqn:grad_estimator}
\end{equation}
An unbiased estimate of the expectation of the DMS collision angle is given by the same procedure, sampling a third collision partner independently, as well as an impact parameter. 
\begin{equation}
    \hat{E}^i_\CTCsub=(N-1)\chi_{DMS}(g_{il},b_\mathrm{max}(g_{il})R^{\frac{1}{2}})\sigma_\mathrm{LJ}(g_{il})g_{il}\frac{W_p\Delta t}{V_\mathrm{cell}}.
\end{equation} The Monte Carlo estimate for the gradient of the loss $\nabla_\omega L$ therefore is
\begin{equation}
    \widehat{\nabla_\omega L}=\sum_i^N\left(\hat{E}^i_{\omega_t}-\hat{E}^i_\CTCsub\right)\hat{\eta}^i.
    \label{eqn:estimator}
\end{equation}
This sampled gradient can now be used in Stochastic Gradient Descent to find the value of $\omega$ for which the expected discrepancy in collision angle is minimised.

Our online optimisation method takes a single stochastic gradient descent step at each DSMC timestep -- updating the collision model in parallel as the DSMC simulation progresses -- during the transient phase of the simulation before averaging begins. At each timestep, the estimator Equation~\ref{eqn:estimator} is sampled for a subset of the particles. We use a sample of size $N_\mathrm{train}=50{,}000$, drawn uniformly from all particles in the domain. For each of these particles, a one-sample estimator of the gradient is used, considering independent collision partners for each term. At timestep $i$, the VHS parameter is updated according to \begin{equation}
    \omega_{i+1}=\omega_i-\alpha_i\widehat{\nabla_\omega L}(\omega_i),
\end{equation}
As in the neural network training, the learning rate $\alpha$ is scheduled to decay according to \begin{equation}
    \alpha_i=\alpha_0\frac{A}{B+Ci},
\end{equation}
with $A=B=100$, $C=1$. In this case, $\alpha_0$ is constant instead of chosen adaptively according to RMSProp.

\subsubsection{Alternative Gradient Estimator}
The above estimator $\hat{\eta}$ for the gradient of the expectation over the DSMC step relies on uniform sampling over possible collision partners. An alternative estimator based on the score function approach, with samples drawn from the possible outcomes of the DSMC procedure, is also natural but in fact has a higher variance. 

In this case, we rewrite the gradient of the expectation, Equation~\ref{eqn:expec_grad}, as, \begin{align}
 \nabla_\omega \mathbb{E}_t[\chi_{\textsc{vhs}, \omega_t}^i]&=\sum_{j \neq i} \pi/2 ( \nabla_{\omega} \log P_{\omega_t}[ \text{real collision of i and j}] )  P_{\omega_t}[ \text{real collision of i and j}],\\
 &=\mathbb{E}_{j\sim P_{\omega_t}[ \text{real collision of i and j}]}[\pi/2 ( \nabla_{\omega} \log P_{\omega_t}[ \text{real collision of i and j}] )],
\end{align}
where the collision partner $j$ is now distributed according to the probability of undergoing a real collision with $i$. We can then sample a Monte Carlo estimate for this expectation as \begin{equation}
    \hat{\eta}'=\frac{\pi}{2}\nabla_{\omega} \log P_{\omega_t}[ \text{real collision of i and }\hat{j}],
\end{equation}
by first uniformly sampling a virtual collision partner $\hat{j}$, with probability $\frac{2N_c}{N}$, and then accepting or rejecting it as in DSMC. With probability $1-\frac{2N_c}{N}$, or if we reject the sample, we have sampled the event ``no real collision'' and the estimate is zero. The variance of this estimator depends on\begin{align}
    \mathbb{E}_t[\hat{\eta}'^2]&=\frac{\pi^2}{4}\sum_{i\neq j}( \nabla_{\omega} \log P_{\omega}[ \text{real collision of i and j}] )^2P_{\omega}[ \text{real collision of i and j}],\\
    &=\frac{\pi^2}{4}\sum_{i\neq j}\left(\frac{\nabla_\omega P_{\omega,ij}}{P_{\omega,ij}}\right)^2P_{\omega,ij},\\
    &=\frac{\pi^2}{4}\sum_{i\neq j}\frac{(\nabla_\omega P_{\omega,ij})^2}{P_{\omega,ij}}> \frac{\pi^2}{4}\sum_{i\neq j}(\nabla_\omega P_{\omega,ij})^2 (N-1).
\end{align}
The final inequality is because we sample possible pairs uniformly and then have a chance to reject, and so $P_{\omega,ij}<\frac{1}{N-1}$. On the other hand, the variance of the first estimator relying on uniform sampling (Equation~\ref{eqn:uniform}) depends on\begin{align}
    \mathbb{E}_t[\hat{\eta}^2]&=(N-1)^2\frac{\pi^2}{4}\sum_{i\neq j}( \nabla_{\omega} P_{\omega,ij} )^2\frac{1}{N-1},\\
    &=(N-1)\frac{\pi^2}{4}\sum_{i\neq j}( \nabla_{\omega} P_{\omega,ij} )^2,
\end{align}
therefore  $\mathbb{E}_t[\hat{\eta}'^2]>\mathbb{E}_t[\hat{\eta}^2]$, and so, \begin{align}
    \mathrm{Var}[\hat{\eta}']&=\mathbb{E}_t[\hat{\eta}'^2]-\mathbb{E}_t[\hat{\eta}']^2\\&=\mathbb{E}_t[\hat{\eta}'^2]-\mathbb{E}_t[\hat{\eta}]^2 \\&> \mathbb{E}_t[\hat{\eta}^2]-\mathbb{E}_t[\hat{\eta}]^2 \\&= \mathrm{Var}[\hat{\eta}],
\end{align}
since both estimators are unbiased and so in particular $\mathbb{E}_t[\hat{\eta}]=\mathbb{E}_t[\hat{\eta}']$. In addition to having higher variance, the sampling procedure for $\hat{\eta}'$ is more complex. Therefore, the uniform sampling method is numerically preferable and used in our implementation.

\subsection{Numerical Results}
\subsubsection{Convergence of the Estimator}
\begin{figure}
    \centering
    \includegraphics[width=0.5\linewidth]{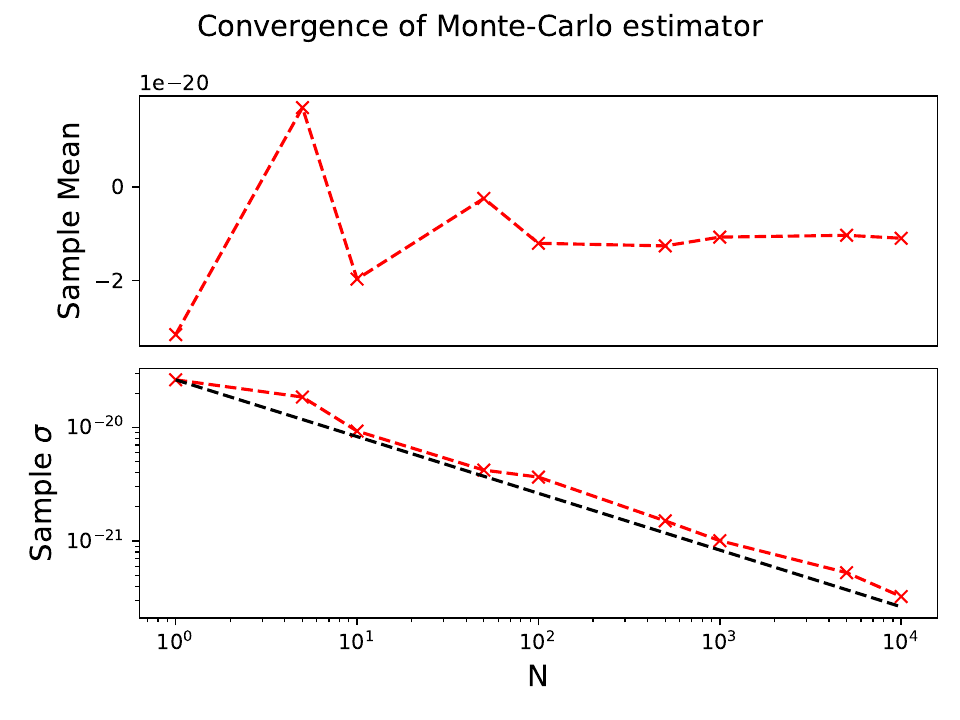}
    \caption{Convergence of the single-particle estimator as the sample size is increased. The black line in the lower plot shows the expected $\frac{1}{\sqrt{n}}$ convergence.}
    \label{fig:MC_convergence}
\end{figure}

Figure~\ref{fig:MC_convergence} displays the convergence of Monte Carlo estimates for the gradient of the single particle expectation (Equation~\ref{eqn:expec_grad}) with increasing sample size $n$, as well as the empirical standard deviation over twenty repeated calculations for each $n$. As expected, the sampled values converge, and the standard deviation of the estimator decreases at a rate proportional to $\frac{1}{\sqrt{n}}$. We take $n=1$, which simplifies the implementation and is sufficient for use in stochastic gradient descent.

\subsubsection{Online Optimisation of DSMC: Numerical Results for 1D Shocks}
\label{sec:VHS_profiles}
\begin{figure}
    \centering
    \begin{subfigure}[t]{0.48\textwidth}
        \includegraphics[width=\textwidth]{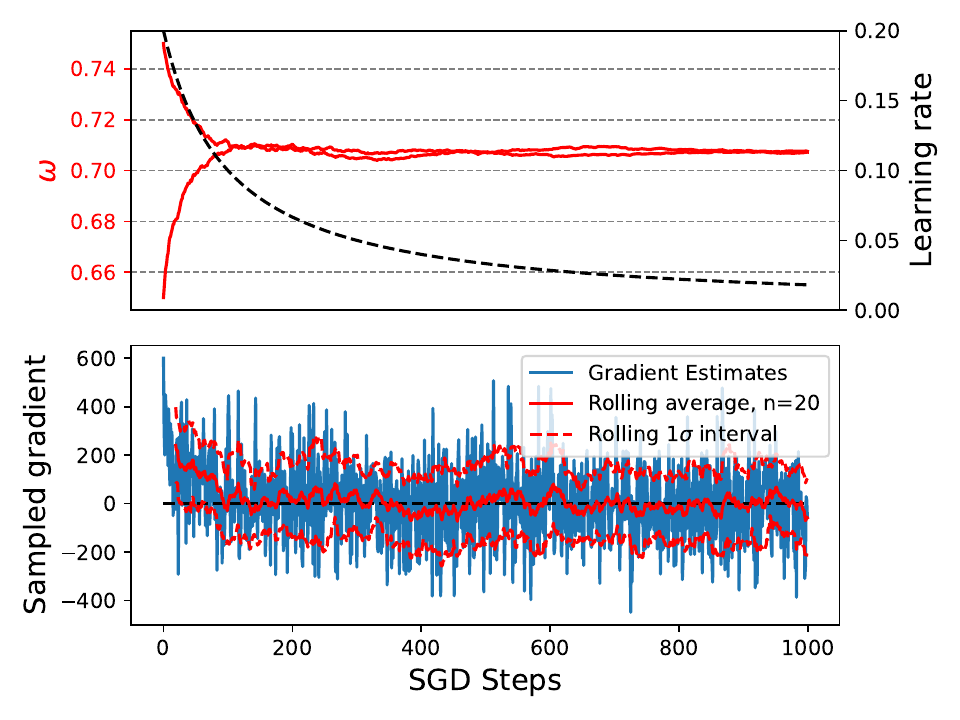}
    \caption{Mach 5, $\alpha_0=0.2$}
    \end{subfigure}
    \begin{subfigure}[t]{0.48\textwidth}
        \includegraphics[width=\textwidth]{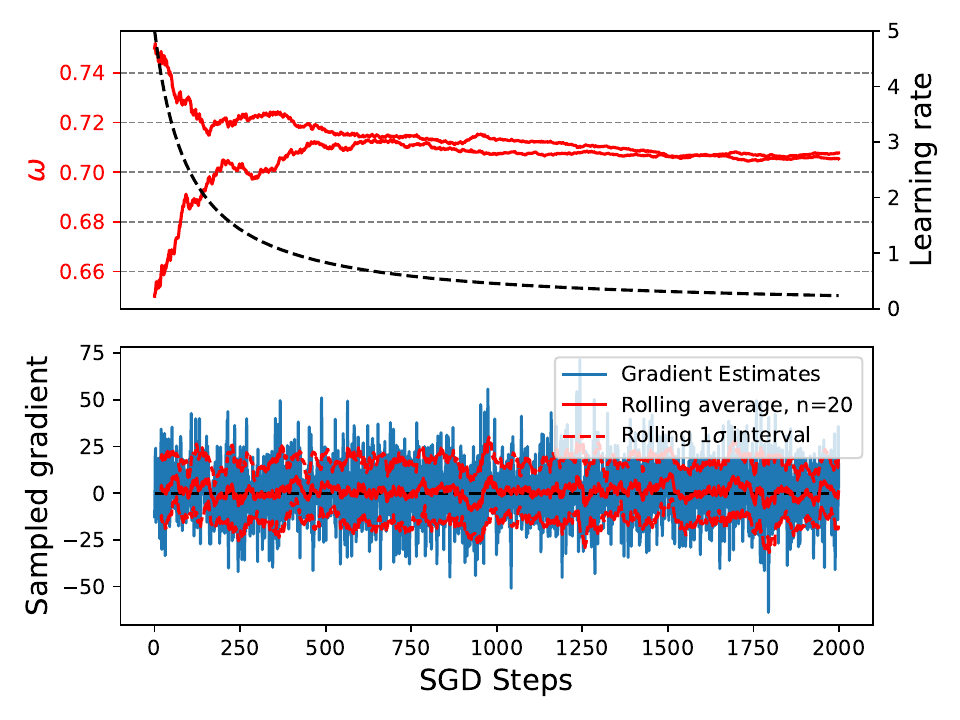}
    \caption{Mach 7.183, $\alpha_0=5$}
    \end{subfigure}
    \begin{subfigure}[t]{0.48\textwidth}
        \includegraphics[width=\textwidth]{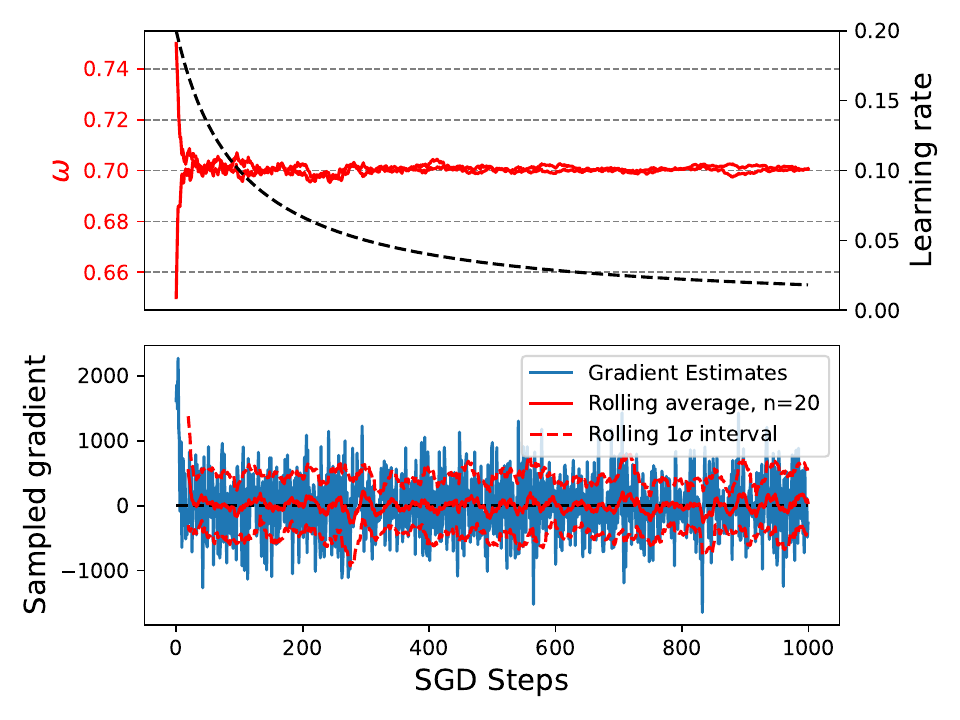}
    \caption{Mach 9, $\alpha_0=0.2$}
    \end{subfigure}
    \begin{subfigure}[t]{0.48\textwidth}
        \includegraphics[width=\textwidth]{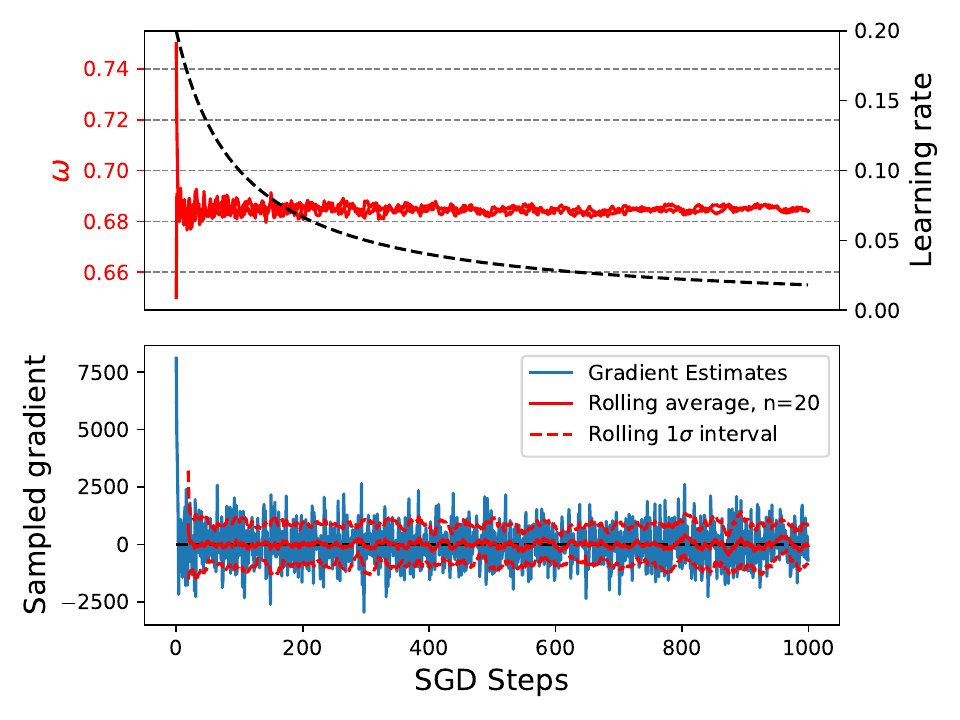}
    \caption{Mach 30, $\alpha_0=0.2$}
    \end{subfigure}
    \caption{Progress of online VHS optimisation at density of $\rho_L=\qty{1}{\kilogram\per\cubic\metre}$, for shocks at Mach 5, 9, 30 ($T_L=\qty{300}{\kelvin}$) and Mach 7.183 ($T_L=\qty{16}{\kelvin}$) using the uniformly sampled Monte Carlo gradient estimator in DSMC. A higher learning rate was required for the lower temperature conditions which have a lower typical relative velocity. }
    \label{fig:mach_5_vhs_MC}
\end{figure}

Figure~\ref{fig:mach_5_vhs_MC} shows the progress of training for shocks at $\rho_L=\qty{1e-3}{\kilogram\per\cubic\metre}$, with Mach 5, 9, 30 at a temperature of $T_L=\qty{300}{\kelvin}$, and Mach 7.183 at $T_L=\qty{16}{\kelvin}$. The results of two training runs are plotted to show that the stochastic gradient descent approaches convergence, and is not simply flattening off due to the decaying learning rate. The plots also show the sampled gradient estimator and its rolling average value. The estimator is very noisy, but nevertheless the stochastic gradient descent converges to an accurate value.

The computational cost of the online VHS DSMC simulation is shown in Table~\ref{table:comp_cost} to be $\sim4\times$ larger than a DSMC simulation with fixed parameters. By comparison, a binary search for the optimal $\omega^*\in[0.5,1]$  based on visual comparison requires 5 DSMC simulations in order to reach $\omega^*\pm0.01$, as well as a relatively costly CTC-DMS run to generate reference profiles. Figure~\ref{fig:mach_5_vhs_MC} shows that when a single stochastic gradient descent step is taken at each DSMC timestep, for most conditions considered, the simulation converges rapidly and most of the training time is spent close to the learned optimal value. It is therefore possible in some cases to reduce the overhead of the online training procedure by training for a smaller number of steps. However, for the cold shock at Mach 7.183, the full length of training is required. We aim here to provide consistent, conservative values of the training hyperparameters which have been observed to perform well in practice, and which in any case result in a relatively small extra computational cost compared to a single DSMC simulation. 

\begin{figure}
    \centering
    \begin{subfigure}[t]{0.49\textwidth}
        \includegraphics[width=\textwidth]{argon_figs/VHS_optimisation/MC/moderate/VHS_comp_M5.0.pdf}
    \caption{Mach 5, $T_L=\qty{300}{\kelvin}$}
    \end{subfigure}
    \centering
    \begin{subfigure}[t]{0.49\textwidth}
        \includegraphics[width=\textwidth]{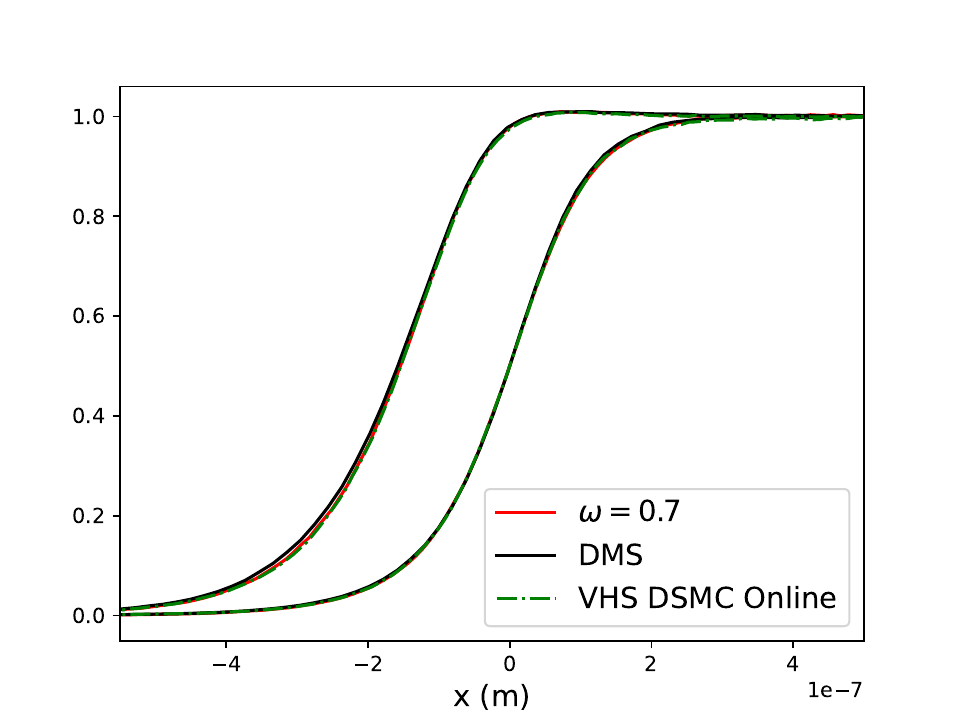}
    \caption{Mach 7.183, $T_L=\qty{16}{\kelvin}$}
    \end{subfigure}
    \centering
    \begin{subfigure}[t]{0.49\textwidth}
        \includegraphics[width=\textwidth]{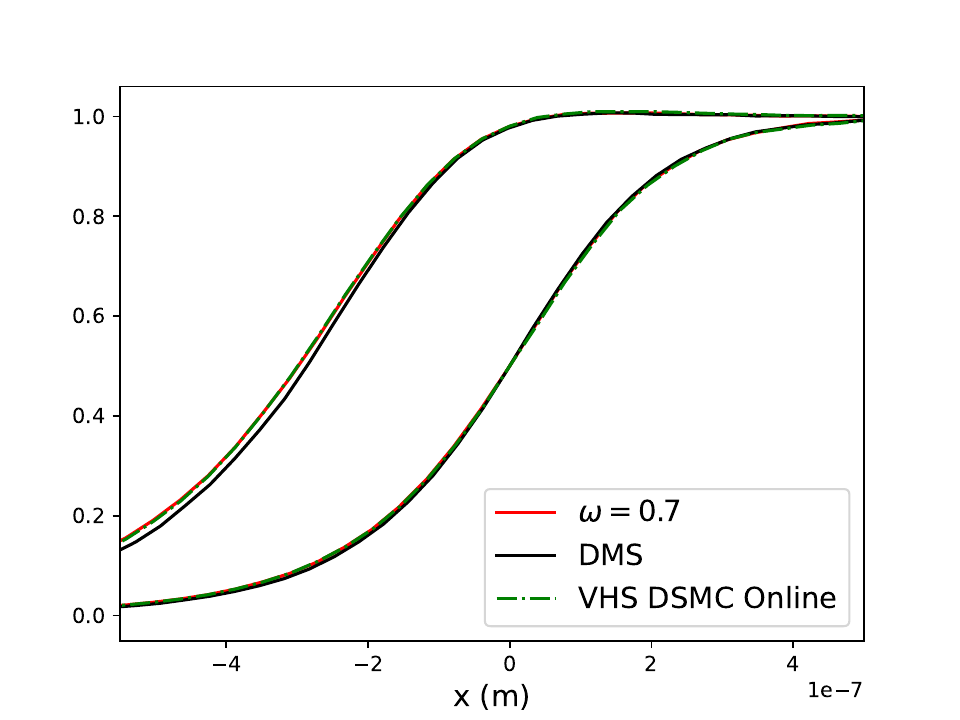}
    \caption{Mach 9, $T_L=\qty{300}{\kelvin}$}
    \end{subfigure}
    \centering
    \begin{subfigure}[t]{0.49\textwidth}
        \includegraphics[width=\textwidth]{argon_figs/VHS_optimisation/MC/moderate/VHS_comp_M30.0.pdf}
    \caption{Mach 30, $T_L=\qty{300}{\kelvin}$}
    \end{subfigure}
    \centering
    \begin{subfigure}[t]{0.49\textwidth}
        \includegraphics[width=\textwidth]{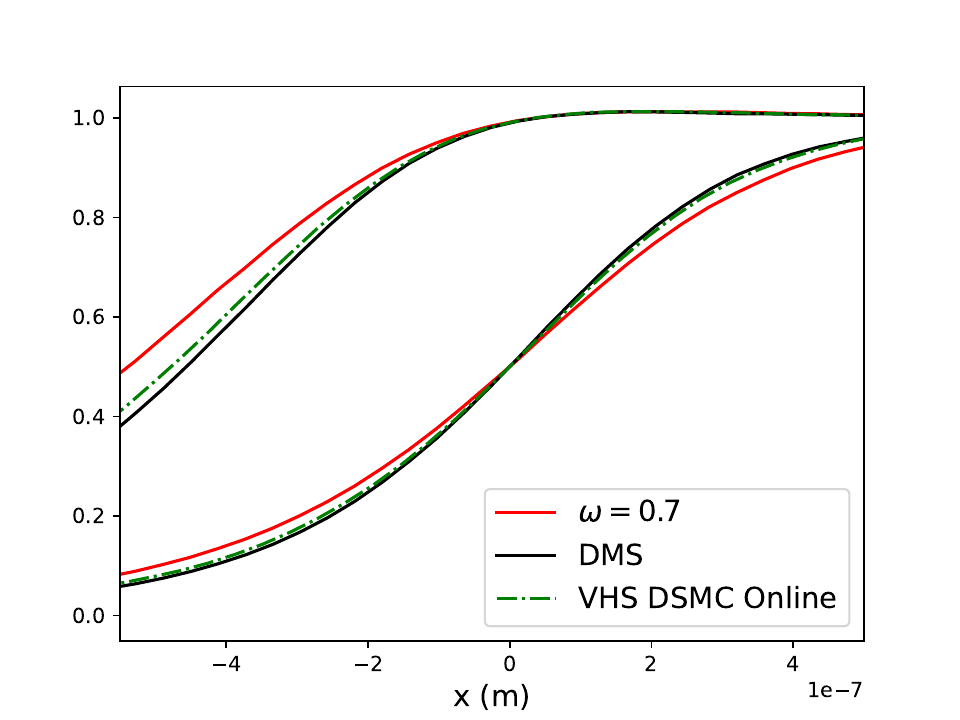}
    \caption{Mach 40, $T_L=\qty{300}{\kelvin}$}
    \end{subfigure}
    \centering
    \begin{subfigure}[t]{0.49\textwidth}
        \includegraphics[width=\textwidth]{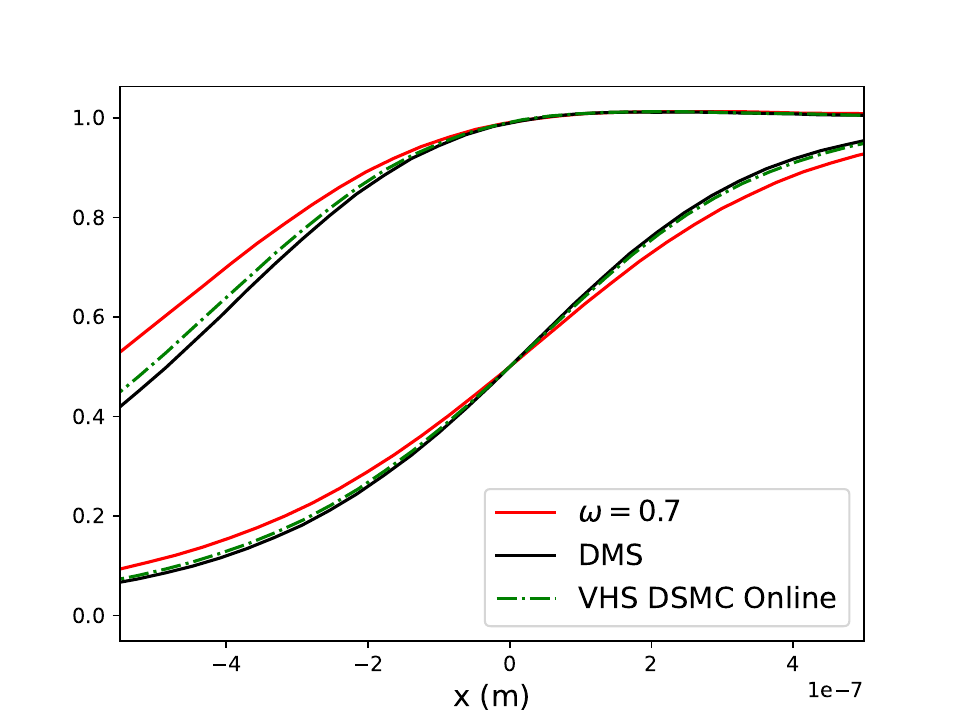}
    \caption{Mach 50, $T_L=\qty{300}{\kelvin}$}
    \end{subfigure}
    \caption{Shock profiles obtained with online VHS optimisation, compared with reference VHS parameters and DMS, at $\rho_L=\qty{1}{\kilogram\per\cubic\metre}$.}
    \label{fig:vhs_MC_profiles}
\end{figure}

The resulting shock profiles are shown in Figure \ref{fig:vhs_MC_profiles}, comparing the online DSMC, CTC-DMS and VHS DSMC with the reference parameters $\omega=0.7$, $d_\mathrm{ref}=\qty{3.974e-10}{\metre}$, $T_\mathrm{ref}=\qty{273}{\kelvin}$. The shock profiles from the online optimised DSMC are able to reproduce or improve upon the results obtained from the reference values. In particular at higher Mach numbers -- where standard VHS DSMC (with the reference value of $\omega = 0.7$) exhibits a large discrepancy with the CTC-DMS profiles -- the optimised parameter value gives a much closer fit to both density and temperature. The key feature of our approach is that the $\omega$ can be calibrated online during a DSMC simulation to a relatively small subset of CTC trajectories. Calibration does not require the (computationally costly) full DMS simulation, which may even be computationally infeasible for higher dimensional configurations with more complex geometries.

\begin{figure}
    \centering
    \begin{subfigure}[t]{0.49\textwidth}
        \includegraphics[width=\textwidth]{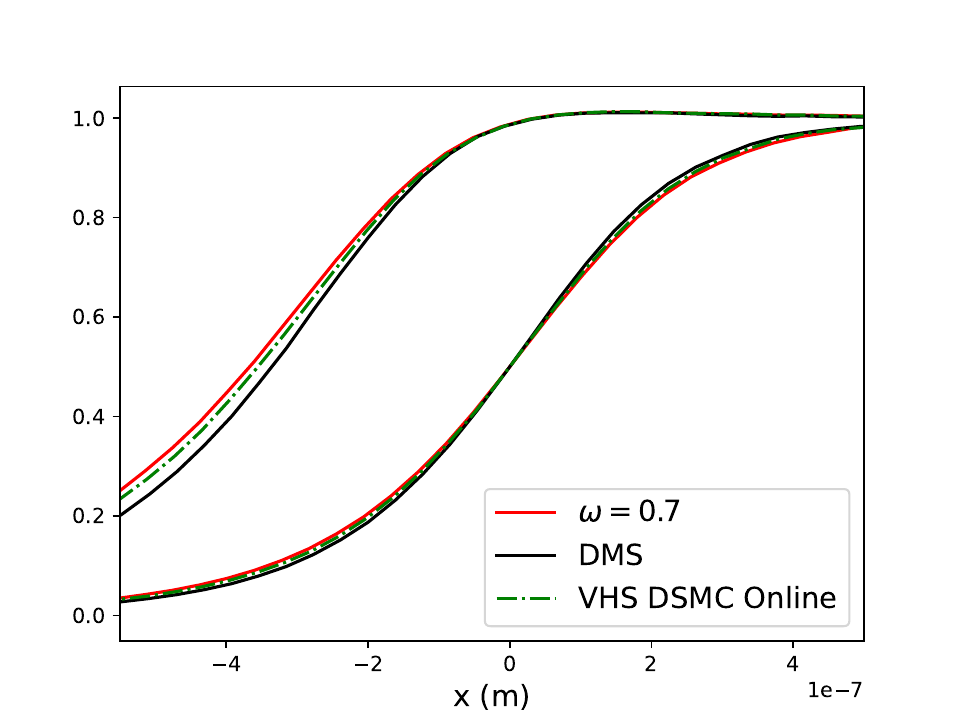}
    \caption{Mach 15}
    \end{subfigure}
    \centering
    \begin{subfigure}[t]{0.49\textwidth}
        \includegraphics[width=\textwidth]{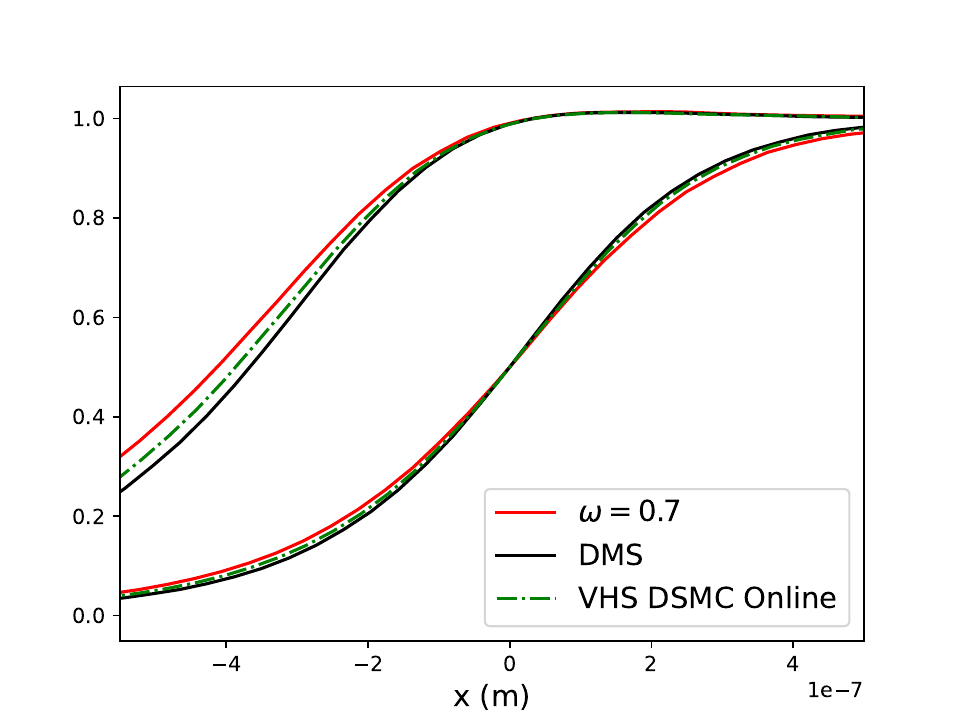}
    \caption{Mach 20}
    \end{subfigure}
    \centering
    \begin{subfigure}[t]{0.49\textwidth}
        \includegraphics[width=\textwidth]{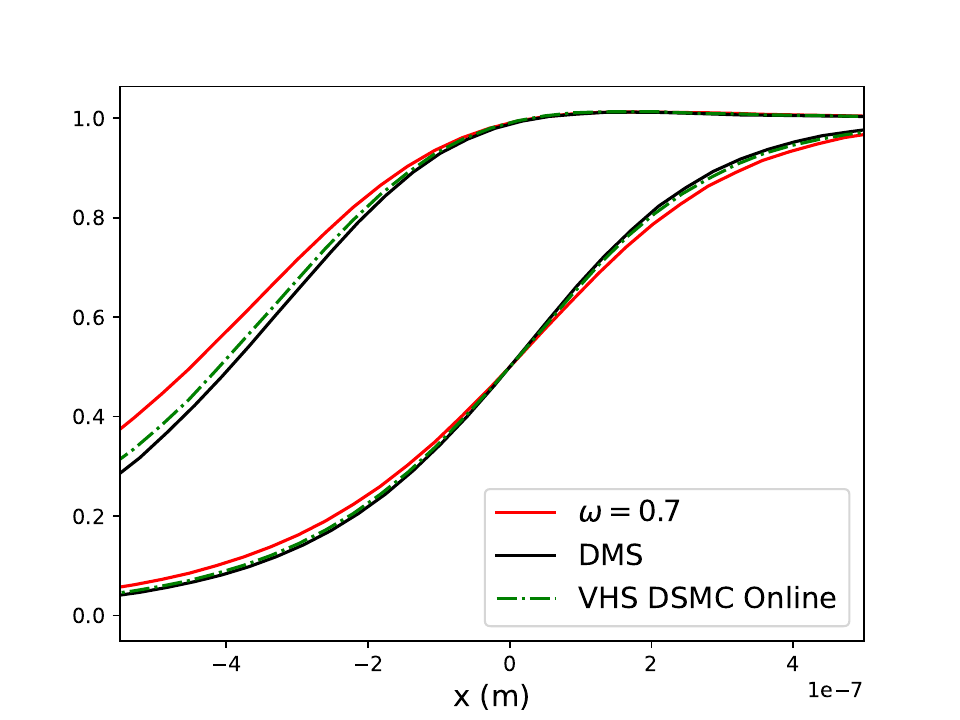}
    \caption{Mach 25}
    \end{subfigure}
    \centering
    \begin{subfigure}[t]{0.49\textwidth}
        \includegraphics[width=\textwidth]{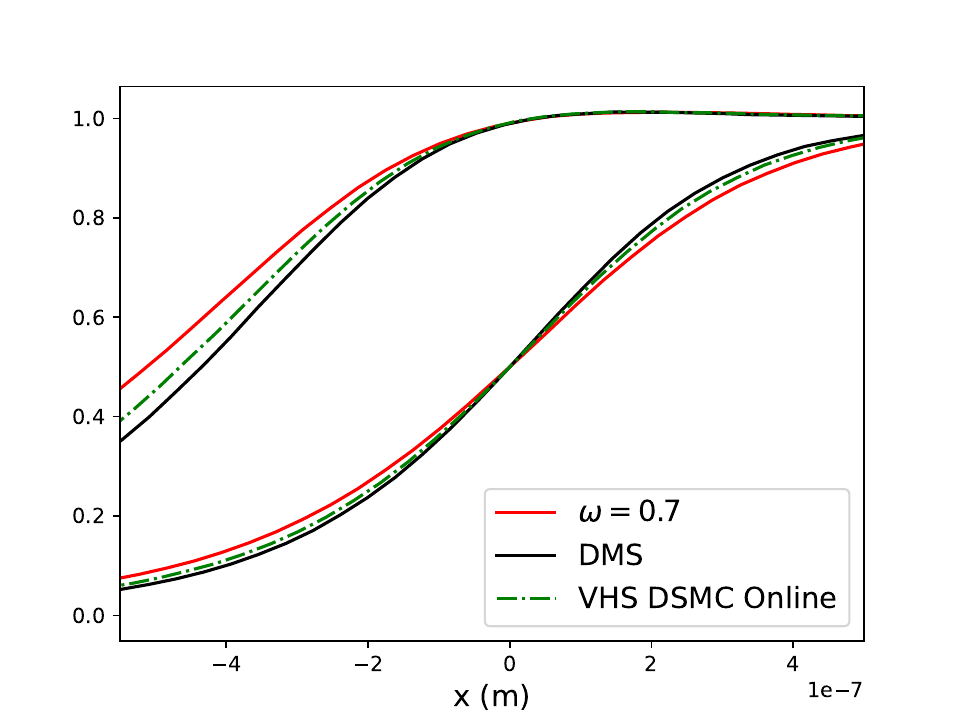}
    \caption{Mach 35}
    \end{subfigure}
    \centering
    \begin{subfigure}[t]{0.49\textwidth}
        \includegraphics[width=\textwidth]{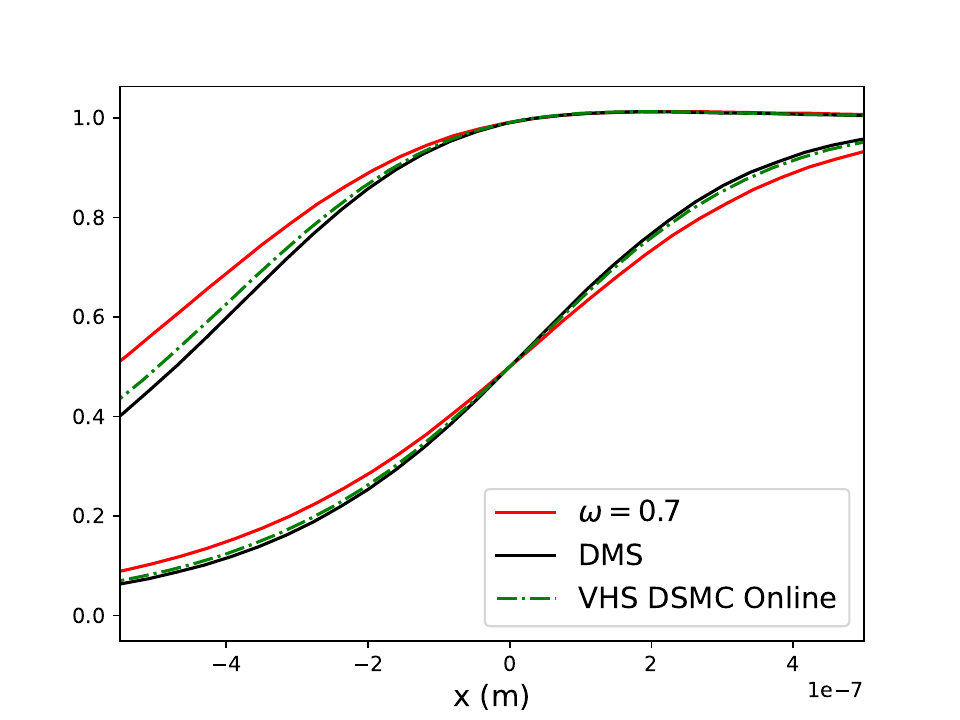}
    \caption{Mach 45}
    \end{subfigure}
    \caption{Additional shock profiles obtained with online VHS optimisation, compared with reference VHS parameters and DMS, at $\rho_L=\qty{1}{\kilogram\per\cubic\metre}$, $T_L=\qty{300}{\kelvin}$.}
    \label{fig:vhs_MC_profiles_additional}
\end{figure}

The optimal values of $\omega$ are observed to depend on Mach number. The results of DSMC with the reference parameters deviate significantly from DMS at high Mach numbers above Mach 9, due to the higher downstream temperatures, and the online trained VHS model is able to much more closely reproduce both the density and temperature profiles of the CTC-DMS. As noted in \cite{MDDilute}, the best value of the VHS $\omega$ is observed to increase for colder temperatures. 

It is interesting that an objective function incorporating the expected collision angle for a given particle should give accurate results. The collision angle is defined in the centre of mass frame of a particular collision, and therefore would seem not to be meaningful without a specific collision partner having been chosen. However, as pointed out above, for VHS Equation~\ref{eqn:coll} is related to the expected number of collisions that a particle will undergo during a timestep. The analogous expectation for the CTC collision angle can then be seen as an effective cross section for the Lennard--Jones interaction under the conditions encountered in the flow, and therefore as a natural candidate for calibration of the VHS model. 

\section{Conclusion}
This work has developed an online optimisation method for calibrating Direct Molecular Simulation and DSMC collision models during simulations using in situ generation of trajectory data. Our numerical results demonstrate that online-optimised neural network collision models are capable of reproducing shock profiles obtained by DMS and MD at a significantly reduced computational cost. The online-optimised neural network collision model is also more accurate (as compared to DMS) than standard VHS DSMC for higher Mach numbers. 

We have also compared offline (the standard approach for existing scientific machine learning models) training to our online training method for calibrating neural network collision models. Training a neural network model offline on trajectory data is adequate if the model is required to run multiple simulations in similar conditions. However, our numerical results show that the offline method is limited by the range and distribution of the training dataset: for prediction cases far from the physical conditions of the training dataset, the offline-trained model becomes less accurate. Therefore, the applicability of offline models to predictions on new physical conditions can be limited by the size of the training dataset. The online-optimised collision models developed in this paper are able to accurately generalise to new prediction cases at low computational cost. 

Although this paper has implemented and evaluated the method for 1D shocks, the online-optimisation method could also be easily applied to more complex, higher-dimensional geometries. Implementation of the method for higher-dimensional spatial flows could be the focus of future research. The method could also be extended to flows with active rotational and vibrational degrees of freedom. The collision processes to be learned in this case are more complex, taking into account the internal energies of the colliding molecules. Existing DMS methods would therefore be substantially more computationally expensive than in the monatomic case, suggesting that the online-optimised ML-DMS method could provide significant computational benefits.

We have also developed a method for online-optimisation of DSMC using a new Monte Carlo estimator for the gradient of the expected scattering angle over a single step. We applied this algorithm to calibrate the VHS model parameter for 1D shocks. The online-optimised DSMC shock profile has significantly higher accuracy than standard VHS DSMC as compared to full CTC-DMS, with computational cost $\sim20 \times$ lower than DMS. 

\section*{Acknowledgements}
This publication is based on work supported by the EPSRC Centre for Doctoral Training in Mathematics of Random Systems: Analysis, Modelling and Simulation (EP/S023925/1), and by the U.S.\ Department of Defense, Office of Naval Research, under Award N00014-22-1-2441. The computations described in this research were performed using the Baskerville Tier 2 HPC service (https://www.baskerville.ac.uk/). Baskerville is funded by the EPSRC and UKRI through the World Class Labs scheme (EP/T022221/1) and the Digital Research Infrastructure programme (EP/W032244/1) and is operated by Advanced Research Computing at the University of Birmingham. We thank Narendra Singh for many helpful discussions on the DMS method and Adam C.~Jones for typographical notes.

\bibliographystyle{my-elsarticle-num} 
\bibliography{biblio}

\end{document}